\documentclass[11pt]{article}
\usepackage{tikz}
\usepackage{xcolor}
\usetikzlibrary{decorations.markings, calc}
\usepackage{mathtools,amssymb,amsthm}
\usepackage[utf8]{inputenc}
\usepackage{enumitem}
\usepackage{amsmath}
\usepackage{amssymb}
\usepackage[all]{xy}
\usepackage{simpler-wick}
\usepackage{lmodern}

\newcommand{\hO}{\hat{\mathcal{O}}}
\newcommand{\mH}{\mathcal{H}}
\newcommand{\ha}{\hat{a}}
\newcommand{\hel}{\hat{\ell}}
\newcommand{\hfa}{\hat{\mathfrak{a}}}
\newcommand{\hF}{\hat{\mathcal{F}}_{\Delta}}
\newcommand{\hh}{\hat{h}}
\newcommand{\appropto}{\mathrel{\vcenter{
  \offinterlineskip\halign{\hfil$##$\cr
    \propto\cr\noalign{\kern2pt}\sim\cr\noalign{\kern-2pt}}}}}

\newcommand{\mh}{\mathcal{H}}

\newcommand{\me}{\hat{\mathcal{E}}}

\newcommand{\be}{\begin{equation}}
\newcommand{\ee}{\end{equation}}

\global\long\def\bra{\langle}%

\global\long\def\dd{{\rm d}}%

\global\long\def\del{\mathcal{\delta}}%

\global\long\def\Del{\Delta}%

\global\long\def\da{\dagger}%

\global\long\def\mf{\mathcal{\mathcal{F}}}%

\global\long\def\ket{\rangle}%

\global\long\def\mh{\mathcal{\mathcal{H}}}%

\global\long\def\mo{\hat{\mathcal{O}}}%

\global\long\def\vp{\varphi}%

\global\long\def\te{\theta}%

\usepackage{mymacros}

\title{Geometry of Chord Intertwiner, Multiple Shocks and Switchback in Double-Scaled SYK}
\author{Sergio E. Aguilar-Gutierrez$^{a}$ and Jiuci Xu$^{b}$}
\affiliation[a]{Qubits and Spacetime Unit, Okinawa Institute of Science and Technology Graduate University,\footnote{\begin{CJK}{UTF8}{min}沖縄科学技術大学院大学\end{CJK}} 1919-1 Tancha, Onna, Okinawa 904 0495, Japan}
\affiliation{$^{b}$Department of Physics, University of California, Santa Barbara, CA 93106, USA.}

\emailAdd{sergio.ernesto.aguilar@gmail.com}
\emailAdd{Jiuci\_Xu@ucsb.edu}
\abstract{We revisit the bulk Hilbert space interpretation of chords in the double-scaled SYK (DSSYK) model and introduce a notion of intertwiner that constructs bulk states from states with fixed boundary conditions. This leads to an isometric map that factorizes the one-particle bulk Hilbert space into a tensor product of two boundary Hilbert spaces without particle insertion. The map enables a systematic derivation of a family of correlation functions with arbitrary finite amount of matter insertions, relevant for capturing the switchback effect—a feature of holographic complexity. We further develop a path integral framework that describes multiple shockwave configurations in the semiclassical limit. For the two-body scattering processes in semi-classical regime, we show it exhibits sub-maximal chaos at finite temperature, consistent with the scramblon dynamics associated with the ``fake disk'' geometry. The effective ``fake temperature'' governing this behavior emerges from the semiclassical limit of the quantum \( 6j \)-symbol associated with the out-of-time-order correlator.  We further analyze multi-shockwave configurations and derive precise conditions under which the switchback effect is realized, both in terms of the total chord number and the Krylov complexity of precursor operators. Our results clarify the structure of correlation functions with multiple operator insertions, their bulk interpretation in terms of shockwave geometries in the semi-classical regime, and provide a microscopic derivation of the switchback effect in the DSSYK model.}

\begin{document}

\maketitle

\section{Introduction}
Remarkable tests of the holographic dictionary have been enabled by the understanding of subregion duality, which asserts that a subregion of the boundary theory encodes complete information about a corresponding subregion of the bulk. The boundary observable algebra associated with such subregions thus plays an essential role in understanding the emergence of bulk spacetime. However, it is usually an arduous and possibly unfeasible task to evaluate arbitrary correlation functions of these operators in the boundary theory at strong coupling, without recurring to the bulk description in the weak gravity regime. However, lower-dimensional models can circumvent some of these technical challenges, making them valuable tools for inferring properties of the dual bulk gravity theory—even when the full gravitational description is not explicitly known. In particular, the double-scaled SYK (DSSYK) model provides a useful toy model for holography: it is ultraviolet (UV) finite, exhibits emergent gravitational features, and is analytically tractable to a significant extent~\cite{Berkooz:2018jqr,Berkooz:2018qkz}. For a modern overview of the model and its recent developments, see~\cite{Berkooz:2024lgq}. Significant progress in understanding the bulk interpretation of the auxiliary chord Hilbert space in this model—especially in the presence of matter insertions—and its associated operator algebra structure has been made through the work of Lin~\cite{Lin:2022rbf}, among others~\cite{Lin:2023trc,Okuyama:2024more,Okuyama:2024yya,Xu:2024gfm,Xu:2024hoc,Ambrosini:2024sre}.  It was also understood from these works that the chord rules, originally introduced as an auxiliary tool to solve the double-scaled SYK (DSSYK) model, naturally give rise to a notion of a bulk-to-boundary map. This map relates states prepared by chord operator insertions at the boundary of an open chord diagram to bulk states defined on the corresponding open slice. These bulk states are characterized by the number of open chords traversing the slice, which serves as a discrete measure of distance between the two boundaries. The inherent discreteness of this structure, as dictated by the chord rules, underlies the UV finiteness of the theory.

On the other hand, this notion of bulk-to-boundary map exists in Jackiw–Teitelboim (JT) gravity~\cite{JACKIW1985343,TEITELBOIM198341} coupled to conformal matter~\cite{Yang18,Kolchmeyer:2023gwa,Penington:2023dql} on a Euclidean disk, which is known to arise as the low-energy limit of DSSYK in the triple-scaling regime~\cite{Lin:2022rbf}. This parallel motivates a natural question: 
\begin{quote}
\emph{Does the bulk-to-boundary correspondence defined by the chord rules extend the JT gravity construction beyond the semi-classical regime, into the full parameter space of DSSYK?}    
\end{quote}
In this paper, we provide an affirmative answer to this question with the notion of chord intertwiner introduced in Sec. \ref{sec:chord Hilbert space}. This provides a concrete, analytically solvable, and UV finite example where the holographic dictionary allows us to gain access to observables in the quantum regime of the bulk theory.

There have also been several proposals to explain DSSYK holography beyond low energies (low temperatures). The ones with the most following identify the bulk dual of the DSSYK with two-dimensional de Sitter (dS$_2$) JT gravity at infinite temperature \cite{Susskind:2021esx,Susskind:2022bia,Lin:2022nss,Rahman:2022jsf,Rahman:2023pgt,Rahman:2024iiu,Rahman:2024vyg,Sekino:2025bsc} (see also \cite{Milekhin:2023bjv,Milekhin:2024vbb,Okuyama:2025hsd,Aguilar-Gutierrez:2024oea,Yuan:2024utc}); dS$_3$ space \cite{Narovlansky:2023lfz,Verlinde:2024znh,HVTalks1,HVTalks2,HVTalks3,HVTalks4,Verlinde:2024zrh,Gaiotto:2024kze,Tietto:2025oxn,Aguilar-Gutierrez:2024nau}; and sine-dilaton gravity \cite{Blommaert:2023opb,Blommaert:2024ydx,Blommaert:2024whf,Blommaert:2025avl,Blommaert:2025rgw} (see also \cite{Almheiri:2024xtw,Mahapatra:2025fpx,Aguilar-Gutierrez:2025pqp}). However, none of the existing approaches have, so far, explicitly reproduced correlation functions involving arbitrary matter insertions with generic boundary conditions\footnote{Here, by boundary conditions, we refer to the conditions imposed on the boundaries of chord diagrams associated with the corresponding correlation function.}. Nor have they provided a concrete dictionary between the observable algebra of DSSYK (often referred to as the double-scaled algebra~\cite{Lin:2022rbf,Xu:2024hoc}) and the corresponding bulk gravitational observable algebras.
However, recently, several works \cite{Blommaert:2024ydx,Narovlansky:2023lfz,Goel:2023svz,Blommaert:2025avl,Blommaert:2024whf,Aguilar-Gutierrez:2025pqp} have realized that path integral methods are promising tools to describe the emergent bulk geometry in the proposed holographic dual theories of the DSSYK model. While these studies have not addressed general situations with arbitrary matter insertions in the bulk Hilbert space, there are natural extensions that allow for such generalizations, as we will discuss in this paper. 

Furthermore, there has been important progress in describing the emergence of dynamical spacetime with matter content from Krylov space methods (see \cite{Nandy:2024htc,Baiguera:2025dkc} for modern reviews on Krylov operator \cite{Parker:2018yvk} and state \cite{Balasubramanian:2022tpr} complexity). Explicit analytical and numerical expressions for Krylov operator and state complexity for a particle state at infinite temperature have been derived in \cite{Ambrosini:2024sre}. The authors previously showed that Krylov complexity is equal to a bulk geodesic in two-dimensional anti-de Sitter (AdS$_2$) black hole at infinite temperature \cite{Rabinovici:2023yex} (even though JT gravity is only supposed to describe the low temperature regime of the DSSYK). Meanwhile, several dynamical observables involving chord matter operators—such as those related to operator spreading—have been studied in~\cite{Xu:2024gfm}. However, most of the developments in Krylov complexity have focused on single matter insertions and are limited to the triple-scaling regime corresponding to JT gravity (with the exception of~\cite{Aguilar-Gutierrez:2025pqp}). Another relevant work, \cite{Berkooz:2022fso} is the first study of shockwaves from correlation functions with matter chord insertions in DSSYK. That analysis, however, was restricted to the triple-scaling limit and appeared before the development of the double-scaled algebra and the chord Hilbert space formalism in~\cite{Lin:2022rbf}. In this work, we extend the study of Krylov complexity and dynamical evolution in DSSYK to include multiple matter insertions and explore regimes beyond the low-energy semi-classical limit

\paragraph{The switchback effect in Krylov complexity}An exciting prospect in understanding Krylov complexity in the DSSYK model and its bulk dual has been highlighted by \cite{Heller:2024ldz}. There are several conjectures regarding the bulk duals of quantum complexity measures defined in boundary theories~\cite{Susskind:2014rva,Stanford:2014jda,Brown:2015bva,Brown:2015lvg,Couch:2016exn,Belin:2021bga,Belin:2022xmt,Myers:2024vve} (see~\cite{Baiguera:2025dkc,Chapman:2021jbh} for recent reviews). These conjectures are motivated by the holographic dictionary, which posits that gauge-invariant observables in the boundary theory should have corresponding realizations as bulk geometric quantities. 
The most updated version of these conjectures, Complexity=Anything (CAny) \cite{Belin:2021bga,Belin:2022xmt,Myers:2024vve}, state that the characteristic of holographic complexity should be (i) it grows linearly at late times, and (ii) it displays a switchback effect, which refers to a decrease in complexity due to a cancellation between early and late time backreaction in the bulk, modeled by several shockwave pulses inserted in an alternating order from the boundary into the bulk. Given that~\cite{Rabinovici:2023yex,Heller:2024ldz} established a concrete correspondence between bulk geodesics and Krylov complexity in the absence of matter insertions, it is natural to expect a similar structure to emerge in the presence of matter chords, as suggested by the findings of~\cite{Xu:2024hoc,Ambrosini:2024sre}. The authors \cite{Ambrosini:2024sre} found an imprint of the switchback effect by analysis the scrambling properties of Krylov operator complexity in the one-particle space of the DSSYK model. However, their study does not include the timefold insertion of arbitrarily many precursor operators, which conditions used to define the switchback effect. This type of analysis has also been improved in \cite{Aguilar-Gutierrez:2025pqp}, which shows that the Hartle-Hawking (HH) state of the DSSYK with single or composite operator insertions indeed describes a bulk geometry (in sine-dilaton gravity) with matter backreaction in a given profile. However, the effect of non-composite operators (see e.g. Fig. \ref{fig:otocprecursor}, \ref{fig:6pointchord}, \ref{fig:2n2_amplitude}), which is necessary to study the switchback effect, has yet to be explored.

\subsection{Purpose of the work}In the light of the lack of a comprehensive study of multi-particle correlation functions in the DSSYK model, and due to the absence of a proof of the switchback effect in Krylov operator complexity, the main goals of our work are:
\begin{itemize}
    \item We study correlation functions involving multiple chord matter insertions that intersect in the bulk. To analyze these configurations, we introduce a chord intertwiner that constructs bulk states defined on a slice from states defined with fixed boundary conditions. Using this intertwiner, we explicitly derive a family of correlation functions by computing inner products between the corresponding bulk states.
    \item We explore the bulk description of the DSSYK model with arbitrary matter chord insertions beyond the triple-scaling limit and identify observables that exhibit the switchback effect. We then investigate a possible complexity-theoretic interpretation of these observables.
\end{itemize}

\noindent  The notion of an intertwiner is motivated by the discussions in~\cite{Yang18,Iliesiu:2024cnh}, where states on a given bulk slice can be constructed by contracting boundary states with interior states that are independent of specific boundary conditions. We show that this structure naturally arises from the chord rules, and that a discrete analogue of this construction holds in the chord Hilbert space. Building on this structure, we demonstrate that the evaluation of a chord diagram with multiple matter insertions can be decomposed into contractions of boundary states (specified by boundary conditions) with interior states that contain the bulk matter crossings. This framework enables us to compute a family of correlation functions involving an arbitrary number of matter insertion operators with fixed boundary energies\footnote{To be more concrete, these are $2m+2, m=1,2,\dots$ correlation functions in DSSYK that corresponds to chord diagrams that contains $m$ matter crossings. They are relevant for later discussions of switchback effect. }, and show that the results factorize into products of crossed four-point functions, with appropriate normalization factors. Our integral expressions are valid for all $q=e^{-\lambda}\in[0,1)$ (the q-deformation parameter of the DSSYK \cite{Berkooz:2018jqr,Berkooz:2018qkz}).  Analytic results are recovered in the semiclassical ($\lambda\rightarrow0^+$) regime. Remarkably, the decomposition of correlation functions indicates spatial connectivity between Euclidean-signature bulk subregions even beyond this limit. Since DSSYK is fully determined by its boundary operator algebra, our finding—that bulk states living on a slice associated with a larger bulk patch can be constructed from correspondingly larger boundary subregion—suggests that a notion of subregion–subalgebra duality persists beyond the semiclassical regime\footnote{The intuition for this relation is supported by the chord diagram representation, as illustrated, for example, in~\eqref{eq:intertwine-0}. However, one might be suspicious of the notion of a subalgebra in DSSYK, since the boundary circle of a chord diagram does not exhibit the usual notion of locality found in conventional boundary CFTs. Nonetheless, several works have attempted to formulate the theory in terms of a boundary propagating particle~\cite{Berkooz:2022mfk,Berkooz:2024ifu}, suggesting that a notion of locality suitable for defining subalgebras may exist in this context. While a precise formulation of this idea lies beyond the scope of the present paper, our results indicate it as a promising direction for future investigation.} and extends into a regime where the bulk geometry is intrinsically quantum and discrete, in line with the perspective proposed in~\cite{Gesteau:2024rpt}. We leave this as a promising direction for future investigation.

For the second goal, we take a seemingly different but connected approach. While the chord diagrams allow us to derive formal correlation functions valid $\forall q\in[0,1)$. We are also interested in the emergent smooth bulk geometry encoded by these observables in the $\lambda\rightarrow0^+$ limit. 
In particular, we are interested in solutions that model shockwaves from one \cite{Shenker:2013pqa} or several particle insertions \cite{Shenker:2013yza}. For this reason, we deduce the saddle point solutions in canonical variables (i.e. the classical phase space) of the path integral of the DSSYK model with one-particle insertions. The path integral we consider prepares the so-called ``wormhole'' density matrices, introduced by \cite{Berkooz:2022fso}, albeit now, in the bulk formulation of the chord Hilbert space by \cite{Lin:2022rbf}. We will show that these semiclassical solutions have the same evolution as the two-sided geodesic lengths in an effective \footnote{It is qualitatively similar, but differs in an important way that gives rise to submaximal chaos at finite temperature.} AdS$_2$ black hole background, crucially without the triple-scaling limit.\footnote{The \emph{triple-scaling limit} refers to taking \( \lambda \to 0 \) while keeping the combination \( \rme^{-\lambda n}/\lambda^2 \) fixed, where \( n \) is an eigenvalue of the total chord number operator (see, e.g.,~\cite{Lin:2022rbf}). In this limit, one typically holds \( \Delta = \mathcal{O}(1) \) and zooms into the edge of the energy spectrum. As a result, the energy spectrum of the limiting theory becomes unbounded and spans the entire real line. By contrast, the \emph{semiclassical limit} is defined by holding \( \lambda n \) fixed as \( \lambda \to 0 \), while also keeping \( \lambda \Delta \) fixed—corresponding to the inclusion of heavy matter operators. In this regime, one typically rescales the Hamiltonian (e.g.~\eqref{eq:H many particles}) so that the energy remains bounded. Importantly, the semiclassical limit \emph{contains} the triple-scaling limit as a special case: it corresponds to zooming into the spectrum edge along with taking a probing limit where \( \lambda \Delta \to 0 \). }  This allows us to determine how the solutions in the bulk dual of the DSSYK model differ from those in the existing literature on shockwave geometries in AdS black holes (e.g. \cite{Shenker:2013pqa,Shenker:2013yza})
\begin{quote}
\emph{Consider the semiclassical limit where $q\rightarrow1^-$. How is the microscopic structure of chords encoded in the dual shockwave geometry? Are there characteristics that differentiate it from the shockwaves in JT gravity on a disk?}
\end{quote}
We find that a combination of $q$ and the conformal dimension of the particle chord characterizes the backreaction of the solution (as illustrated in Tab. \ref{tab:dictionary} and in the main text (\ref{eq:alpha sw})). We also encounter notable differences in the shockwave solutions compared to JT gravity \cite{Shenker:2013pqa,Berkooz:2022fso}. In particular, we do not need to restrict the analysis of closed form semiclassical solutions to perturbations of the thermofield double (TFD) states (which was originally considered by \cite{Shenker:2013pqa}). Similar to other studies on the DSSYK \cite{Susskind:2021esx,Narovlansky:2023lfz,Blommaert:2024ydx} we find a ``fake'' temperature that controls the energy dependence of the saddle points. This allows us to compare the growth of crossed four-point functions computed from the path integral with the chaos bound \cite{Maldacena:2015waa}. Our result is consistent with earlier findings that DSSYK exhibits a submaximal chaos. \cite{Lin:2023trc,Blommaert:2024ydx}. Moreover, the ``fake temperature'' extracted from the semiclassical limit of the quantum \( 6j \)-symbol agrees with the scramblon analysis in~\cite{Lin:2023trc}, thereby motivating an interpretation of scramblon dynamics as shockwave scattering in a corresponding ``fake geometry''.

After deducing the bulk shockwave interpretation for the DSSYK with matter chords, we move on to analyze multiple-shock geometries \cite{Shenker:2013yza} by introducing many precursor operator insertions in the boundary side. We illustrate the bulk geometry in Fig. \ref{fig:multi_shock}. This allows us to answer a natural question motivated by \cite{Xu:2024gfm,Ambrosini:2024sre,Heller:2024ldz,Aguilar-Gutierrez:2025pqp} in our second goal above:
\begin{quote}
    \emph{Is there any observable in the DSSYK model that displays a switchback effect in the sense of \cite{Stanford:2014jda}?}
\end{quote}
We show that under the timefold conditions (introduced in the context of holographic complexity by \cite{Stanford:2014jda}), namely, when the operator insertion is performed at very early and late times in an alternating order, the expectation value of the total chord number evaluated on a chord state with $m$-particle insertions can be expressed as a sum of correlation functions with one or zero-particle insertions. This technical simplification allows us to show that the total chord number displays the same switchback effect as AdS$_2$ bulk geodesics lengths under shockwave perturbations \cite{Stanford:2014jda}.

Later, we connect our results with the developments in the Krylov complexity literature detailed above. Motivated by a future prospect in \cite{Heller:2024ldz}, we ask
\begin{quote}
    \emph{Is Krylov complexity $=$ holographic complexity in this model?}
\end{quote}
We answer this affirmatively. Our results show that the total chord number in the semiclassical limit can be expressed as the sum of the Krylov operator complexity for each of the precursor operators (in the Heisenberg picture), and with additional cancellation the spread complexity of the HH state evolving between the precursors. We extend the definition of Krylov complexity for multiple operator insertions, so that it obeys an additive property similar to those encountered in the Complexity=Volume (CV) conjecture \cite{Susskind:2014rva} (and more generally in the CAny conjectures \cite{Belin:2021bga,Belin:2022xmt,Myers:2024vve}). The results indicate that this notion of Krylov complexity is holographic complexity, in the sense that it displays the switchback effect; while previous works have shown that Krylov operator complexity without precursors lead to late time linear growth \cite{Ambrosini:2024sre,Aguilar-Gutierrez:2025pqp}.
\begin{table}
    \centering
    \begin{tabular}{|p{7.5cm}|p{7.5cm}|}\hline
    \textbf{Bulk}\vspace{0.2cm} &\textbf{Boundary}\vspace{0.2cm}\\\hline
    {AdS$_2$ shockwave geometry}&$\hat{\rho}_{\Delta}$ (\ref{eq:rho LR}): ``Wormhole'' density matrix\vspace{0.1cm}\\
    $u_{L/R}$: Asymptotic boundary time &$t_{L/R}$: DSSYK Hamiltonian time\vspace{0.1cm}\\   
    $L_{\rm AdS}(u_L,u_R)$ (\ref{eq:AdS shockwave length}): Single shock two-sided geodesic distance&$\lambda\expval{\hat{N}(t_L,t_R)}$ (\ref{eq:shockwave approx}): Expectation value of the total chord number \vspace{0.1cm}\\
$1/\beta_{\rm fake}$: Black hole temperature&$\tfrac{J\sin\theta}{\pi}$ (\ref{eq:fake temperature}): Fake temperature\vspace{0.1cm}\\
$L_{\rm reg}$: Length regularization constant&$\ell_*(\theta)=\lambda \expval{\hat{N}(0,0)}$ (\ref{eq:length0 Delta many}): Initial total chord number\vspace{0.1cm}\\
$\alpha_{\rm sw}$: Shockwave shift parameter&$\alpha_i=\frac{1}{2}\qty(1-\frac{q^{\Delta_i}\rme^{-\ell_*(\theta)}}{\sin^2\theta})\rme^{2J\sin\theta \abs{t_i}}$ (\ref{eq:alpha sw})\vspace{0.1cm}\\

(\ref{eq:switchback property0}): Geodesic distance with multiple shocks&\eqref{eq:total chord number shocks}: Total chord number with multiple precursor operator insertions\\\hline
\end{tabular}
\caption{Holographic dictionary between the DSSYK model and its dual geometry. $t_i$ is the insertion time of a precursor operator (taken $\abs{t_i}\gg \abs{t_{L/R}}$ to generate shockwaves); $\theta$ is the parametrization of the DSSYK energy spectrum $E(\theta)$, $\Delta$ the conformal dimension of a matter chord operator $\hat{\mathcal{O}}_\Delta$, $J$ is a coupling in the DSSYK Hamiltonian (\ref{eq:H many particles}).}\label{tab:dictionary}
    \end{table}

\paragraph{Organization of the paper} In Sec. \ref{sec:chord Hilbert space} we revisit the bulk Hilbert space of DSSYK and explore its structure using the chord intertwiner, which enables the construction of bulk states with or without matter insertions from boundary states. In Sec. \ref{sec:single shock}, we set up a path integral formalism associated with dynamical evolution induced by the left/right chord Hamiltonians with one-particle insertion. We show its saddle point solution describes Krylov operator complexity growth of a wormhole density matrix in the semiclassical limit. We match it to a geodesic length in an effective AdS$_2$ black hole shockwave geometry, which we relate to the ``fake'' disk in \cite{Lin:2023trc}, and we derive the holographic dictionary for the DSSYK model with multiple operator insertions. In Sec. \ref{sec:mutiple shocks} we generalize the results with two and arbitrarily more particle chords. We show there is a switchback effect in the total chord number, and we express it in terms of Krylov complexity for the precursor operators and the HH state. It is manifested in the bulk as a length geodesic in a multiple shockwave background. In addition, we summarize our results and point out future directions, in Sec. \ref{sec:discussion}. 

We include App. \ref{app:notation} to summarize the notation in the manuscript; App. \ref{app:recursion} with  technical details for the evaluation of the six-point correlation function in the main text, and App. \ref{app:q-calculus} for additional technical details on the $q$-calculus employed in Sec. \ref{sec:chord Hilbert space}

\section{Revisiting the Bulk Hilbert Space of Chords}\label{sec:chord Hilbert space}
In this section we work on correlation functions with arbitrarily many operator insertions. We express correlation functions involving multiple sub-diagrams in terms of simpler factors by identifying wavefunctions associated to subregions in the chord diagram which intercept at a given location. This allows to express these observables in an integral form that contains chord diagrams associated to each subregion, in terms of zero and one-particle states.

\paragraph{Outline}In Sec. \ref{ssec: trinagle identity} we explain our definition of an intertwining wavefunction, illustrated in the simplest case where there are no matter chords. In Sec. \ref{eq:chord fusion} we explore the chord intertwining with a one-particle chord insertion. In Sec. \ref{ssec:fixed energy wavefunctions}, we generate intertwiners using the energy basis. This allows for a simpler and more convenient evaluation of the chord diagram correlation functions. In Sec. \ref{ssec:comments factorization} we explain how to incorporate a map relating one-particle and the zero-particle states to evaluate crossed four-point functions in the . In Sec. \ref{eq:crossed chord diagram}, we apply some of our results to deduce the integral representation of diagrams with two or more particle insertions. At last, in Sec. \ref{sec:chord blocks} we derive a family of six and higher-point correlation functions with crossed four-point  function sub-diagrams.

\subsection{Triangle Identity of Chords}\label{ssec: trinagle identity}
A state in the chord Hilbert space \(\mathcal{H}_0\) can be viewed in two equivalent ways: either as a bulk state defined on a spatial slice containing a fixed number of \(H\)-type chords, or as a boundary state labeled by definite energy. The former is naturally specified by the chord number eigenstate \(|n)\), while the latter is characterized by the energy eigenstate \(|\theta)\). We adopt the convention of using round brackets \((\cdot|\cdot)\) to denote inner products in \(\mathcal{H}_0\), distinguishing them from states with multiple types of chords in later sections.

The overlap between the chord number basis and energy eigenbasis is given by the \(q\)-Hermite polynomials:
\be\label{eq:wavefunctions E basis}
\phi_{E(\te)}\left(n\right)=\frac{(n|\te)}{\sqrt{\left(q;q\right)_{n}}}=\frac{H_{n}\left(\cos\te|q\right)}{\left(q;q\right)_{n}},\quad (a;q)_n:=\prod_{k=0}^{n-1}(1-aq^k)~,
\ee
where the normalization is chosen such that the chord number states are orthonormal:
\be
(n | m) = \delta_{mn},
\ee
while the energy eigenstates are normalized with respect to the \(q\)-deformed measure:
\be\label{eq:zero particle identities}
(\theta_1 | \theta_2) = \frac{\delta(\theta_1 - \theta_2)}{\mu(\theta_1)}, \quad \mu(\theta) = \frac{(e^{\pm 2i\theta}, q; q)_\infty}{2\pi}, \quad \hH|\theta) = E(\theta)|\theta).
\ee
Here, \(\hH\) denotes the chord Hamiltonian, which acts by either adding or removing an \(H\)-type chord. It is normalized such that the energy eigenvalue is given by:
\be\label{eq:conserved energies}
E(\theta) = \frac{2J\cos \theta}{\sqrt{\lambda(1 - q)}}.
\ee
The spectral measure \(\mu(\theta)\) plays the role of a normalized \(q\)-deformed Gaussian distribution. We adopt the shorthand:
\be\label{eq:measure theta}
\int d\mu(\theta) := \int_0^\pi \mu(\theta)\, d\theta = 1,
\ee
which we use when integrating over intermediate energy states. As an analogue to pure JT gravity, the chord number in the double-scaled SYK model can be interpreted as representing the (discrete) geodesic length between two boundary points, while the parameter \(\theta\) labels the boundary energy. Consequently, we can interpret the wavefunction \(\phi_E(n)\) as describing a quantum state defined on a bulk slice with fixed energy on the asymptotic boundary and fixed chord number on the geodesic segment anchoring that boundary interval. In the JT gravity context, such a wavefunction corresponds to a gravitational path integral over geometries bounded by the specified boundary conditions.

In JT gravity, suppose we are given two such wavefunctions: one associated with a geodesic segment connecting boundary points 1 to 2, and another connecting points 2 to 3. By gluing along the asymptotic boundaries at point 2, we can ``\emph{intertwine}'' the two geometries into a new one that prepares a state defined on the slice from point 1 to point 3. A discrete version of this identity holds in the double-scaled SYK model, and it is expressed as:
\be \label{eq:fusion-1}
(q;q)_{m_{13}}\delta(E_1-E_2) \phi_{E_1}(m_{13}) = \sum_{m_{12}, m_{23} = 0}^{\infty} \textcolor{blue}{\phi_{E_1}}(m_{12}) \textcolor{red}{\phi_{E_2}}(m_{23}) I_{m_{12}, m_{23}, m_{13}},
\ee
where the coefficient \(I_{m_{12}, m_{23}, m_{13}}\) encodes the overlap or the \emph{intertwining} correlation functions between the intermediate states. Diagrammatically, this identity is represented as:
\begin{equation} \label{eq:intertwine-0}
\begin{tikzpicture}
    % Define radius
    \def\r{2}

    % Fill the disk
    \fill[gray!20] (-\r,0) arc[start angle=-180,end angle=0, radius=\r];

    % Draw the circle
    \draw[thick] (-\r,0) arc[start angle=-180,end angle=0, radius=\r];

    % Boundary points
    \coordinate (A) at (-\r,0);
    \coordinate (B) at (\r,0);
    \coordinate (O) at (0,0);
    \coordinate (M) at (0,0);
    \coordinate (X) at ($(\r*cos{30},-\r*sin{30})$);

    % Define marked points
    \coordinate (M1) at (-.45,-.8);
    \coordinate (M2) at (-2.2,0);
    \coordinate (M3) at ($(1.1*\r*cos{60},-1.1*\r*sin{60})$);
    \coordinate (M4) at (2.2,0); 
    \coordinate (M23) at (1.5, -.8);
    \coordinate (M13) at (0,0.2);
    \coordinate (I) at (.5, -0.5 );

    %Draw Geodesics
        \draw[thick,blue] (A) arc[start angle=90,end angle=30, radius=2*1.73];
        \draw[thick,red] (B) arc[start angle=120,end angle=180, radius=2];   
        \draw[thick] (A)--(B);
    % Draw nodes
    \node at (M1) {$m_{12}$};
    \node at (M2) {$1$};
    \node at (M3) {$2$};
    \node at (M4) {$3$};
    \node at (M23) {$m_{23}$};
    \node at (M13) {$m_{13}$};
    \node at (I) {$I$};
\end{tikzpicture}
\end{equation}
The kernel $I$ can be understood as the scattering amplitude from two universes into one universe \cite{Yang18}. Its value can be calculated explicitly as:
\be \label{eq:I-def-sum}
I_{m_{1},m_{2},m_{3}}	=\sum_{\{k_{ij}\}}\prod_{i=1}^{3}\binom{m_{i}}{k_{i1},k_{i2},k_{i3}}\prod_{ij\in\{12,13,23\}}\left(q;q\right)_{k_{ij}},
\ee
where the sum involves symmetric matrices $k_{ij}$ with $k_{ii}=0$ and $\sum_{j=1}^3 k_{ij}= m_i$.  The identity above can be derived by evaluating the overlap of wavefunctions on both sides of equation~\eqref{eq:fusion-1}. This calculation involves the Ismail–Stanton–Viennot integral~\cite{Ismail:1987}:
\be \label{eq:I-def}
I_{m_1,m_2,m_3} = \int^{\pi}_0 \dd \mu(\theta) \prod_{i=1}^{3} H_{m_i} (\cos \te|q),
\ee
which computes the triple product of $q$-Hermite polynomials over the energy spectrum. To make contact with thermodynamic intuition, we introduce a thermal state $|\beta)$ via
\be\label{eq:HH state beta}
|\beta) = e^{-\beta H}|\Omega) = \int\dd\mu(\te) e^{-\beta E(\te)}|\te),
\ee
where $|\Omega)$ is the chord empty state and $E(\te)$ is the energy eigenvalue. Defining the thermal wavefunction
\begin{equation}\label{eq:thermal wave}
    \phi_\beta(n) := (n|\beta)=\int^{\pi}_0 \dd \mu(\theta) \rme^{-\beta E(\theta)}\frac{H_{n}\left(\cos\te|q\right)}{\sqrt{\left(q;q\right)_{n}}}~,
\end{equation}
we can re-express equation~\eqref{eq:fusion-1} in the following convolution form:
\be
\phi_{\beta_1 +\beta_2}(m_{13}) = \sum_{m_{12}, m_{23} = 0}^{\infty} \phi_{\beta_1}(m_{12})\, \phi_{\beta_2}(m_{23})\, I_{m_{12}, m_{23}, m_{13}},
\ee
which admits a natural geometric interpretation: it describes the intertwining of two thermal segments with inverse temperatures $\beta_1$ and $\beta_2$ into a longer thermal system with total length $\beta_1 + \beta_2$, joined at the intermediate point $2$.

Note that the $q$-Hermite polynomials describe fixed-energy ``microstates'', with $\dd\mu(\theta)$ serving as the spectral energy density. The expression~\eqref{eq:I-def} can thus be interpreted as defining an interior GHZ-like \cite{Greenberger:1989tfe} state for three bulk slices:
\be\label{eq:first instance}
|I\rangle = \int \dd\mu(\theta)\, |\theta, \theta, \theta),
\ee
where the same energy eigenstate is assigned to all three legs of the trivalent junction. 

{{In the following, let us introduce the notation 
$$|x_1,x_2,\dots x_n):=|x_1)\otimes|x_2)\otimes\dots\otimes|x_n)$$ for either the energy or chord basis in the zero particle chord space, such that the inner product $(y_1,y_2,\dots y_n|x_1,x_2,\dots x_n)=(y_1|x_1)(y_2|x_2)\dots (y_n|x_n)$, so that the inner product is determined by the zero particle chord space.}} As a direct application of the viewpoint {{in \eqref{eq:first instance}}}, we can compute the DSSYK partition function as:
\be\label{eq:DSSYK Z}
Z_{\mathrm{DSSYK}}(\beta) = \sum_{n_1, n_2, n_3 = 0}^{\infty} \phi_{\beta_1}(n_1)\, \phi_{\beta_2}(n_2)\, \phi_{\beta_3}(n_3)\, I_{n_1, n_2, n_3},
\ee
with total inverse temperature $\beta = \beta_1 + \beta_2 + \beta_3$. This expression can be understood as the overlap between the tripled thermal state and the entangled state $|I\rangle$, and is naturally represented by the following diagram:
\be
Z_{{\rm DSSYK}}(\beta)= \begin{tikzpicture}[scale=0.6,baseline={([yshift=-.5ex]current bounding box.center)}]
% Define radius and key points on the circle
\def\r{2}
% Draw and fill the disk
\draw[thick] (0,0) circle (\r);
% Define boundary points using angles
\coordinate (A) at ({\r*cos(90)}, {\r*sin(90)}); 
\coordinate (B) at ({\r*cos(210)}, {\r*sin(210)}); 
\coordinate (C) at ({\r*cos(330)}, {\r*sin(330)}); 
\coordinate (A1) at ({\r*cos(30)}, {\r*sin(30)}); 
\coordinate (B1) at ({\r*cos(150)}, {\r*sin(150)}); 
\coordinate (C1) at ({\r*cos(270)}, {\r*sin(270)}); 
\draw[thick] (A) arc[start angle=0, end angle= -60, radius= \r*1.73];
\draw[thick] (B) arc[start angle=120, end angle= 60, radius= \r*1.73];
\draw[thick] (C) arc[start angle=240, end angle= 180, radius= \r*1.73];
% Label boundary points
\node[above right] at (A1) {$\beta_{1}$};
\node[above left] at (B1) {$\beta_{2}$};
\node[below] at (C1) {$\beta_3$};
\node at (0,0) {$I$}; 
\end{tikzpicture}~.
\ee
In the triple-scaled limit:
\be
q=e^{-\lambda}~, \quad \lambda\to0~,\quad n_{ij}\to\infty~,\quad\lambda n_{ij}=\lambda^{2}e^{-l_{ij}}~,\quad\te=\lambda k~.
\ee
The continuum limit of \eqref{eq:I-def} becomes the corresponding interior state of pure JT gravity on the disk \cite{Lin:2022rbf, Xu:2024hoc}:
\be\label{eq:cont version of In1n2n3}
I\left(n_{12},n_{23},n_{13}\right)\stackrel{n_{ij}\to l_{ij}}{\Longrightarrow}\int_{0}^{\infty}dk~k\sinh(2\pi k)K_{2ik}\left(2e^{-l_{12}/2}\right)K_{2ik}\left(2e^{-l_{13}/2}\right)K_{2ik}\left(2e^{-l_{23}/2}\right),
\ee
{{where $K_{n}(x)$ is the modified Bessel functions of second kind and \eqref{eq:cont version of In1n2n3}}} is the value of which calculates the volume of hyperbolic triangle \cite{Iliesiu:2024cnh}:
\be
I(l_{12},l_{23},l_{13})=\exp(-2e^{-(l_{12}+l_{23}+l_{13})/2}(e^{l_{12}}+e^{l_{23}}+e^{l_{13}})).
\ee

\subsection{Intertwiners and Matter Operators}\label{eq:chord fusion}
We revisit matter chord operators in this section, and we discuss their geometric interpretation from the intertwining property of last section. There are two notions of matter chord operators: the closed ones are inserted as a bi-local operator from the boundary point of view and put weights on state on the slice according to their total $H$-chord number. The open matter chord operator creates an open matter chord to the state, transforming an $n$-particle state into $n+1$ one. We will discuss them separately below.

\paragraph{The closed matter chord operator} This operator can be understood as inserting a closed matter chord horizontally that intersects with all open chords living on the slice. Therefore, if a state has definite amount of open $H$-chords, it acts simply as:
\be
q^{\Delta \hat{n}} |n) = q^{\Delta n}|n).
\ee
However, usually the state on the slice is not a fixed chord number state but instead prepared by fusing states with certain number of chords on the bulk slice with the same boundary energy. Such states encode the information of matter correlation at fixed energies where the correlation functions can be reproduced by the action of $q^{\Delta \hat{n}}$ on them. To see exactly how this works, we introduce the state $\mf_{mn}$ as:
\be\label{eq:Fmn state}
|\mf_{mn})	=\int\dd\mu\left(\te\right)|\te)(\te,\te|m,n),
\ee
which can be viewed as the state on the (13) slice of \eqref{eq:fusion-1} through the intertwining of state on (12) and (23) with the same energy. Note that it is a superposition of fixed length state:
\be\label{eq:coefficients extra}
|\mf_{mn})=\sum_{k=0}^{{\rm min}(m,n)}\frac{\left(q;q\right)_{m}\left(q;q\right)_{n}}{\left(q;q\right)_{m-k}\left(q;q\right)_{n-k}(q;q)_{k}}|m+n-2k)=:\sum_{k=0}^{{\rm min}(m,n)}\left[\begin{array}{c}
m,n\\
k
\end{array}\right]_{q}|m+n-2k),
\ee
{{where $$\begin{bmatrix}m, n\\k\end{bmatrix}_{q}\equiv\frac{(q;q)_{m}(q;q)_n}{(q;q)_{m-k}(q;q)_{n-k}(q;q)_{k}} . $$ }~ } and the coefficients in {{\eqref{eq:coefficients extra}}} are related to the components interior state $I$ as:
\be\label{eq:I coefficients}
I_{n_{1},n_{2},n_{3}}=(n_{3}|\mf_{n_{1}n_{2}})\prod_{i=1}^{3}\left(q;q\right)_{n_{i}}^{1/2}.
\ee
The state $|\mf_{mn})$ is a superposition of fixed chord number states that serve as remaining open chords that have not been contracted from the two sides. The corresponding coefficients in front of the state calculates the $q$-weights that accounts for contracting $k$ chords out of the $(m,n)$ left/right $H$ chords. Explicitly, we have:
\be
|\mf_{mn})=\sum_{k=0}^{{\rm min}(m,n)}\frac{\left(q;q\right)_{m}\left(q;q\right)_{n}}{\left(q;q\right)_{m-k}\left(q;q\right)_{n-k}(q;q)_{k}}|m+n-2k)=:\sum_{k=0}^{{\rm min}(m,n)}\left[\begin{array}{c}
m,n\\
k
\end{array}\right]_{q}|m+n-2k).
\ee
Now we examine the action of the closed $M$-chord operator, for which we define:
\be
|\mf_{mn}^{\Del})=q^{\Del\hat{n}}|\mf_{mn}).
\ee
Note that $|\mf_{mn})$ is a state living on the bulk slice, where the action of the closed $M$-chord simply weights the state according to the total number of open chords it contains. However, it is more insightful to understand its action from the boundary perspective, where the state $\mf_{mn}$ is constructed by fusing two states with fixed boundary energies. In this context, we ask how the state $|\mf^{\Delta}_{mn})$ relates to the original basis $|\mf_{m^\prime n^\prime})$. To answer this, we introduce coherent states $|s)$:
\be\label{eq:DS coherent state}
|s)=\sum_{n=0}^{\infty}\frac{s^n}{\sqrt{(q;q)_n}} |n),\quad (\te|s)= \frac{1}{(se^{i \theta},s e^{-i\theta};q)_\infty} =: \frac{1}{(se^{\pm i \theta};q)_\infty}.
\ee
where we use the shorthand notation $(a_1,a_2;q)_\infty = (a_1;q)_\infty (a_2;q)_\infty$. Using this, we define a coherent version of the $\mf$-state as:
\be \label{eq:def-F-no-matter}
|\mf_{st})=\sum_{n,m}\frac{s^{m}t^{n}}{\sqrt{\left(q;q\right)_{m}\left(q;q\right)_{n}}}|\mf_{mn})=\int\dd\mu\left(\te\right)\frac{1}{\left(se^{\pm i\te},t e^{\pm i\te};q\right)_{\infty}}|\te).
\ee
Now we consider the state 
\begin{equation}\label{eq:qDelta F coherent}
|\mf_{st}^{\Del})=q^{\Del\hat{n}}|\mf_{st})~,    
\end{equation}
and compute its overlap with a fixed-energy boundary state $|\phi)$ \eqref{eq:AW integral}:
\be
(\phi|\mf_{st}^{\Del})=\int\dd\mu\left(\te\right)\frac{\left(q^{2\Del};q\right)_{\infty}}{\left(q^{\Del}e^{\pm i\te\pm i\phi},se^{\pm i\te},te^{\pm i\te};q\right)_{\infty}}=\frac{\left(q^{2\Del}st;q\right)_{\infty}}{\left(q^{\Del}se^{\pm i\phi},q^{\Del}te^{\pm i\phi},st;q\right)_{\infty}},
\ee
which implies that
\be \label{eq:def-F-matter}
|\mf_{st}^{\Del})=\int\dd\mu\left(\te\right)\frac{\left(q^{2\Del}st;q\right)_{\infty}}{\left(st,q^{\Del}se^{\pm i\te},q^{\Del}te^{\pm i\te};q\right)_{\infty}}|\te).
\ee
Comparing \eqref{eq:def-F-matter} with \eqref{eq:def-F-no-matter}, and expanding both in powers of $s^m t^n$, we find:
\be \label{eq:F-relation}
|\mf_{mn}^{\Del})=\sum_{k=0}^{{\rm min}(m,n)}q^{\Del(m+n-2k)}\frac{\left(q^{2\Del};q\right)_{k}}{\left(q;q\right)_{k}}|\mf_{m-k,n-k}),
\ee
where we have used the identity:
\be 
\frac{\left(q^{2\Del}st;q\right)_{\infty}}{\left(st;q\right)_{\infty}}=\sum_{k=0}^{\infty}\frac{\left(q^{2\Del};q\right)_{k}}{\left(q;q\right)_{k}}s^{k}t^{k}.
\ee
The relation \eqref{eq:F-relation} matches exactly the combinatoric factor derived in \cite{Berkooz:2018jqr} accounting for the number of $H$-chords entering the $M$-region, from which the boundary matter correlator can be extracted. In our current formulation, this $q$-combinatorics emerges from solving for the bulk state $|\mf_{mn})$ prepared by fusing two fixed-energy boundary states. The action of $q^{\Delta \hat{n}}$ on the bulk slice then leads to a superposition of intertwined states with lower chord numbers. This relation can also be written in terms of a two-sided operator expression:
\be\label{eq:inter eq}
|\mf_{mn}^{\Del})=\int\dd\mu\left(\te\right)|\te)(\te,\te|\me_{0}^{-1}q^{\Del(\hat{n}_{L}+\hat{n}_{R})}\me_{0}|m,n),
\ee
where $\me_{0}^{-1}=(\ha_{L}\otimes\ha_{R};q)_{\infty}$ {{and we defined $\hat{a}_L=\hat{a}\otimes\mathbb{1}$, and $\hat{a}_R=\mathbb{1}\otimes\hat{a}$ with $\ha^\dagger$ and $\ha$ being the creation and annihilation operators in the zero particle chord space respectively, $$\hat{a}^\dagger|n)=\sqrt{[n+1]_q}|n+1)~,\quad \hat{a}|n)=\sqrt{[n]_q}|n-1)~,$$
where $[n]_q:= \frac{1-q^n}{1-q}$ and the relation above can be used to prove that $\me_{0}^{-1}$ is an invertible operator (see e.g. \cite{Okuyama:2024yya,Okuyama:2024more,Xu:2024gfm})}}. We will revisit such operators in our later discussion of crossed configurations.

\paragraph{The Open Matter Chord Operator}We now consider the open matter chord operator, which we denote $\hO_\Delta$. It adds an open matter chord to a given state. In previous work \cite{Lin:2022rbf,Xu:2024hoc}, these operators are defined by attaching an $M$-chord at either the leftmost or the rightmost position of a bulk state with a fixed chord configuration. However, in JT gravity coupled with matter, one can insert boundary operators at arbitrary points on the boundary, thereby preparing bulk states that encode the backreaction from these insertions. This setup has been explored in detail in \cite{Penington:2023dql,Iliesiu:2024cnh}. Motivated by this, we aim to extend the definition of matter chord operators to allow insertions at arbitrary boundary locations in a manner analogous to JT+matter theories.

The challenge in directly inserting a matter chord at a specific bulk location stems from the fact that chord-number eigenstates $|n)$ do not possess intrinsic bulk locality. To enable matter insertion at specific locations, we apply the inverse of the intertwining map developed earlier, decomposing a bulk state into a tensor product of two boundary states with fixed chord numbers, and then inserting a chord operator between them.

To this end, we rewrite \eqref{eq:fusion-1} as
\be \label{eq:fusion-2}
\mu^{-1}(\te_L)\delta(\te_L - \te_R)(\te_L,\te_R|\me_0^{-1}|m_L,m_R) = \sum_{n_L,n_R} \phi_{E_L}(n_L)\phi_{E_R}(n_R) \frac{I_{n_L,n_R,m_L + m_R}}{\sqrt{(q;q)_{n_L}(q;q)_{n_R}}},
\ee
where on the LHS, we used the linearization formula for $q$-Hermite polynomials:
\be
H_{m+n}(\cos\te|q) = \sum_{k=0}^{\min(m,n)} \frac{(-1)^k q^{\binom{k}{2}}(q;q)_m (q;q)_n}{(q;q)_{m-k}(q;q)_{n-k}(q;q)_k} H_{m-k}(\cos\te|q) H_{n-k}(\cos\te|q),
\ee
which organizes into the operator identity:
\be
\frac{H_{m+n}(\cos\te|q)}{\sqrt{(q;q)_m (q;q)_n}} = \sqrt{\binom{m+n}{m}_q}(\te|m+n) = (\te,\te|\me_0^{-1}|m,n), \quad \me_0^{-1} = (\ha_L \ha_R;q)_\infty.
\ee
Inserting a matter chord between $|m_L)$ and $|m_R)$ prepares a one-particle state, which we denote by
\be\label{eq:wavef Energy 1 part}
\phi^{\Delta}_{E_L E_R}(m_L, m_R) = \langle \Delta; \te_L, \te_R | \Delta; m_L, m_R \rangle,
\ee
where $|\Delta;m_L,m_R\rangle$ is the one-particle state living in the Hilbert space $\mh_1$ associated to the bulk slice, defined such that there are $m_L$ $H$-chords to the left of the open matter chord, and $m_R$ to the right of it{{; which we stress is different from tensor product state $|m_L,m_R)$ belonging to the doubled zero-particle Hilbert space}}. $|\Delta;\te_L,\te_R\rangle$ is defined as eigenstates of one-particle chord Hamiltonian $\hH_{L/R}$, which we explicitly construct in next section. In the following discussion, we compute their overlap $\phi^{\Delta}_{E_L E_R}(m_L,m_R)$ with the one-particle intertwiner $I^{\Delta}$.   

We thus expect \eqref{eq:fusion-2} to generalize to
\be \label{eq:fusion-3}
\phi^{\Delta}_{E_L E_R}(m_L, m_R) = \sum_{n_L, n_R = 0}^{\infty} \phi_{E_L}(n_L)\phi_{E_R}(n_R) I^{\Delta}_{n_L, n_R; m_L, m_R}.
\ee
\begin{equation}\label{eq:geodesics intertwining}
\begin{tikzpicture}[scale=1]
    % Define radius
    \def\r{2}

    % Fill the disk
    \fill[gray!20] (-\r,0) arc[start angle=-180,end angle=0, radius=\r];

    % Draw the circle
    \draw[thick] (-\r,0) arc[start angle=-180,end angle=0, radius=\r];

    % Boundary points
    \coordinate (A) at (-\r,0);
    \coordinate (B) at (\r,0);
    \coordinate (O) at (0,0);
    \coordinate (M) at (0,0);
    \coordinate (X) at ($(\r*cos{30},-\r*sin{30})$);
    \coordinate (B2) at ($(1.2*\r *cos{20}, - 1.2*\r*sin{20} )$);
    \coordinate (B1) at ($(1.2*\r* cos{155}, - 1.2*\r*sin{155})$);

    % Define marked points
    \coordinate (M1) at (-.45,-.8);
    \coordinate (M2) at (-2.2,0);
    \coordinate (M3) at ($(1.1*\r*cos{60},-1.1*\r*sin{60})$);
    \coordinate (M4) at (2.2,0); 
    \coordinate (M23) at (1.5, -.8);
    \coordinate (M13) at (0,0.2);
    \coordinate (I) at (.5, -0.5 );

    %Draw Geodesics
        \draw[thick,blue] (A) arc[start angle=90,end angle=30, radius=2*1.73];
        \draw[thick,red] (B) arc[start angle=120,end angle=180, radius=2];   
        \draw[thick] (A)--(B);
    % Draw nodes
    \node at (M1) {$m_{12}$};
    \node at (M2) {$1$};
    \node at (M3) {$2$};
    \node at (M4) {$3$};
    \node at (M23) {$m_{23}$};
    \node at (M13) {$m_{13}$};
    \node at (I) {$I^{\Delta}$};
    \node at (B1) {$E_L$};
    \node at (B2) {$E_R$};
    \draw[thick,black] (M3) .. controls (0.4,-0.4)  .. (0.3,0); 
\end{tikzpicture}
\end{equation}

In this context, the delta function constraint on $\te_L = \te_R$ no longer applies, as general states with matter insertions may possess unequal left and right energies. Nevertheless, since $\hH_L$ and $\hH_R$ commute, the state remains specified by simultaneous eigenvalues. Prior studies \cite{Xu:2024hoc,Okuyama:2024more} evaluated $\phi^{\Delta}_{E_L E_R}(m_L, m_R)$ and showed that it admits the following operator form:
\be \label{eq:phi-Delta-def}
\phi^{\Delta}_{E_L E_R}(m_L, m_R) = (\te_L, \te_R|\mo_{\Delta} \me_0^{-1} |m_L, m_R),
\ee
where
\be
\mo_{\Delta} = \frac{(q^{\Delta} \ha_L \ha_R; q)_\infty}{(\ha_L \ha_R; q)_\infty} = \sum_{k=0}^{\infty} \frac{(q^{\Delta};q)_k}{(q;q)_k} \ha_L^k \ha_R^k~.
\ee
Clearly, $\mo_{\Delta} \to \mathbb{1}$ as $\Delta \to 0$. Combining \eqref{eq:fusion-3} and \eqref{eq:phi-Delta-def}, we obtain the relation between $I^\Delta$ and $I$:
\be
I^{\Delta}_{n_L, n_R; m_L, m_R} = \sum_{k=0}^{\min(m_L, m_R)} \frac{(q^{\Delta};q)_k}{(q;q)_k} \frac{I_{n_L, n_R, m_L + m_R - 2k}}{\sqrt{(q;q)_{m_L - k}(q;q)_{m_R - k}}}.
\ee
This kernel satisfies an intertwining property: the action of the zero-particle chord Hamiltonians, {{$\hh_L:=\hH\otimes\mathbb{1}$ and $\hh_R:=\mathbb{1}\otimes\hH$,\footnote{{{The zero particle Hamiltonian was defined in \eqref{eq:zero particle identities}.}}}}}
on the $n_L$, $n_R$ indices is equivalent to the action of the one-particle chord Hamiltonians $\hH_L$, $\hH_R$ (defined in \eqref{eq:def-HLR}) on the $m_L$, $m_R$ indices. This makes $I^\Delta$ a discrete analogue of the intertwining kernel $\tilde{\phi}_\Delta(l_L, l_R; l, u)$ introduced in \cite{Iliesiu:2024cnh}. To see this, introduce
\be\label{eq:one particle entangler}
\me_{\Delta}^{-1} = \mo_{\Delta} \me_0^{-1} = (q^{\Delta} \ha_L \ha_R; q)_\infty,
\ee
and express $I^\Delta$ as the matrix elements of $\me_\Delta^{-1}$:
\be
I^{\Delta}_{n_L, n_R; m_L, m_R} = \sqrt{(q;q)_{n_L}(q;q)_{n_R}} \, (n_L, n_R | \me_\Delta^{-1} | m_L, m_R).
\ee
The intertwining property can then be understood as examining the action of $\hh_L$ in the following expression:
\be
(n_L, n_R | \hh_L \me_\Delta^{-1} | m_L, m_R ),
\ee
where $\hh_L$ acts on the bra states as zero-particle Hamiltonian
\be \label{eq:action-H}
\hh_L |n_L, n_R) = \sqrt{[n_L+1]_q} \, |n_L+1, n_R) + \sqrt{[n_L]_q} \, |n_L-1, n_R),
\ee
but acts on the ket state as one-particle Hamiltonian:
\be \label{eq:action-H2}
\begin{aligned}
    \hh_L \me^{-1}_\Delta |n_L, n_R) = &\sqrt{[n_L+1]_q} \, |n_L+1, n_R) + \sqrt{[n_L]_q} \, |n_L-1, n_R) \\
    &+ q^{\Delta + n_L} \sqrt{[n_R]_q} \, |n_L, n_R-1)~,
\end{aligned}
\ee
{{where the notation $[n]_q:=\frac{1-q^n}{1-q}$ was introduced below \eqref{eq:inter eq}.}} From \eqref{eq:action-H2}, it is tempting to identify the state \( |\Delta; n_L, n_R \rangle \) with \( \me_\Delta^{-1} |n_L, n_R) \), as the latter reproduces the action of one-particle Hamitlonian. However, this map is not an isometry, as the inner product on \( \mathcal{H}_1 \) differs from that on \( \mathcal{H}_0 \otimes \mathcal{H}_0 \), the latter being ignorant of matter insertions\footnote{Here, the inner product on \( \mathcal{H}_0 \otimes \mathcal{H}_0 \) is the natural tensor product induced by the inner product on \( \mathcal{H}_0 \). }. In this sense, relations such as \eqref{eq:fusion-3} should be understood as expressing an intertwine property: one can think of bulk slice states as being constructed by contracting left and right boundary states with an intertwine operator, which plays the role of an interior state encoding matter insertions but not boundary conditions.

In later discussion, we will make this intertwining structure more precise by constructing an isometric factorization map
\begin{equation}
\hF : \mathcal{H}_1 \to \mathcal{H}_0 \otimes \mathcal{H}_0,
\end{equation}
which not only identifies states on both sides but also preserves the inner product.

We emphasize that \eqref{eq:fusion-3} only computes the overlap between states, without specifying the individual definitions of \( |\Delta; n_L, n_R \rangle \) and \( |\Delta; \theta_L, \theta_R \rangle \). While states in the chord number basis can be defined unambiguously using the chord rules~\cite{Lin:2023trc}, the construction of energy eigenstates has not yet been presented—we turn to this in the following discussion.

\subsection{Fixed-Energy Wavefunctions with Matter}\label{ssec:fixed energy wavefunctions}

In this section, we present a derivation for the energy eigenstates of the one-particle bulk Hamiltonian $\hH_{L/R}$, which acts on states $|\Delta;m_L,m_R\rangle\in\mh_1$ as :
\begin{equation} 
\begin{split} \label{eq:def-HLR}
\frac{\sqrt{\lambda}}{J}\hat{H}_L |\Delta ; m_L, m_R \rangle  =&\sqrt{\left[n_L+1\right]_q} |\Delta ; n_L+1, n_R \rangle+\sqrt{ [n_L]_q} |\Delta, n_L-1, n_R \rangle\\
&+q^{\Delta+n_L} \sqrt{[n_R]_q}|\Delta, n_L, n_R-1\rangle  \\
\frac{\sqrt{\lambda}}{J}\hat{H}_R |\Delta ; m_L, m_R \rangle 
=&\sqrt{ [n_R+1]_q} |\Delta ; n_L, n_R+1 \rangle+\sqrt{ [n_R]_q} |\Delta, n_L, n_R-1 \rangle\\
&+q^{\Delta+n_R} \sqrt{ [n_L ]_q} |\Delta, n_L-1, n_R \rangle .
\end{split}
\end{equation}
We denote the energy eigenstate as $|\Delta;\te_L,\te_R\rangle$, such that\footnote{The energy eigenvalues are determined by analyzing the asymptotic form of the recursion relation in the large chord number limit. For a more detailed treatment of the asymptotic analysis for similar recursion relations, we refer the reader to~\cite{Berkooz20}. }
\be
\hH_{L/R}|\Delta;\te_L,\te_R \rangle = E_{L/R} |\Delta; \te_L,\te_R\rangle,\quad E_{L/R}=\frac{2J}{\sqrt{\lambda(1-q)}}\cos \te_{L/R}~.
\ee
Now we build these eigenstates by superposing the bulk states:
\be\label{eq: def zeta}
|\Delta;\theta_{L},\theta_{R}\rangle=\sum_{n_{L},n_{R}=0}^{\infty}\zeta_{n_{L},n_{R}}(\te_L,\te_R)|\Delta;n_{L},n_{R}\rangle~,
\ee 
Now acting {{with $\hH_L/\hH_R$}} on the states, and match{{ing}} the coefficients of two sides, \footnote{Here, we used the fact that for general values of $0\leq q <1$ and $\Delta>0$, the one-particle states $|\Delta;n_L,n_R\rangle \in \mh_1$ are linearly independent. } we deduce the recursion relation of $\zeta_{n_L,n_R}$ as:
\be\label{eq:recursion-Q} 
\begin{aligned}
2 \cos \theta_L \zeta_{n_L, n_R} & =\sqrt{1-q^{n_L}} \zeta_{n_L-1, n_R}+\sqrt{1-q^{n_L+1}} \zeta_{n_L+1, n_R} \\
& +q^{\Delta+n_L} \sqrt{1-q^{n_R+1}} \zeta_{n_L, n_R+1}~, \\
2 \cos \theta_R \zeta_{n_L, n_R} & =\sqrt{1-q^{n_R}} \zeta_{n_L, n_R-1}+\sqrt{1-q^{n_R+1}} \zeta_{n_L, n_R+1} \\
& +q^{\Delta+n_R} \sqrt{1-q^{n_L+1}} \zeta_{n_L+1, n_R}~,
\end{aligned}
\ee
{{where we have rescaled $\zeta_{n_L,n_R}$ in \eqref{eq: def zeta} by a overall constant $\sqrt{1-q}$ appearing in the factors $[n]_q$ in \eqref{eq:def-HLR}.}}

With boundary condition such that 
\be
\begin{split}
    \zeta_{0,0}(\te_L,\te_R) & = (\te_L|q^{\Delta \hat{n}}|\te_R) = \sum^{\infty}_{n=0} \frac{q^{\Delta n}}{(q;q)_n} H_n(\cos\te_L;q) H_n(\cos\te_R;q)\\ 
    & = \frac{(q^{2\Delta};q)_\infty}{(q^{\Delta} e^{\pm i\te_L\pm i \te_R};q)_\infty} = \mathcal{N}_q \frac{\Gamma_q(\Delta \pm i\te_L/\lambda \pm i \te_R/\lambda)}{\Gamma_q (2\Delta)}~,
\end{split}
\ee
The first equality reflects our choice of normalization coefficient \( \zeta_{0,0} \), while the second follows from inserting a resolution of identity in the chord number basis. The third equality uses the bi-variate generating function of \( q \)-Hermite polynomials, which is then rewritten in a more familiar form involving \( \Gamma_q \) functions in the fourth equality, where we define
\begin{equation}
\Gamma_q(x) = (1 - q)^{1 - x} \frac{(q; q)_\infty}{(q^x; q)_\infty}.
\end{equation}
The alternating signs in the argument indicate a shorthand for multiplying over all possible sign configurations, and $\mathcal{N}_q$ is an overall constant independent of energy $\te_{L/R}$ and matter weight $\Delta$.

This normalization is the \( q \)-deformed analogue of the matter two-point function in JT gravity coupled to conformal matter. The choice of \( \zeta_{0,0} \) ensures that the inner product between the zero-\( H \)-chord one-particle state \( |\Delta; 0, 0\rangle \) and the fixed-energy state \( |\Delta; \theta_L, \theta_R \rangle \) reproduces the chord diagram corresponding to those fixed boundary energies.

\be \label{eq:normalization-zeta}
\zeta_{0,0}(\te_L,\te_R)= \langle\Delta;0,0|\Delta; \te_L,\te_R\rangle =  \begin{tikzpicture}[baseline={([yshift=-0.1cm]current bounding box.center)}, line width=1pt, scale=0.75]

% Radius of the disk
\def\r{1}

% Coordinates
\coordinate (T) at (0,\r);   % Top
\coordinate (B) at (0,-\r);  % Bottom
\coordinate (L) at (-\r,0);  % Left
\coordinate (R) at (\r,0);   % Right
\coordinate (C) at (0,0);    % Center

% Disk boundary
\draw (C) circle (\r);

% Vertical geodesic
\draw[blue] (T) -- (B);

% Horizontal geodesic

% Four curved black geodesics forming diamond

% Dots at endpoints
\foreach \pt in {T, B, L, R}
  \fill[black] (\pt) circle (0.6pt);

\node at (-1.3*\r, 0 ) {\small $\te_L$};
\node at (1.3*\r, 0 ) {\small $\te_R$};
\node at (0.25, 0) {\tiny $\Delta$}; 
\end{tikzpicture}.
\ee
An analogue of \eqref{eq:normalization-zeta} in \( \mathcal{H}_0 \) is that the overlap between the empty state \( |\Omega) \) and any fixed-energy state \( |\theta) \) is unity: 
\begin{equation}
(\Omega | \theta) = 1.
\end{equation}
This follows from the fact that the empty state \( |\Omega) \) is prepared by superposing all energy eigenstates with equal weight. This is also why \( |\Omega) \) is identified with the infinite-temperature state in various gravity dual proposals of DSSYK~\cite{lin2022hot,Blommaert:2023opb,Blommaert:2024ydx,Blommaert:2025avl}.
As we will show below, the one-particle state with no \( H \)-chords can likewise be viewed as a uniform superposition of all one-particle wavefunctions with fixed left and right boundary energies.

To solve the recursion relation \eqref{eq:recursion-Q}, we apply the following ansatz:
\be
\zeta_{n_{L},n_{R}}(\te_L,\te_R)=\frac{a_{n_{L},n_{R}} (\te_L,\te_R)}{\sqrt{\left(q;q\right)_{n_{L}}\left(q;q\right)_{n_{R}}}},\quad a_{0,0}=\zeta_{0,0}~,
\ee
which simplifies \eqref{eq:recursion-Q} into:
\be\label{eq:anLnR relation}
\begin{split}
   & 2\cos\te_{L}a_{n_{L},n_{R}}	=a_{n_{L}+1,n_{R}}+q^{\Del+n_{L}}a_{n_{L},n_{R}+1}+(1-q^{n_{L}})a_{n_{L}-1,n_{R}}, \\
& 2\cos\te_{R}a_{n_{L},n_{R}}	=a_{n_{L},n_{R}+1}+q^{\Del+n_{R}}a_{n_{L}+1,n_{R}}+(1-q^{n_{R}})a_{n_{L},n_{R}-1}.
\end{split}
\ee
The recursion relation can be solved by the bi-variate symmetric kernel:
\be\label{eq:overlap}
\begin{split}
    a_{n_{L},n_{R}}&=(\te_L|q^{\Delta\hat{n}}|\te_R) Q_{n_{L},n_{R}}\left(\cos\te_{L},\cos\te_{R}|q^{\Del},q\right)   \\
 &:=\sum_{k=0}^{\infty}q^{\Del k}\frac{H_{n_{L}+k}\left(\cos\te_{L}|q\right)H_{n_{R}+k}\left(\cos\te_{R}|q\right)}{\left(q;q\right)_{k}},
\end{split}
\ee 
where we verify that it does satisfy the recursion relation in \ref{app:recursion}.  The symmetric kernel exhibits the following operator expression:
\begin{equation}
\frac{Q_{m_{L},m_{R}}\left(\cos\theta_{L},\cos\theta_{R} \mid q^{\Delta},q\right)}{\sqrt{(q;q)_{n_{L}}(q;q)_{n_{R}}}} = \frac{(m_{L},m_{R}|\me_{\Delta}|\theta_{L},\theta_{R})}{(\theta_{L}|q^{\Delta \hat{n}}|\theta_{R})},
\end{equation}
by which, we find the energy eigenstates are proportional to the factorized states \( |\theta_{L},\theta_{R}) \):
\begin{equation} \label{eq:tempt}
|\Delta;\theta_{L},\theta_{R}\rangle = \sum_{m_{L},m_{R}=0}^{\infty} (m_{L},m_{R}|\me_{\Delta}|\theta_{L},\theta_{R})  |\Delta;m_L,m_R\rangle~.
\end{equation}
This completes the construction of energy eigenstates in the one-particle Hilbert space. In \cite{Xu:2024hoc},  the inner product between states with given left and right chord number is expressed as:
\be
\begin{aligned}
    &\langle \Delta; n_L, n_R |\Delta; m_L,m_R\rangle \\
&= \int \dd\mu(\te_L)\dd \mu(\te_R) (\te_L|q^{\Delta \hat{n}}|\te_R) (n_L,n_R| \me^{\da -1}_\Delta |\te_L,\te_R) (\te_L,\te_R| \me^{-1}_\Delta |m_L,m_R)~.
\end{aligned}
\ee
We have identified the bivariate \( q \)-Hermite polynomial with the matrix elements of \( \me^{-1}_\Delta \). One can directly verify that
\begin{equation} \label{eq:inner-product-2} 
\langle \Delta; \theta_L, \theta_R | \Delta; \phi_L, \phi_R \rangle = (\theta_L | q^{\Delta \hat{n}} | \theta_R) \frac{\delta(\theta_L - \phi_L)\, \delta(\theta_R - \phi_R)}{\mu(\theta_L)\, \mu(\theta_R)},
\end{equation}
which shows that the inner product between two one-particle wavefunctions with fixed energy contains the two-point function with corresponding fixed-energy boundary conditions. One can thus think of \eqref{eq:inner-product-2} as sewing two open chord diagram into one:
\be \label{eq:diagram-2pt}
\left\langle \begin{tikzpicture}[baseline={([yshift=-0.1cm]current bounding box.center)}, line width=1pt, scale=0.6]

\draw (0,1) arc[start angle= 90, end angle = 270, radius=1];

\draw[thick, blue] (-1,0) -- (0,0);
\fill[blue] (-1,0) circle (1.5pt); 

\node at (-0.3,1.2) {\tiny $\te_L$};
\node at (-0.3,-1.2) {\tiny $\te_R$};

\end{tikzpicture}
\middle|
\begin{tikzpicture}[baseline={([yshift=-0.1cm]current bounding box.center)}, line width=1pt, scale=0.6]

\draw (0,1) arc[start angle= 90, end angle = -90, radius=1];

\draw[thick, blue] (0,0) -- (1,0);
\fill[blue] (1,0) circle (1.5pt); 

\node at (0.3,1.2) {\tiny $\phi_L$};
\node at (0.3,-1.2) {\tiny $\phi_R$};

\end{tikzpicture}
\right\rangle = \begin{tikzpicture}[baseline={([yshift=-0.1cm]current bounding box.center)}, line width=1pt, scale=0.6]

\draw (0,0) circle (1);
\draw[thick, blue] (-1,0) -- (1,0);

\node at (0,1.2) {\tiny $\te_L$};
\node at (0,-1.2) {\tiny $\te_R$};

\end{tikzpicture}\times \frac{\delta(\te_L-\phi_L)\delta(\te_R-\phi_R)}{\mu(\te_L)\mu(\te_R)},
\ee
where the $\delta$-function contribution arises whenever two open chord diagrams are sewn together along fixed-energy boundaries, enforcing energy conservation at each contact point on the boundary.

\subsection{Comments on the Isometric Factorization of \texorpdfstring{$\mh_1$}{}}\label{ssec:comments factorization}

In the preceding discussion, we have distinguished states in \( \mathcal{H}_1 \) from those in \( \mathcal{H}_0 \otimes \mathcal{H}_0 \) based on their differing inner product structures. In particular, we have chosen to treat the states \( |\Delta; n_L, n_R \rangle \) and \( \me_\Delta^{-1} |n_L, n_R) \) as distinct, even though they lead to the same recursion relations given in \eqref{eq:action-H2} and \eqref{eq:def-HLR}.  In this section, we first demonstrate the inconsistency that arises in the inner product structure if one were to identify \( |\Delta; n_L, n_R \rangle \) with \( \me_\Delta^{-1} |n_L, n_R) \). We then introduce an isometric map \( \mathcal{F}_\Delta : \mathcal{H}_1 \to \mathcal{H}_0 \otimes \mathcal{H}_0 \) under which the states on both sides can be identified. Finally, we examine the chord number operator and its action on boundary states, leading to a derivation of the crossed four-point function and the quantum \( 6j \)-symbol.

To clarify, let us compare the one-particle Hamiltonians as \(\hat{H}_{L/R} \) and the zero-particle ones as \( \hh_{L/R} \), with
\begin{equation}\label{eq:rescaled hLR}
\hH_{L/R} = \frac{J}{\sqrt{\lambda}}\qty(\hfa_{L/R}^\dagger + \hfa_{L/R}), \quad \hh_{L/R} = \frac{J}{\sqrt{\lambda}}\qty(\ha_{L/R}^\dagger + \ha_{L/R})~,
\end{equation}
where the curly ladder operators are defined consistently with the chord rules \cite{Lin:2023trc}:
\begin{subequations}\label{eq:mathfrak a dagger}
    \begin{align}
\hfa_{L}^\dagger |\Delta;n_{L},n_{R}\rangle &= \sqrt{[n_L+1]_q} |\Delta;n_{L}+1,n_{R}\rangle~,\\
\hfa_{L} |\Delta;n_{L},n_{R}\rangle &= \sqrt{[n_L]_q} |\Delta;n_{L}-1,n_{R}\rangle + q^{\Delta + n_L} \sqrt{[n_R]_q} |\Delta;n_{L},n_{R}-1\rangle~,
\end{align}
\end{subequations}
and similarly for the right operators $\hat{\mathfrak{a}}_{R}/\hat{\mathfrak{a}}_{R}^\dagger$. One can verify that \( \hh_{L/R}\) act the same as \( \hH_{L/R} \) on \( \me_\Delta^{-1} |n_L,n_R) \), but the individual ladder operators differ. In fact, we have
\begin{subequations}\label{eq:a adagger hLR}
    \begin{align}
\ha_L^\dagger \me^{-1}_\Delta |n_L,n_R) &= \sqrt{[n_L+1]_q} \me^{-1}_\Delta |n_L+1,n_R) + q^{\Delta + n_L} \me^{-1}_\Delta |n_L,n_R-1), \\
\ha_L \me^{-1}_\Delta |n_L,n_R) &= \sqrt{[n_L]_q} \me^{-1}_\Delta |n_L-1,n_R).
\end{align}
\end{subequations}
{{which follows directly from the definition of the zero-particle creation/annihilation operators $\ha_L$, $\ha^\dagger_L$ below \eqref{eq:inter eq}, and the definition of $\mathcal{E}_\Delta^{-1}$ in \eqref{eq:one particle entangler}.}}

The chord inner product in $\mh_1$ is defined so that \( \hfa_{L/R} \) and \( \hfa_{L/R}^\dagger \) are Hermitian conjugates, which leads to the recursion:
\begin{align}
\sqrt{[m_L]_q} \langle \Delta;m_L,m_R | \Delta;n_L,n_R \rangle &= \sqrt{[n_L]_q} \langle \Delta;m_L-1,m_R | \Delta;n_L-1,n_R \rangle \nonumber \\
& \quad + q^{\Delta + m_L} \sqrt{[n_R]_q} \langle \Delta;m_L-1,m_R | \Delta;n_L,n_R-1 \rangle~.
\end{align}
This is not compatible with the recursion obtained from requiring that \( \ha_L \) and \( \ha_L^\dagger \) are Hermitian conjugates:
\begin{equation}
    \begin{aligned}
\sqrt{[m_L]_q} (m_L-1,m_R |\me^{-1\da}_\Delta \me^{-1}_\Delta   & |n_L-1,n_R \rangle = \sqrt{[n_L]_q} ( m_L,m_R | \me^{-1\da}_\Delta \me^{-1}_\Delta |n_L,n_R )  \\
& \quad + q^{\Delta + n_L} \sqrt{[n_R]_q} (m_L,m_R | \me^{-1\da}_\Delta \me^{-1}_\Delta  |n_L,n_R-1 )~.
\end{aligned}
\end{equation}
In summary, although one can reproduce the recursion relation of $\hH_{L/R}$ acting on $|\Delta;n_L,n_R\rangle$  via the intertwining operator, the Hilbert spaces have distinct inner products. When computing correlation functions in DSSYK, one must use the chord inner product defined in the bulk Hilbert space. In our notation, \( \langle \cdot | \cdot \rangle \) denotes the bulk inner product, while \( (\cdot | \cdot) \) refers to the doubled Hilbert space. We continue using the latter as it is often more practical and easier to compute.

It might be surprising at first that the one-particle chord Hamiltonian $\hH_{L/R}$ has exactly the same eigenvalues as $\hh_{L/R}$, which does not depend on the matter weight $\Delta$.   The independence of $\Delta$ follows from the existence of the intertwining operator which allows us to construct each state in \( \mathcal{H}_1 \) with a state in the doubled Hilbert space \footnote{Technically, we have thus far shown how the intertwiner \( I^\Delta \) relates the wavefunctions (overlap of states) corresponding to fixed-energy states on both sides. In what follows, we will demonstrate that a relation between the states themselves also exists.}, which itself carries no explicit matter content. The energy spectrum of \( \hH_{L/R} \) is determined by the asymptotic form of the recursion relation \eqref{eq:def-HLR}, which approaches the factorized recursion relation induced by \( \hh_{L/R} \) in the large chord number limit. This behavior arises because, for the left/right chord Hamiltonian to annihilate an \( H \)-chord located to the right/left of the matter insertion, it must first cross all \( H \)-chords to its left/right. Such processes are exponentially suppressed in the asymptotic limit, effectively vanishing as the total chord number becomes large.

We can also offer a boundary perspective for why the one-particle Hamiltonian $\hH_{L/R}$ has the same eigenvalues as the zero-particle Hamiltonian.  Note that from the boundary point of view, $\hH_{L/R}$ can be viewed as limiting value of the operator \( e^{-\beta \hH_{L/R}} \), which adds a left/right boundary segment of length \( \beta \), regardless of whether matter chords are inserted in the intermediate region. The vacuum case corresponds to inserting matter of zero weight. Hence, from the boundary perspective, the left and right chord Hamiltonians act locally on the boundary and are independent of local matter chord insertions:
\begin{subequations}
\begin{align}
&\rme^{-\beta \hH_L}  \left| \begin{tikzpicture}[baseline={([yshift=-0.1cm]current bounding box.center)}, line width=1pt, scale=0.6]
\draw (1,0) arc[start angle= 0, end angle = -180, radius=1];
\draw[thick, blue] (0,-1) -- (0,0);
\fill[blue] (-1,0) circle (1.5pt); 
\node at (-1.3,-0.8) {\tiny $\beta_L$};
\node at (1.3,-0.8) {\tiny $\beta_R$}; 
\node at (0,0.2) {\tiny $\Delta$}; 
\end{tikzpicture}\right\rangle
= \left| \begin{tikzpicture}[baseline={([yshift=-0.1cm]current bounding box.center)}, line width=1pt, scale=0.6]
\draw (1,0) arc[start angle= 0, end angle = -180, radius=1];
\draw[thick, blue] (0,-1) -- (0,0);
\fill[blue] (-1,0) circle (1.5pt); 
\node at (-1.2,-0.8) {\tiny $\beta+\beta_L$};
\node at (1.3,-0.8) {\tiny $\beta_R$}; 
\node at (0,0.2) {\tiny $\Delta$}; 
\end{tikzpicture}\right\rangle~,\\ &\rme^{-\beta \hH_R}  \left| \begin{tikzpicture}[baseline={([yshift=-0.1cm]current bounding box.center)}, line width=1pt, scale=0.6]
\draw (1,0) arc[start angle= 0, end angle = -180, radius=1];
\draw[thick, blue] (0,-1) -- (0,0);
\fill[blue] (-1,0) circle (1.5pt); 
\node at (-1.3,-0.8) {\tiny $\beta_L$};
\node at (1.3,-0.8) {\tiny $\beta_R$}; 
\node at (0,0.2) {\tiny $\Delta$}; 
\end{tikzpicture}\right\rangle
= \left| \begin{tikzpicture}[baseline={([yshift=-0.1cm]current bounding box.center)}, line width=1pt, scale=0.6]
\draw (1,0) arc[start angle= 0, end angle = -180, radius=1];
\draw[thick, blue] (0,-1) -- (0,0);
\fill[blue] (-1,0) circle (1.5pt); 
\node at (-1.2,-0.8) {\tiny $\beta_L$};
\node at (1.3,-0.8) {\tiny $\beta_R+\beta$}; 
\node at (0,0.2) {\tiny $\Delta$}; 
\end{tikzpicture}\right\rangle~.
\end{align}
\end{subequations}
The fact that the one-particle chord Hamiltonian has the same spectrum as its zero-particle counterpart enables an isometric factorization of the Hilbert space into a doubled Hilbert space. We make this precise by introducing an isometric factorization map \( \hF : \mathcal{H}_1 \to \mathcal{H}_0 \otimes \mathcal{H}_0 \), defined by
\begin{equation}\label{eq:isometric linear map}
\hF |\Delta; \theta_L, \theta_R \rangle = \gamma_\Delta(\theta_L, \theta_R) | \theta_L, \theta_R ),
\end{equation}
where the normalization factor \( \gamma_\Delta \) is chosen so that the inner product between one-particle energy eigenstates matches the two-point function with fixed energy boundary conditions. This is diagrammatically represented in \eqref{eq:diagram-2pt}.

Requiring \( \hF \) to be an isometry determines
\begin{equation}\label{eq:matrix element in energy basis}
\gamma_\Delta(\theta_L, \theta_R) = \sqrt{ (\theta_L | q^{\Delta \hat{n}} | \theta_R ) } = \sqrt{ \frac{ \Gamma_q(\Delta \pm i \theta_L / \lambda \pm i \theta_R / \lambda) }{ \Gamma_q(2\Delta) } },
\end{equation}
where the \( q \)-Pochhammer symbols have been rewritten in terms of \( q \)-Gamma functions. The \( \pm \) notation indicates multiplication over all sign combinations.

The action of \( \hF \) on the length basis is less trivial but has been worked out previously \cite{xu2024chord}. We summarize the result:
\begin{equation} \label{eq:F-def}
\hF |\Delta; n_L, n_R \rangle = \int_0^\pi d\mu(\theta_L) d\mu(\theta_R) \gamma_\Delta(\theta_L, \theta_R) | \theta_L, \theta_R ) ( \theta_L, \theta_R | \me_\Delta^{-1} | n_L, n_R ),
\end{equation}
from which it is straightforward to show that the one-particle chord Hamiltonian maps to the zero-particle one:
\begin{align}
\hh_L \hF |\Delta; n_L, n_R \rangle &= \int_0^\pi d\mu(\theta_L) d\mu(\theta_R) \gamma_\Delta(\theta_L, \theta_R) E(\theta_L) | \theta_L, \theta_R ) ( \theta_L, \theta_R | \me_\Delta^{-1} | n_L, n_R ) \\
&= \int_0^\pi d\mu(\theta_L) d\mu(\theta_R) \gamma_\Delta(\theta_L, \theta_R) | \theta_L, \theta_R ) ( \theta_L, \theta_R | \hh_L \me_\Delta^{-1} | n_L, n_R ),
\end{align}
where we used that \( \hh_L \) is Hermitian with respect to the inner product on \( \mathcal{H}_0 \otimes \mathcal{H}_0 \), allowing us to move it to the bra.

Combining the prior results, the operator \( \hh_L \me_\Delta^{-1} \) acts as
\begin{equation}
\begin{aligned}
\hh_L \me_\Delta^{-1} | n_L, n_R ) =& \frac{J}{\sqrt{\lambda}}\bigg(\sqrt{[n_L + 1]_q} | n_L + 1, n_R ) + \sqrt{[n_L]_q} | n_L - 1, n_R ) \\
&+ q^{\Delta + n_L} \sqrt{[n_R]_q} | n_L, n_R - 1 )\bigg)~.
\end{aligned}
\end{equation}
Plugging into the expression above, we find
\begin{equation}
\hh_L \hF| \Delta; n_L, n_R \rangle = \hF \hH_L | \Delta; n_L, n_R \rangle~,
\end{equation}
with \( \hH_L \) defined as the one-particle Hamiltonian:
\begin{equation}
\begin{aligned}
\hH_L | \Delta; n_L, n_R \rangle = \frac{J}{\sqrt{\lambda}}\bigg(&\sqrt{[n_L + 1]_q} | \Delta; n_L + 1, n_R \rangle + \sqrt{[n_L]_q} | \Delta; n_L - 1, n_R \rangle\\ &+ q^{\Delta + n_L} \sqrt{[n_R]_q} | \Delta; n_L, n_R - 1 \rangle\bigg)~.    
\end{aligned}
\end{equation}
This relation is expected from the isometric definition of \( \hF \). It highlights how the statistics of chords organizing one-particle bulk states from boundary configurations can be fully captured by the isometry $\hF$. 

\paragraph{Quantum \( 6j \)-Symbol from Chord Number Operator}
We now analyze the relation between the total chord number operator and the crossed four-point function with fixed energy boundary condition, the latter was shown in \cite{Berkooz:2018jqr} to be proportional to a quantum $6j$-symbol. Following our previous discussion, one can slice the chord diagram with crossed matter chords open and evaluates it as the expectation value of a closed chord operator $q^{\Delta_1 \hat{N}}$ in one-particle states $\left|\Delta ; \theta_1, \theta_2\right\rangle$ and $\left|\Delta ; \theta_3, \theta_4\right\rangle$. Thus, we expect:
\begin{equation} \label{eq:slicing-4pt-cr}
\langle \Delta; \theta_1, \theta_2 | q^{\Delta_1 \hat{N}} | \Delta; \theta_3, \theta_4 \rangle 
=
    \left\langle \begin{tikzpicture}[baseline={([yshift=-0.1cm]current bounding box.center)}, line width=1pt, scale=0.6]

\draw (0,1) arc[start angle= 90, end angle = 270, radius=1];

\draw[thick, blue] (-1,0) -- (0,0);
\fill[blue] (-1,0) circle (1.5pt); 

\node at (-0.3,1.2) {\tiny $\te_1$};
\node at (-0.3,-1.2) {\tiny $\te_2$};
\node at (-0.3,0.3) {\tiny $\Delta$};

\end{tikzpicture}
\right| q^{\Delta_1 \hat{N}} \left| 
\begin{tikzpicture}[baseline={([yshift=-0.1cm]current bounding box.center)}, line width=1pt, scale=0.6]

\draw (0,1) arc[start angle= 90, end angle = -90, radius=1];

\draw[thick, blue] (0,0) -- (1,0);
\fill[blue] (1,0) circle (1.5pt); 

\node at (0.3,1.2) {\tiny $\te_3$};
\node at (0.3,-1.2) {\tiny $\te_4$}; 
\node at (0.3,0.3) {\tiny $\Delta$};

\end{tikzpicture}
\right\rangle 
=\begin{tikzpicture}[baseline={([yshift=-0.1cm]current bounding box.center)}, line width=1pt, scale=0.6]

\draw (0,0) circle (1);

\draw[thick, blue] (-1,0) -- (1,0);
\draw[thick,red] (0,-1) -- (0,1);

\node at (-0.6,1.2) {\tiny $\te_1$};
\node at (-0.6,-1.2) {\tiny $\te_2$};
\node at (0.6,1.2) {\tiny $\te_3$};
\node at (0.6,-1.2) {\tiny $\te_4$};
\node at (0.6,-0.3) {\tiny $\Delta$};
\node at (-0.3, 0.6) {\tiny $\Delta_1$};

\end{tikzpicture}\propto \left\{\begin{array}{ccc}
\Delta & \theta_1 & \theta_2 \\
\Delta_1 & \theta_4 & \theta_3
\end{array}\right\}_q
\end{equation}
where the total chord number operator \( \hat{N} \) is defined on one-particle states by\footnote{This differs from \cite{Lin:2023trc} by a constant $\Delta$. In terms of crossed-four-point function, this amounts to neglect a constant coefficient $q^{\Delta\Delta_1}$. }
\begin{equation}
\hat{N} | \Delta; n_L, n_R \rangle = (n_L + n_R) | \Delta; n_L, n_R \rangle.
\end{equation}
To compute the expectation value in the energy basis, we apply the isometry \( \hF \):
\begin{equation}
\hF\hat{N} | \Delta; n_L, n_R \rangle = \int d\mu(\theta_1) d\mu(\theta_2) \gamma_{\Delta}(\theta_1, \theta_2) | \theta_1, \theta_2 ) ( \theta_1, \theta_2 | \me_\Delta^{-1} ( \hat{n}_L + \hat{n}_R ) | n_L, n_R ).
\end{equation}
To understand how the operator \( \hat{N} \) commutes with the factorization map \( \hF \), we rewrite \( |n_L, n_R) \) in the right-hand side above in terms of \( \hF|\psi\rangle \), where \( |\psi\rangle \) lives in the one-particle Hilbert space. Since \( \hF \) is invertible, this is possible. Using the definition \eqref{eq:F-def}, we have:
\begin{equation}
|n_L, n_R) = \int d\mu(\theta_1) d\mu(\theta_2) \gamma_{12}^{-1} \mathcal{E}_\Delta | \theta_L, \theta_R ) ( \theta_L, \theta_R | \hF | \Delta; n_L, n_R \rangle.
\end{equation}
Substituting into the previous expression yields
\begin{equation}
\begin{aligned}
    &\hF \hat{N} | \Delta; n_L, n_R \rangle \\
    &= \int \prod_{i=1}^4 d\mu(\theta_i) \frac{\gamma_\Delta(\te_1,\te_2)}{\gamma_\Delta(\te_3,\te_4)} |\te_1,\te_2)(\theta_1, \theta_2 | \me_\Delta^{-1} (\hat{n}_L + \hat{n}_R) \mathcal{E}_\Delta | \theta_3, \theta_4 ) ( \theta_3, \theta_4 | \hF| \Delta; n_L, n_R \rangle.
\end{aligned}
\end{equation}
Since \( | \Delta; n_L, n_R \rangle \) form a complete basis, we deduce that
\begin{equation} \label{eq:FNF}
\hF \hat{N} \hF^{-1} = \int \prod_{i=1}^4 d\mu(\theta_i) 
\frac{\gamma_\Delta(\te_1,\te_2)}{\gamma_\Delta(\te_3,\te_4)}
(\theta_1, \theta_2 | \me_\Delta^{-1} (\hat{n}_L + \hat{n}_R) \me_\Delta | \theta_3, \theta_4 ) ( \theta_3, \theta_4 |,
\end{equation}
as an operator acting on the factorized Hilbert space.

This specific combination, \( \me_\Delta^{-1} (\hat{n}_L + \hat{n}_R) \me_\Delta \), was previously proposed in \cite{Okuyama:2024more} to reproduce the chord-counting calculations in \cite{Berkooz:2018jqr}. Here, we derive it directly from the definition of the factorization map \( \hF \), which itself is motivated by the intertwining property.

\begin{figure}
    \centering
\begin{tikzpicture}[line width=0.8 pt,scale=0.6]

% Define radius
\def\r{3}

% Draw the circle
\draw[black] (\r, -.5) arc[start angle=0, end angle=-180, radius=\r];
\draw[black] (\r, .5) arc[start angle=0, end angle=180, radius=\r];

% Points
\coordinate (top) at (0, 0.5+\r);
\coordinate (bottom) at (0, -0.5 -\r);

% Blue vertical geodesics
\draw[blue] (top) -- (0, 0.5);
\draw[blue] (bottom) -- (0, -0.5);

% Red horizontal geodesic (approximate straight)
\draw[red] (-\r, 0) -- (\r, 0);

% Blue dots at endpoints
\fill[blue] (top) circle (3pt);
\fill[blue] (bottom) circle (3pt);

%labels 

\node at ($(1.2*\r * cos{45}, 1.2*\r * sin{45})$)  {$\te_4$};
\node at ($(1.2*\r * cos{135}, 1.2*\r * sin{135})$)  {$\te_3$};
\node at ($(1.2*\r * cos{225}, 1.2*\r * sin{225})$)  {$\te_1$};
\node at ($(1.2*\r * cos{315}, 1.2*\r * sin{315})$)  {$\te_2$};

\node at (0.3,-1.5) {$\Delta$};
\node at (0.3,1.5)  {$\Delta$}; 
\node at (2,0.3) {$\Delta_1$};

\end{tikzpicture}
    \caption{Two-sided two-point function with two light or heavy operators $\hat{\mathcal{O}}_{\Delta}$ (red), and two light operators $\hat{\mathcal{O}}_{\Delta_{1}}$ (blue) that generates the operator $q^{\Delta_1\hat{N}}$. \cite{Xu:2024hoc,Berkooz:2018jqr} in (\ref{eq:twosidedtwopoint}). For clarity, we omit $H$-chords in the diagram.}
    \label{fig:otocprecursor}
\end{figure}
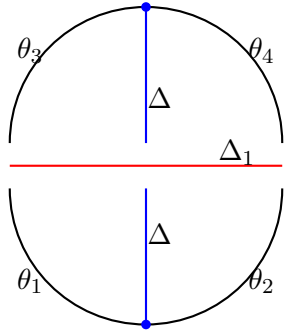

The asymmetry in the \( (\te_1,\te_2)/(\te_3,\te_4) \)-dependence of \eqref{eq:FNF} is also meaningful, as it correctly encodes the appropriate normalization of the quantum \( 6j \)-symbol that arises from computing crossed four-point functions.  The matrix elements of \( q^{\Delta_1 \hat{N}} \) involved in \eqref{eq:slicing-4pt-cr} are more transparent in the factorized description. Applying \( \hF \), the expectation value becomes an inner product between two states:
\begin{equation}
\hF | \Delta; \theta_1, \theta_2 \rangle = \gamma_{\Delta}(\te_1,\te_2) | \theta_1, \theta_2 ),
\end{equation}
and 
\begin{equation}
\hF q^{\Delta_1 \hat{N}} | \Delta; \theta_3, \theta_4 \rangle = \int \prod_{i=1}^2 d\mu(\theta_i) \gamma_{\Delta}(\te_1,\te_2) | \theta_1, \theta_2 ) ( \theta_1, \theta_2 | \me_{\Delta}^{-1} q^{\Delta_1 (\hat{n}_L + \hat{n}_R)} \me_{\Delta} | \theta_3, \theta_4 ).
\end{equation}
Hence,
\begin{align}\label{eq:6j energy basis}
\langle \Delta; \theta_1, \theta_2 | q^{\Delta_1 \hat{N}} | \Delta; \theta_3, \theta_4 \rangle &= (\theta_1 | q^{\Delta \hat{n}} | \theta_2)( \theta_1, \theta_2 | \me_{\Delta}^{-1} q^{\Delta_1 (\hat{n}_L + \hat{n}_R)} \me_{\Delta} | \theta_3, \theta_4 ) \\
&= \left\{ \begin{array}{ccc} \Delta & \theta_1 & \theta_2 \\ \Delta_1 & \theta_4 & \theta_3 \end{array} \right\}_q \Gamma^{\Delta}\Gamma^{\Delta_1},
\end{align}
where we denote the density function \( \Gamma^{\Delta_i} \) to be the matter densities associated to the intersecting vertex between matter chord $\Delta_i$ and the boundary,
\begin{equation}
\Gamma^{\Delta} =\gamma_{\Delta}(\te_1,\te_2)\gamma_{\Delta}(\te_3,\te_4),\quad \Gamma^{\Delta_1} =\gamma_{\Delta_1}(\te_1,\te_4)\gamma_{\Delta_1}(\te_2,\te_3)~.
\end{equation}
We represent (\ref{eq:6j energy basis}) in Fig. \ref{fig:otocprecursor}.
The first equality in \eqref{eq:6j energy basis} can be derived by inserting $\text{id}=\hF^{-1}\hF$ in the middle and use the definition of $\hF$ to convert it to the right-hand side in terms of quantities defined on the doubled boundary Hilbert space. Meanwhile, the second equality was derived explicitly in the appendix of \cite{xu2024chord}. While in \cite{xu2024chord} the factor \( (\theta_1 | q^{\Delta \hat{n}} | \theta_2) \) was inserted by hand to match the symmetric normalization of the quantum $6j$-symbol, we now see that it follows naturally from the isometry \( \hF \) and the structure of \( \hF \hat{N} \hF^{-1} \). We stress that the internal structure of the theory \emph{uniquely determines the normalization} and thereby yields a properly normalized quantum \( 6j \)-symbol.

Furthermore, due to the explicit form of (\ref{eq:6j energy basis}), we may simply find the crossing symmetry of the $6j$-symbol as:
\begin{equation}
    \langle \Delta; \theta_1, \theta_2 | q^{\Delta_1 \hat{N}} | \Delta; \theta_3, \theta_4 \rangle=\langle \Delta_1; \theta_1, \theta_3 | q^{\Delta \hat{N}} | \Delta_1; \theta_2, \theta_4 \rangle~,\label{eq:swtich property}
\end{equation}
which in terms of chord rules, is a manifestation of the independence of bulk slicing:
\be
\left\langle \begin{tikzpicture}[baseline={([yshift=-0.1cm]current bounding box.center)}, line width=1pt, scale=0.6]

\draw (0,1) arc[start angle= 90, end angle = 270, radius=1];

\draw[thick, blue] (-1,0) -- (0,0);
\fill[blue] (-1,0) circle (1.5pt); 

\node at (-0.3,1.2) {\tiny $\te_1$};
\node at (-0.3,-1.2) {\tiny $\te_2$};
\node at (-0.3,0.3) {\tiny $\Delta$};

\end{tikzpicture}
\right| q^{\Delta_1 \hat{N}} \left| 
\begin{tikzpicture}[baseline={([yshift=-0.1cm]current bounding box.center)}, line width=1pt, scale=0.6]

\draw (0,1) arc[start angle= 90, end angle = -90, radius=1];

\draw[thick, blue] (0,0) -- (1,0);
\fill[blue] (1,0) circle (1.5pt); 

\node at (0.3,1.2) {\tiny $\te_3$};
\node at (0.3,-1.2) {\tiny $\te_4$}; 
\node at (0.3,0.3) {\tiny $\Delta$};

\end{tikzpicture}
\right\rangle = 
\begin{tikzpicture}[baseline={([yshift=-0.1cm]current bounding box.center)}, line width=1pt, scale=0.6]

\draw (0,0) circle (1);

\draw[thick, blue] (-1,0) -- (1,0);
\draw[thick,red] (0,-1) -- (0,1);

\node at (-0.6,1.2) {\tiny $\te_1$};
\node at (-0.6,-1.2) {\tiny $\te_2$};
\node at (0.6,1.2) {\tiny $\te_3$};
\node at (0.6,-1.2) {\tiny $\te_4$};
\node at (0.6,-0.3) {\tiny $\Delta$};
\node at (-0.3, 0.6) {\tiny $\Delta_1$};

\end{tikzpicture}
=
\left\langle \begin{tikzpicture}[baseline={([yshift=-0.1cm]current bounding box.center)}, line width=1pt, scale=0.6]

\draw (0,1) arc[start angle= 90, end angle = 270, radius=1];

\draw[thick, red] (-1,0) -- (0,0);
\fill[red] (-1,0) circle (1.5pt);

\node at (-0.3,1.2) {\tiny $\te_3$};
\node at (-0.3,-1.2) {\tiny $\te_1$};
\node at (-0.3,0.3) {\tiny $\Delta_1$};

\end{tikzpicture}
\right| q^{\Delta \hat{N}} \left| 
\begin{tikzpicture}[baseline={([yshift=-0.1cm]current bounding box.center)}, line width=1pt, scale=0.6]

\draw (0,1) arc[start angle= 90, end angle = -90, radius=1];

\draw[thick, red] (0,0) -- (1,0);
\fill[red] (1,0) circle (1.5pt); 

\node at (0.3,1.2) {\tiny $\te_4$};
\node at (0.3,-1.2) {\tiny $\te_2$}; 
\node at (0.3,0.3) {\tiny $\Delta_1$};

\end{tikzpicture}
\right\rangle.
\ee
The role of $\hO_{\Delta}$ and $\hO_{\Delta_1}$ are inverted under the crossing transformation.

\subsection{Matter Crossing as an Interior State}\label{eq:crossed chord diagram}
In the previous section, we noted that bulk states can be understood as arising from the contraction of boundary states with interior states. One might naturally think that if there exists a consistent Hilbert space description of chord diagrams associated with the double-scaled SYK model—including arbitrary matter insertions along the boundary—then any such chord diagram should admit a slicing in which all matter crossings occur in the interior. This slicing also defines the interior state with all matter crossing in the bulk.

By contracting this interior state with multiple fixed boundary condition wavefunctions in \( \mathcal{H}_0 \), one should recover the full chord diagram correlation function corresponding to those boundary conditions. In this section, we elaborate on this idea by examining the crossed four-point function and connecting components of the interior state to the inner product structure of one-particle states.

Motivated by the discussion above, we begin with the crossed four-point function \eqref{eq:6j energy basis}. We decompose it into contraction between the interior state that contains the matter crossing, and states with fixed energy boundary conditions:
\begin{equation}\label{eq:6j chord number basis}
\Gamma^{\Delta_1}\Gamma^{\Delta_2}\left\{ \begin{array}{ccc}
\Delta_1 & \theta_1 & \theta_2 \\
\Delta_2 & \theta_4 & \theta_3
\end{array} \right\}_q = \sum_{m_1,\dots,m_4} \left\{ \begin{array}{ccc}
\Delta_1 & m_1 & m_2 \\
\Delta_2 & m_4 & m_3
\end{array} \right\}_q \prod_{i=1}^{4} \phi_{E_i}(m_i),
\end{equation}
which defines the interior state with fixed chord number boundary conditions in the right-hand side. The above equation can be illustrated as:
\be
\begin{tikzpicture}[baseline={([yshift=-0.1cm]current bounding box.center)}, line width=1pt, scale=0.75]

\draw (0,0) circle (1);

\draw[thick, blue] (-1,0) -- (1,0);
\draw[thick,red] (0,-1) -- (0,1);

\node at (-0.6,1.2) {\tiny $\te_1$};
\node at (-0.6,-1.2) {\tiny $\te_2$};
\node at (0.6,1.2) {\tiny $\te_3$};
\node at (0.6,-1.2) {\tiny $\te_4$};
\node at (0.6,-0.3) {\tiny $\Delta_1$};
\node at (-0.3, 0.6) {\tiny $\Delta_2$};
\end{tikzpicture}
=
\sum_{m_1\dots m_4 =0}^{\infty}\begin{tikzpicture}[baseline={([yshift=-0.1cm]current bounding box.center)}, line width=1pt, scale=0.75]

\def\r{1}

\draw[thick, blue] (-1,0) -- (1,0);
\draw[thick,red] (0,-1) -- (0,1);

\draw[black,dashed] (\r,0)  .. controls (0.565*0.8,0.565*0.8) .. (0,\r);
\draw[black,dashed] (-\r,0)  .. controls (-0.565*0.8,0.565*0.8) .. (0,\r);
\draw[black,dashed] (\r,0)  .. controls (0.565*0.8,-0.565*0.8) .. (0,-\r);
\draw[black,dashed] (-\r,0)  .. controls (-0.565*0.8,-0.565*0.8) .. (0,-\r);

\node at (-0.8,0.8) {\tiny $m_1$};
\node at (0.8,0.8) {\tiny $m_2$};
\node at (0.8,-0.8) {\tiny $m_3$};
\node at (-0.8,-0.8) {\tiny $m_4$};
\node at (0.3,-0.3) {\tiny $\Delta_1$};
\node at (-0.3, 0.3) {\tiny $\Delta_2$};
\end{tikzpicture}
\prod_{i=1}^{4} 
\begin{tikzpicture}[baseline={([yshift=-0.1cm]current bounding box.center)}, line width=1pt, scale=0.75]

\draw (1,0) arc[start angle = 0, end angle = -180, radius=1];
\draw[dashed,thick] (-1,0)--(1,0); 
\node at (0,0.3) {\tiny $m_i$};
\node at (0, -1.3) {\tiny $E_i$};
\end{tikzpicture}~.
\ee  
We denote the interior state in \eqref{eq:6j chord number basis} as $\left\{\begin{array}{lll}
\Delta_1 & m_1 & m_2 \\
\Delta_2 & m_4 & m_3
\end{array}\right\}_q$ because it corresponds to expanding the quantum $6j$-symbol in terms of complete basis of chord, and we will show how the inner product in $\mh_1$ computes this kernel. 

Note that the left-hand side of \eqref{eq:6j chord number basis} has a crossing symmetry, which implies that the kernel $\left\{\begin{array}{lll}
\Delta_1 & m_1 & m_2 \\
\Delta_2 & m_4 & m_3
\end{array}\right\}_q$  should be invariant under the crossing transformation. We show this by introducing the following coherent basis,
\begin{equation}\label{eq:6j symbol coherent state basis}
\left\{ \begin{array}{ccc}
\Delta_1 & s_1 & s_2 \\
\Delta_2 & s_4 & s_3
\end{array} \right\}_q = \sum_{m_1,\dots,m_4} \left\{ \begin{array}{ccc}
\Delta_1 & m_1 & m_2 \\
\Delta_2 & m_4 & m_3
\end{array} \right\}_q \prod_{i=1}^{4} \frac{s_i^{m_i}}{\sqrt{(q;q)_{m_i}}}.
\end{equation}
We leave the explicit derivation to the appendix ~\ref{app:6j-derivation} and present the result here:
\begin{equation}
\left\{ \begin{array}{ccc}
\Delta_1 & s_1 & s_2 \\
\Delta_2 & s_4 & s_3
\end{array} \right\}_q = \frac{(q^{2\Delta_1 + 2\Delta_2}s_1 s_2 s_3 s_4;q)_\infty}{(q^{\Delta_1+\Delta_2}s_3 s_2, q^{\Delta_1+\Delta_2}s_1 s_4, q^{\Delta_1}s_1 s_2, q^{\Delta_2}s_2 s_4, q^{\Delta_2}s_1 s_3, q^{\Delta_1}s_3 s_4;q)_\infty},
\end{equation}
which is manifestly invariant under the crossing transformation \( (\Delta_1, s_2) \leftrightarrow (\Delta_2, s_3) \). Since the expansion \eqref{eq:6j symbol coherent state basis} is invariant under permutations of the pairs \( (s_i, m_i) \) for \( i = 1, 2, 3, 4 \), this symmetry is inherited by the kernel in the number basis as well.

The coherent basis serves as generating function for chord inner product. To connect the $6j$-symbol to the chord inner product of number basis, we introduce
\be
\langle \Delta;s_1,s_2|\Delta;s_3,s_4 \rangle = \sum^{\infty}_{m_1,\dots,m_4=0} \langle \Delta;m_1,m_2|\Delta;m_3,m_4 \rangle \prod^{4}_{i=1} \frac{s^{m_i}}{\sqrt{(q;q)_i}}.
\ee
The inner product between these coherent states has been worked out in \cite{Xu:2024hoc} (see also \cite{Okuyama:2024more}):
\begin{equation}
\langle \Delta; s_1, s_2 | \Delta; s_3, s_4 \rangle = \frac{(q^{2\Delta} s_1 s_2 s_3 s_4;q)_\infty}{(q^{\Delta}s_1 s_4, q^{\Delta}s_2 s_3, s_1 s_3, s_2 s_4;q)_\infty},
\end{equation}
which allows us to rewrite the coherent component of the \( 6j \)-symbol as
\begin{equation}
\left\{ \begin{array}{ccc}
\Delta_1 & s_1 & s_2 \\
\Delta_2 & s_4 & s_3
\end{array} \right\}_q = \frac{\langle \Delta_1; s_1, s_2 | \Delta_1; q^{\Delta_2} s_3, q^{\Delta_2} s_4 \rangle}{(q^{\Delta_1}s_1 s_2, q^{\Delta_1}s_3 s_4;q)_\infty}.
\end{equation}
Using this identity, together with the Taylor series expansion of $q$-exponential:
\be
\frac{1}{(x;q)_\infty} = \sum^{\infty}_{n=0} \frac{x^n}{(q;q)_n}~.
\ee
We thus obtain a simple relation between the \( 6j \)-symbol in the number basis and the one-particle chord inner product:
\begin{equation}
\left\{ \begin{array}{ccc}
\Delta_1 & m_1 & m_2 \\
\Delta_2 & m_4 & m_3
\end{array} \right\}_q = \sum_{k_1,k_2=0}^{\infty} q^{\Delta_2(m_3+m_4-2k_2) + \Delta_1(k_1+k_2)} \langle \Delta_1; m_1 - k_1, m_2 - k_1 | \Delta_1; m_3 - k_2, m_4 - k_2 \rangle~,
\end{equation}
where states with negative chord number in above summand should be understood as zero. In other words, the right-hand side above is a truncated sum with $k_1 $ ranges from $0$ to $\min(m_1,m_2)$ and $k_2$ ranges from $0$ to $\min(m_3,m_4)$. This explicitly shows that the one-particle chord inner product computes the matrix elements of the quantum \( 6j \)-symbol in the number basis, thereby establishing a direct connection between the two structures.

\subsection{Evaluating a \texorpdfstring{$2m+2$}{} Point Function by Chord Inner Product} \label{sec:chord blocks}
Previously, we showed that a bulk crossing can be understood as the insertion of a closed chord operator \( q^{\Delta \hat{N}} \) into a one-particle state. From this perspective, we can construct a class of \( 2m+2 \)-point functions in which one matter chord of weight \( \Delta \) intersects all the others. This configuration can be evaluated by slicing along the direction of the \( \Delta \) chord and inserting \( m \) closed chord operators successively. 

A subtlety arises in how to impose boundary conditions between each pair of insertions. In this section, we propose a prescription for implementing fixed boundary energy conditions between successive insertions and derive the resulting \( 2m+2 \)-point function using the isometry map \( \hF \).

\begin{figure}
    \centering
    \includegraphics[width=0.5\textwidth]{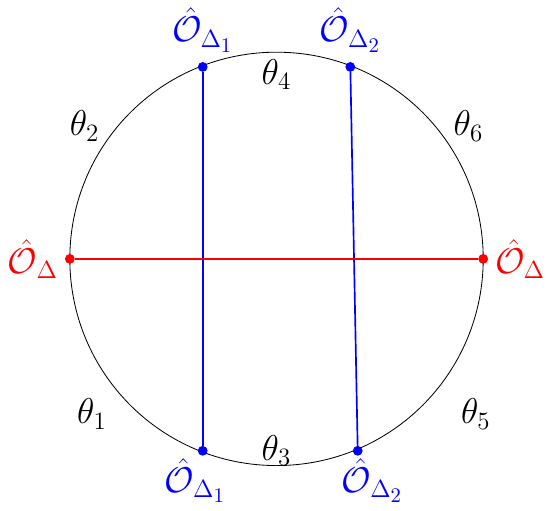}
    \caption{Chord diagram of the six-point correlation function. We again omit the H-chords.}
    \label{fig:6pointchord}
\end{figure}

Before studying the most general case, consider the six-point function e.g. shown in Fig. \ref{fig:6pointchord}. 
By slicing the middle of the correlation function in Fig. \ref{fig:6pointchord} , we can interpret the result as an inner product between two one-particle states, each prepared with a bulk matter crossing and specified boundary energies. A natural starting point is the inner product between the states
\begin{equation}
q^{\Delta_1 \hat{N}} | \Delta; \theta_1, \theta_2 \rangle, \quad \text{and} \quad q^{\Delta_2 \hat{N}} | \Delta; \theta_5, \theta_6 \rangle.
\end{equation}
However, to fix the intermediate boundary energies to be \( \theta_3 \) and \( \theta_4 \), we must project onto \( |\theta_3, \theta_4) \), which lives in \( \mathcal{H}_0 \otimes \mathcal{H}_0 \). To translate this projection into \( \mathcal{H}_1 \), we insert
\begin{equation} \label{eq:projection}
\hF^{-1} |\theta_3, \theta_4) (\theta_3, \theta_4| \hF^{-1\dagger} = \frac{1}{(\theta_3 | q^{\Delta \hat{n}} | \theta_4)} | \Delta; \theta_3, \theta_4 \rangle \langle \Delta; \theta_3, \theta_4 |~,
\end{equation}
into the inner product, yielding
\begin{equation}
I_6 = \frac{\langle \Delta; \theta_3, \theta_4 | q^{\Delta_1 \hat{N}} | \Delta; \theta_1, \theta_2 \rangle\langle \Delta; \theta_5, \theta_6 | q^{\Delta_2 \hat{N}} | \Delta; \theta_3, \theta_4 \rangle}{(\theta_3 | q^{\Delta \hat{n}} | \theta_4)}.
\end{equation}
Evaluating the expectation values gives
\begin{equation}
    \begin{aligned}\label{eq:I6 energy basis}
I_6 &= \gamma_{\Delta}(\theta_5, \theta_6) \gamma_{\Delta_2}(\theta_5, \theta_3) \gamma_{\Delta_1}(\theta_3, \theta_1) \gamma_{\Delta}(\theta_1, \theta_2) \gamma_{\Delta_1}(\theta_2, \theta_4) \gamma_{\Delta_2}(\theta_4, \theta_6) \\
& \quad \times \left\{ \begin{array}{ccc} \Delta & \theta_3 & \theta_4 \\ \Delta_1 & \theta_2 & \theta_1 \end{array} \right\}_q\left\{ \begin{array}{ccc} \Delta & \theta_5 & \theta_6 \\ \Delta_2 & \theta_4 & \theta_3 \end{array} \right\}_q~.
\end{aligned}
\end{equation}
The coefficient in front can be understood as the normalization factor ensuring that the inserted projector is normalized with respect to the inner product in \( \mathcal{H}_1 \).

\begin{figure}
    \centering
    \includegraphics[width=0.55\textwidth]{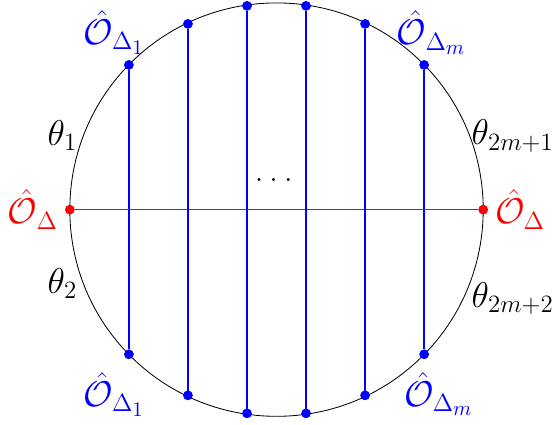}
    \caption{Chord diagram with $m$ particle insertions (blue) and only four crossings allowed in (\ref{eq:general m}), $\Delta$ (red) denotes the conformal weight in the total chord number factor $q^{\Delta\hat{N}}$.}
    \label{fig:2n2_amplitude}
\end{figure}

This structure generalizes to the case of \( m \) closed chord insertions with weights \( (\Delta_2,\dots, \Delta_{m+1}) \), and fixed energies \( \theta_1, \dots, \theta_{2m+2} \). The corresponding \( (2m+2) \)-point function is
\begin{align}\label{eq:2m2 point function}
I_{2m+2} &= \langle \Delta; \theta_{1}, \theta_{2} | q^{\Delta_{1} \hat{N}} \left[ \prod_{i=1}^{m-1} \frac{ | \Delta; \theta_{2i+1}, \theta_{2i+2} \rangle \langle \Delta; \theta_{2i+1}, \theta_{2i+2} | }{(\theta_{2i+1} | q^{\Delta \hat{n}} | \theta_{2i+2})} q^{\Delta_{i+1} \hat{N}} \right] | \Delta; \theta_{2m+1}, \theta_{2m+2} \rangle \nonumber\\
&= \gamma_{\Delta}(\theta_1, \theta_2) \gamma_{\Delta}(\theta_{2m+1}, \theta_{2m+2}) \prod_{j=1}^m \gamma_{\Delta_{j}}(\theta_{2j-1}, \theta_{2j+1}) \gamma_{\Delta_{j}}(\theta_{2j}, \theta_{2j+2}) \\
& \quad \times \prod_{j=1}^m \left\{ \begin{array}{ccc} \Delta & \theta_{2j-1} & \theta_{2j} \\ \Delta_{j} & \theta_{2j+2} & \theta_{2j+1} \end{array} \right\}_q~,\nonumber
\end{align}
where the product in the middle means in between $q^{\Delta_i \hat{N}}$ and $q^{\Delta_{i+1} \hat{N}}$, we impose the boundary condition $E_L = E(\te_{2i+1}), E_R = E(\te_{2i+2})$ for the intermidiate one-particle state with $\Delta$ by inserting the projection  \eqref{eq:projection}
See Fig. \ref{fig:2n2_amplitude} for the corresponding chord diagram.

\section{Single shockwave geometries from the DSSYK path integral}\label{sec:single shock}
In this section, we study dynamical properties of the model under time evolution generated by the two-sided chord Hamiltonian, with a focus on the one-particle Hilbert space. Using path integral methods, we derive the classical phase space structure of the theory. The path integral prepares the so-called wormhole density matrix, originally introduced in the \( N \)-site SYK model~\cite{Berkooz:2022fso}. To simplify the evaluation, we construct a double-scaled version of the corresponding states in the extended chord space~\cite{Lin:2022rbf}. Our analysis reveals smooth wormhole geodesics connecting the asymptotic boundaries of an effective AdS\(_2\) black hole geometry \footnote{The relevant notion of real time geometry for the dynamical process under consideration is only effective as the states should be naturally thought as prepared by a path integral on a fake disk \cite{Lin:2023trc}. We will discuss this in the last part of this section.}, exhibiting sub-maximal chaotic behavior characteristic of this background.
\paragraph{Outline}
In Sec. \ref{ssec:chord states} we briefly review the general construction of many particle chord states. 
In Sec. \ref{ssec:double scaled wormhole} we formulate double-scaled wormhole density matrices. We derive their corresponding Heisenberg equations, and the initial conditions for expectation values for observables evaluated in these states. In Sec. \ref{ssec:PI} we then develop a path integral to prepare the density matrices. The classical phase space solutions correspond to the expectation values of the total chord number of the DSSYK with a one-particle insertion, and its conjugate momenta, both evaluated with respect to the wormhole density matrix. In Sec. \ref{ssec:sshockwaves 1p} we study the shockwave limit for the saddle point solution, which corresponds to the expectation value of the total chord number. This follows from introducing double-scaled precursor operators (based on the quantum circuit definition by \cite{Stanford:2014jda}) and working out the limiting behavior for the total chord number in the shockwave limit as defined by \cite{Shenker:2013pqa}. This corresponds to an early perturbation of the TFD state. We study the conditions for the total chord number/dual geodesic length to exhibit scrambling behavior (a transition from exponential to linear growth).  In Sec. \ref{ssec:Krylov operator complexity} we develop the Krylov space of the wormhole density matrices. We derive the Krylov basis and Lanczos coefficient. We show that the saddle point solutions we found correspond to Krylov operator complexity of the wormhole density matrices. In Sec. \ref{ssec:holo dictionary} we derive the holographic dictionary relating Krylov operator complexity of the wormhole density matrices, and minimal geodesic lengths in the shockwave geometry. Later in Sec. \ref{ssec:probing chaos} we examine our results in terms of quantum chaos in the dual fake disk geometry of the DSSYK model \cite{Lin:2023trc}.

\subsection{Chord states with many particles}\label{ssec:chord states}
We consider the explicit DSSYK two-sided Hamiltonian with arbitrary many matter chords \cite{Lin:2022rbf} in canonical variables \cite{Aguilar-Gutierrez:2025pqp}
\begin{subequations}\label{eq:H many particles}
    \begin{align}\label{eq:H LR many particles1}
&\tfrac{\sqrt{\lambda(1-q)}}{J}\hH_{L}=\rme^{-\rmi \hat{P}_0}+\sum_{i=0}^m\rme^{\rmi \hat{P}_i}(1-\rme^{-\hat{\ell}_i})\rme^{-\hat{\ell}_i^<}\quad\text{where}\quad\hat{\ell}_i^<=\sum_{j=0}^{i-1}\qty(\hat{\ell}_j+\lambda\Delta_{j+1})~,\\
&\tfrac{\sqrt{\lambda(1-q)}}{J}\hH_{R}=\rme^{-\rmi \hat{P}_m}+\sum_{i=0}^m\rme^{\rmi \hat{P}_i}(1-\rme^{-\hat{\ell}_i})\rme^{-\hat{\ell}_i^>}\quad\text{where}\quad\hat{\ell}_i^>=\sum_{j={i+1}}^{m}\qty(\hat{\ell}_j+\lambda\Delta_{j+1})~.\label{eq:H LR many particles2}
\end{align}
\end{subequations}
The chord Hilbert space with $m$-particle insertions  $\mh_m$ is constructed as \cite{Lin:2022rbf},
\begin{equation}\label{eq:Fock space with matter}
    \mathcal{H}_{m}=\bigoplus_{n_0,n_1,\cdots,n_m=0}^\infty\mathbb{C}\ket{\tilde{\Delta};n_0,n_1,\cdots,n_m}~,
\end{equation}
where $\tilde{\Delta}=\qty{\Delta_1,\cdots,\Delta_m}$ represents a string of matter operator insertions
\begin{equation}\label{eq:reference pets}
\hat{\mathcal{O}}_{\tilde{\Delta}}=\qty{\hat{\mathcal{O}}_{\Delta_{1}},\dots,~\hat{\mathcal{O}}_{\Delta_{m}}}~,  \quad m\in\mathbb{Z}_{\geq0}~;
\end{equation}
$n_0$ is the number of open $H$-chords to the left of all matter chords, $n_1$ the number between the first two particles, all the way up to the number of chords between all the $m$ particles.

In the following, we denote a class of states built by acting operators in the double-scaled algebra to the empty chord state as
\begin{equation}\label{eq:multiple particles}
    \ket{\Psi^{L/R}_{\tilde{\Delta}}(\tau_0,\dots,\tau_m)}=\rme^{-\tau_0\hH_{L/R}}\hat{\mathcal{O}}^{L/R}_{\Delta_1}\rme^{-\tau_1\hH_{L/R}}\hat{\mathcal{O}}^{L/R}_{\Delta_2}\cdots\hat{\mathcal{O}}^{L/R}_{\Delta_m}\rme^{-\tau_m\hH_{L/R}}\ket{0}~,
\end{equation}
where $\tau_i\in\mathbb{C}$, while the matter chord operator $\hat{O}_{\Delta}$ either adds or deletes a new matter open chord with weight $\Delta$:
\begin{equation}\label{eq:DS creation annih}
    \hat{\mathcal{O}}^{L/R}_{\Delta}=b_{L/R}^\dagger+b_{L/R}~,
\end{equation}
\begin{equation}
    \begin{aligned}
        &\hat{b}_{L}^{\dagger}\ket{\tilde{\Delta};n_{0},\cdots,n_{m}}=\ket{\Delta,\tilde{\Delta};0,n_{0},\cdots,n_{m}}~,\\
    &\hat{b}_{R}^{\dagger}\ket{\tilde{\Delta};n_{0},\cdots,n_{m}}=\ket{\Delta,\tilde{\Delta};n_{0},\cdots,n_{m},0}~.
    \end{aligned}
\end{equation}
In case that all the matter chords are of the same type with equal weights $\Delta_1=\dots=\Delta_m:=\Delta_*$, we can express the left matter annihilation operator $\hat{b}_L$ as:
\begin{equation}\label{eq:property operator2}
    \hat{b}_{L}\ket{\Delta^{(m)};n_{0},\cdots,n_{m}}	=\sum_{j=1}^{m}q^{(j-1)\Delta_*^2}q^{\Delta\sum_{l<j}n_{l}}\ket{\Delta_*^{(m-1)};n_{0},\cdots,n_{j-2},n_{j-1}+n_{j},\cdots,n_{m}}~,
\end{equation}
where $\Delta^{(m)}$ represents a string of the same $\Delta$ $m$-times. If some of the conformal weights $\Delta_{1\leq i\leq m}$ are different from each other, we only need to account for those with the same conformal weight as $\hat{\mathcal{O}}_{\Delta}$.

\subsection{Double-scaled wormhole density matrices}\label{ssec:double scaled wormhole}
We now introduce a double-scaled version of the wormhole density matrices based on \cite{Berkooz:2022fso}. The motivation for doing so is that they showed that two-sided two-point function in the triple scaling limit reproduce geodesic lengths in JT gravity. We want to find out if the path integral of the DSSYK knows about these types of solutions without assuming a triple scaling limit. For simplicity, this section focuses on one-particle states. We will generalize it later with arbitrary finite number of particles.

Based on the finite $N$ SYK original construction in \cite{Berkooz:2022fso} (and its double-scaling limit), we define wormhole density matrices (in operator basis) as
\begin{equation}\label{eq:rho LR}
    \hat{\rho}^{(\rm op.\,basis)}_{\Delta_i}=\rme^{-\frac{\beta_L}{4}\hH}\hO^R_{\Delta_i}(t_R)\rme^{-\qty(\frac{\beta_L}{2}\hH_L+\frac{\beta_R}{2}\hH_R)}\hO^L_{\Delta_i}(t_L)\rme^{-\frac{\beta_R}{4}\hH}~,
\end{equation}
where
\begin{align}\label{eq:double scaled precursor shockwave}
        \hat{\mathcal{O}}^{L}_{\Delta_{i}}(t_i)=\rme^{\rmi t_i \hH_{L}}\hat{\mathcal{O}}^{L}_{\Delta_{i}}\rme^{-\rmi t_i\hH_{L}}~,\quad \hat{\mathcal{O}}^{R}_{\Delta_{i}}(t_i)=\rme^{-\rmi t_1 \hH_{R}}\hat{\mathcal{O}}^{R}_{\Delta_{i}}\rme^{\rmi t_i\hH_{R}}~,\quad\qty(\hat{\mathcal{O}}_\Delta^\dagger=\hO_\Delta)~.
        \end{align}
Meanwhile, it's also useful to think of the density matrix in terms of its canonical purification. In terms of \eqref{eq:multiple particles}, we have:
\begin{equation}\label{eq:rho LR2}
    \hat{\rho}^{(\rm st.\,basis)}_{\Delta_i}=\ket{\Psi^{R}_{{\Delta}_{i}}(\tau_R^*,\tau_R)}\bra{\Psi^{L}_{{\Delta}_{i}}(\tau_L,\tau_L^*)}~,
\end{equation}
where $\tau_{L/R}=\frac{\beta_{L/R}}{2}+\rmi t_{L/R}$, and $^*$ refers to complex conjugation. This state in the one-particle Hilbert space \( \mathcal{H}_1 \) reproduces the same two-sided two-point correlation functions as those computed in~\cite{Berkooz:2022fso}, where the authors instead define states in the finite-\( N \) SYK Hilbert space and subsequently take the double-scaling limit. Namely, we define the expectation values in the wormhole density matrix in state \eqref{eq:rho LR} or in the operator basis \eqref{eq:rho LR2} as
\begin{align}\label{eq:two-sided correlator}
   &\frac{\Tr[\hat{\rho}_{\Delta_{i}}q^{\Delta\hat{N}}]_q}{Z_{\Delta_i}(\beta_L,\beta_R)}=\frac{(\Omega|\rme^{-\frac{\beta_L}{2}\hH_L}\hat{\mathcal{O}}^{L}_{\Delta_{i}}(t_L)\rme^{-\frac{\beta_L}{2}\hH_L}q^{\Delta\hat{N}}\rme^{-\frac{\beta_R}{2}\hH_R}\hat{\mathcal{O}}^{R}_{\Delta_{i}}(t_R)\rme^{-\frac{\beta_R}{2}\hH_R}|\Omega)}{(\Omega|\rme^{-\frac{\beta_L}{2}\hH_L}\hat{\mathcal{O}}^{L}_{\Delta_{i}}\rme^{-\frac{\beta_L}{2}\hH_L-\frac{\beta_R}{2}\hH_R}\hat{\mathcal{O}}^{R}_{\Delta_{i}}\rme^{-\frac{\beta_R}{2}\hH_R}|\Omega)}~,
\end{align}
where
\begin{align}
        Z_{\Delta_i}(\beta_L,\beta_R):=\bra{\Delta_i;0,0}\rme^{-\beta_L\hH_L-\beta_R\hH_R}\ket{\Delta_i;0,0}~.\label{eq:partition function 1p}
    \end{align}
Note that (\ref{eq:two-sided correlator}) is a two-sided correlation function, where the operators at each side evolve unitarily. The trace involved here is the standard Type I trace in $\mh_1$, such that the $\text{Tr} (\hat{\rho}_\Delta(\cdot))$ is the same as taking expectation value under state $|\Psi_{\Delta}\rangle$.

\paragraph{Heisenberg picture}
Whilst not necessary, we find it convenient to work in the Heisenberg picture, where we define the two-sided evolving operators:
\begin{equation}\label{eq:new operators Heisenberg wormhole}
\begin{aligned}
    \hat{\ell}(\tau_L,\tau_R)&:=\rme^{-\tau_L^*\hH_L-\tau_L\hH_R}\hat{\ell}~\rme^{-\tau_R\hH_L-\tau_R^*\hH_R}~,\\
   \rme^{-\rmi\hat{P}_{0/m}(\tau_L,\tau_R)}&:=\rme^{-\tau_L^*\hH_L-\tau_L\hH_R}\rme^{-\rmi\hat{P}_{0/m}}\rme^{-\tau_R\hH_L-\tau_R^*\hH_R}~,
\end{aligned}
\end{equation}
{{where $\hel:=\hel_L+\hel_R$ and $\hel_{L}:=\hel_0$, $\hel_R:=\hel_m$ are the operators in the Hamiltonian \eqref{eq:H many particles}. The corresponding}} expectation values of {{\eqref{eq:new operators Heisenberg wormhole}}} can be expressed as
\begin{equation}\label{eq:equiv to H and S pictures}
    \begin{aligned}
       &\bra{\Delta;0,0}\hat{\ell}(\tau_L,\tau_R)\ket{\Delta;0,0}= \bra{\Psi_{{\Delta}}(\tau_L,\tau^*_L)}\hat{\ell}\ket{\Psi_{{\Delta}}(\tau^*_R,\tau_R)}~,\\
   &\bra{\Delta;0,0}\rme^{-\rmi\hat{P}_{0/m}(\tau_L,\tau_R)}\ket{\Delta;0,0}= \bra{\Psi_{{\Delta}}(\tau_L,\tau^*_L)}\rme^{-\rmi\hat{P}_{0/m}}\ket{\Psi_{{\Delta}}(\tau_R^*,\tau_R)}~.
\end{aligned}
\end{equation}
We now deduce the Heisenberg equation of motion by taking derivatives in (\ref{eq:new operators Heisenberg wormhole}),
\begin{subequations}\label{eq:Heisenberg wormhole}
\begin{align}
    &\begin{aligned}\label{eq:length variable LR}
        -\rmi&\qty(\partial_{t_L}+\partial_{t_R})\hat{\ell}(\tau_L,\tau_R)=\qty[\hH_L-\hH_R,~\hat{\ell}(\tau_L,\tau_R)]\\
        &=\lambda~\rme^{-(\tau^*_L+\tau_R)\hH_L-(\tau^*_R+\tau_L)\hH_R}(\hH_L-\hH_R)-\tfrac{2J\sqrt{\lambda}}{\sqrt{1-q}}\qty(\rme^{-\rmi \hat{P}_0(\tau_L,\tau_R)}-\rme^{-\rmi \hat{P}_m(\tau_L,\tau_R)})~,
    \end{aligned}\\
&\begin{aligned}
-\rmi\qty(\partial_{t_L}+\partial_{t_R})\rme^{-\rmi P_{0/m}(\tau_L,\tau_R)}&=\qty[\hH_L-\hH_R,~\rme^{-\rmi P_{0/m}(\tau_L,\tau_R)}]\\
=&J\sqrt{\tfrac{1-q}{\lambda}}\Bigg(\rme^{-(\tau^*_L+\tau_R)\hH_L-(\tau^*_R+\tau_L)\hH_R}+\rme^{-2\rmi \hat{P}_{0/m}(\tau_L,\tau_R)}\\
&\qquad\quad-\tfrac{\sqrt{\lambda(1-q)}}{J}\rme^{-\rmi \hat{P}_{0/m}(\tau_L,\tau_R)}\hH_{L/R}-\rme^{-\hat{\ell}(\tau_L,\tau_R)}q^{\Delta_{\rm S}}\Bigg)~,
\end{aligned}
\end{align}
\end{subequations}
{{where we have used the commutation relations \cite{Lin:2023trc}
\begin{equation}
 \begin{aligned}
 \qty[\hH_{L/R},~\hat{N}]&=\hH_{L/R}-\frac{2}{\sqrt{\lambda}}\ha^\dagger_{L/R}~,\quad \qty[\hH_{L/R},~\ha^\dagger_{R/L}]=\frac{q^{\Delta}+\hat{N}}{\sqrt{\lambda}}~.\\
     \qty[\hH_{L/R},~\ha^\dagger_{L/R}]&=\frac{J^2}{\sqrt{\lambda}}+\frac{(1-q)}{\sqrt{\lambda}}(\ha_{L/R}^\dagger)^2-(1-q)\ha_{L/R}^\dagger\hH_{L/R}
 \end{aligned}
\end{equation}
with the total chord number being $\hat{N}:= \hel/\lambda$ and $\ha^\dagger_{L/R}:=\frac{J}{\sqrt{1-q}}\rme^{-\rmi \hat{P}_{0/m}}$ in our notation. }}
Alternatively, one finds the same equations of motion for the expectation values of the above operators in the Schrödinger picture. Note that due to the term $\hH_L-\hH_R$ in (\ref{eq:length variable LR}), we look for solutions where $E_L=E_R$ to avoid explicit time dependent terms in the equations of motion. This means that we describe a system in thermodynamic equilibrium despite a particle insertion. This corresponds to a shockwave perturbation of the TFD state \cite{Shenker:2013yza}, as we will clarify below.

Next, we deduce the initial conditions that the expectation values of the canonical operators must satisfy when $\beta_L=\beta_R\equiv\beta/2$

\begin{subequations}\label{eq:initial quantum picture}
    \begin{align}\label{eq:initial length exp val}
    &\hspace{4cm}\bra{\Delta;0,0}\hat{\ell}\qty(\tfrac{\beta}{2},\tfrac{\beta}{2})\ket{\Delta;0,0}= \ell_*~,\\
    &\label{eq:initial velocity}
    {{\eval{\partial_{t_{L}}\bra{{\Delta};0,0}\hat{\ell}(\tau_L,\tau_R)\ket{{\Delta};0,0}}_{\tau_{L/R}=\frac{\beta}{2}}=\eval{\partial_{t_{R}}\bra{{\Delta};0,0}\hat{\ell}(\tau_L,\tau_R)\ket{{\Delta};0,0}}_{\tau_{L/R}=\frac{\beta}{2}}=0~,}}
\end{align}
\end{subequations}
where in the first condition, $\ell_*$ is a constant (determined by $\beta$), {{and the last one\footnote{{{One can express \eqref{eq:initial velocity} as
\begin{equation}
    \rmi\eval{\partial_{t_{L}}\bra{{\Delta};0,0}\hat{\ell}(\tau_L,\tau_R)\ket{{\Delta};0,0}}_{\tau_{L/R}=\frac{\beta}{2}}=\bra{{\Delta};0,0}\rme^{-\frac{\beta}{2}(\hH_L+\hH_R)}[\hH_{L},~\hat{\ell}]\rme^{-\frac{\beta}{2}(\hH_L+\hH_R)}\ket{{\Delta};0,0}~,
\end{equation}
and similarly for $\eval{\partial_{t_{R}}\bra{{\Delta};0,0}\hat{\ell}(\tau_L,\tau_R)\ket{{\Delta};0,0}}_{\tau_{L/R}=\frac{\beta}{2}}$.}}}, is proved in \cite{Aguilar-Gutierrez:2025pqp} (2.21b).}} Below, we study the path integral that prepares the wormhole density matrix based on the results above.

\subsection{A Path Integral Description}\label{ssec:PI}
We work with the path integral in the canonical variables of the Hamiltonian (\ref{eq:H many particles}) with one particle insertion\footnote{The analysis can be easily extended with composite operators \cite{Aguilar-Gutierrez:2025pqp}; however, we focus on non-composite matter chord operators for later discussion on the multiple shocks and the switchback effect (Sec. \ref{ssec:switchback many}).}
\begin{equation}\label{eq:Hamilton PI}
    \int\prod_{i=L,~R}[\rmd \ell_i][\rmd P_i]\exp\qty[\int\rmd\tau_L\rmd\tau_R\qty(\frac{\rmi}{\lambda}\sum_{i=L,~R}P_i\qty(\partial_{\tau_L}+\partial_{\tau_R})\ell_i-(H_L-H_R))]~,
\end{equation}
The saddle point ($\lambda
\rightarrow0$) equations of motion are \footnote{Note that the Hamiltonian contains a normalization factor dependent on $\lambda$ such that the equations of motion remains non-trivial in the $\lambda\to0$ limit. }
\begin{subequations}
\label{eq:ell P evol many}
\begin{align}
\frac{1}{\lambda}\qty(\partial_{t_{L}}+\partial_{t_{R}})\ell&=\sum_{i=0}^m\pdv{(H_{L}-H_{R})}{P_{i}}~,\label{eq:ell evol many}\\
\frac{1}{\lambda}\qty(\partial_{t_{L}}+\partial_{t_{R}}){P_{i}}&=-\pdv{(H_{L}-H_{R})}{\ell_{i}}~,\label{eq:P evol many}
\end{align}
\end{subequations}
where $\ell=\ell_L+\ell_R$. This reproduces the Heisenberg picture ones in (\ref{eq:Heisenberg wormhole}) after replacing the commutators for Poisson brackets with the $\rmi$ factor, leading to the classical limit of the Heisenberg picture system in (\ref{eq:Heisenberg wormhole}). 

We look for a class of solutions where we can split the time dependence of the system (\ref{eq:ell P evol many}) into the left/right Hamiltonian sectors, as
\begin{subequations}\label{eq:ell P evol many split}
    \begin{align}
\frac{1}{\lambda}\pdv{\ell}{t_{L}}&=\sum_{i=L,~R}\pdv{H_{L}}{P_{i}}~,&&\frac{1}{\lambda}\pdv{\ell}{t_{R}}=-\sum_{i=L,~R}\pdv{H_{R}}{P_{i}}\label{eq:ell evol many split}\\
\frac{1}{\lambda}\pdv{P_{i}}{t_{L}}&=-\pdv{H_{L}}{\ell_{i}}~,&&\frac{1}{\lambda}\pdv{P_{i}}{t_{R}}=\pdv{H_{R}}{\ell_{i}}~,\label{eq:P evol many split}
\end{align}
\end{subequations}
as explained below the Heisenberg equations (\ref{eq:Heisenberg wormhole}), the solutions with conserved energy are those where $E_L=E_R=E$, which we parametrize as in (\ref{eq:conserved energies}). Then, (\ref{eq:ell P evol many split}) with (\ref{eq:conserved energies}) and the classical version of the Hamiltonians (\ref{eq:H many particles}) leads us to
\begin{subequations}\label{eq:shockwave EOM}
    \begin{align}
-\frac{1-q}{\lambda}\dv{\ell}{\tilde{t}_{R}}&=2\rmi \qty(\cos\theta-\rme^{-\rmi P_R(t_L,t_R)})~,&&\frac{1-q}{\lambda}\dv{\ell}{\tilde{t}_{L}}=2\rmi \qty(\cos\theta-\rme^{-\rmi P_L(t_L,t_R)})~,\label{eq:eq1many}\\
-\dv{P_{L}}{\tilde{t}_{R}}&=-{ q^{\Delta_{i}}}\rme^{-\ell(t_L,t_R)}\rme^{\rmi P_{L}(t_L,t_R)}~,&&\dv{P_{R}}{\tilde{t}_{L}}=-{q^{\Delta_{i}}}\rme^{-\ell(t_L,t_R)}\rme^{\rmi P_{R}(t_L,t_R)}~,\label{eq:eq2many}\\
-\dv{P_{R}}{\tilde{t}_{R}}&=2\qty(\cos\theta-\cos P_{R}(t_L,t_R))~,&&\dv{P_{L}}{\tilde{t}_{L}}=2\qty(\cos\theta-\cos P_{L}(t_L,t_R))~,\label{eq:eq3many}
\end{align}
\end{subequations}
where $\tilde{t}_{L/R}\equiv J\sqrt{\frac{1-q}{\lambda}}t_{L/R}$. 

Since we want to prepare wormhole density matrices that obey the Heisenberg picture, our set of initial conditions is (\ref{eq:initial quantum picture})\footnote{Note that the Hamiltonian \eqref{eq:H many particles} is Hermitian for the particle chord inner product introduced in \cite{Lin:2023trc} (also summarized in Sec 2.5 in \cite{Ambrosini:2024sre}; and extended with multiple particles in \cite{Xu:2024hoc}). This choice of the inner product is in fact encoded in the initial values \eqref{eq:in cond many} that are implemented to solve the classical system. The reason is that the inner product of the chord Hilbert space is determined by the chord algebra \cite{Lin:2023trc,Xu:2024hoc}, which has its own definition of Hermitian conjugate (i.e. the $*$-structure of the corresponding $*$-algebra) as well as commutation relations. The chord algebra is the input used to deduce the initial conditions for the expectation values that are then applied to define the initial conditions in the saddle point equations of motion of the path integral in (\ref{eq:in cond many}). This guarantees that the Hamiltonians $\hH_{L/R}$ implemented in this work are Hermitian. Furthermore, the results for the shockwave limit (see Sec. \ref{ssec:sshockwaves 1p}) are consistent with previous literature where the Hamiltonian (with chord inner product) is manifestly Hermitian (for instance it matches the renormalized wormhole geodesic length in \cite{Berkooz:2022fso}).} in terms of classical fields: \footnote{We are seeking for solutions with conserved energy $E_L=E_R=E(\theta)$ given by \eqref{eq:conserved energies}. Therefore, the initial length $\ell_* = \ell_* (\theta)$ is also parameterized in terms of $\theta$, which we specify below.}
\begin{equation}\label{eq:in cond many}
\ell{(t_L=t_R=0)}=\ell_*(\theta)\in\mathbb{R}~,\quad \eval{\pdv{\ell}{t_{L/R}}}_{t_L=t_R=0}=0~.
\end{equation}
The solutions to the system above are then
\begin{subequations}\label{eq:solutions shockwave}
    \begin{align}\label{eq:womrhole semiclassical answer many}
&\ell(t_L,t_R)=2\log \qty(\cosh(\gamma t_L)\cosh(\gamma t_R)-\frac{q^{\Delta_{i}}\rme^{-\ell_*}}{\sin^2\theta}\sinh(\gamma t_L)\sinh(\gamma t_R))+\ell_*~,\\
&P_{L}(t_L,t_R)=\rmi\log(\cos\theta+\rmi\sin\theta\frac{\rme^{\ell_*}\sin^2\theta\tanh(\gamma t_{L})-q^{\Delta_{i}}\tanh(\gamma t_{R})}{\rme^{\ell_*}\sin^2\theta-q^{{\Delta_{i}}}\tanh(\gamma t_R)\tanh(\gamma t_L)})~,\label{eq:canonical momenta0 many}\\
&P_{R}(t_L,t_R)=\rmi\log(\cos\theta+\rmi\sin\theta\frac{q^{\Delta_{i}}\tanh(\gamma t_{L})-\rme^{\ell_*}\sin\theta\sin\theta\tanh(\gamma t_{R})}{\rme^{\ell_*}\sin^2\theta-q^{{\Delta_{i}}}\tanh(\gamma t_R)\tanh(\gamma t_L)})~,\label{eq:canonical momentam many}
\end{align} 
\end{subequations}
where $\gamma=J\sin\theta$.

Introducing (\ref{eq:solutions shockwave}) in the Hamiltonians $H_{L/R}$, which have a conserved energy spectrum (\ref{eq:conserved energies}), we recover
\begin{equation}\label{eq:l 0m}
    \rme^{-\ell_{L/R}(t_L,t_R)}=\frac{2\cos P_{L/R}(t_L,t_R)-2\cos\theta_{L/R}-q^{\Delta_{i}}\rme^{-\ell(t_L,t_R)}\rme^{\rmi P_{L/R}(t_L,t_R)}}{\rme^{\rmi P_{L/R}(t_L,t_R)}-q^{\Delta_{i}}\rme^{\rmi P_{R/L}}}~.
\end{equation}
We can then determine the constant $\ell_*(\theta)$ in (\ref{eq:womrhole semiclassical answer many}) by demanding that $\ell=\ell_L+\ell_R$ with the solutions in (\ref{eq:l 0m}), which means\footnote{It would be interesting to deduce how the above expressions are modified for other configurations beyond the one-particle irrep. However, this will not be needed for the shockwave analysis.}
\begin{equation}\label{eq:length0 Delta many}
\rme^{-\frac{1}{2}\ell_*(\theta)}=\frac{q^{-\Delta_{i}}}{2}\qty(-1+q^{\Delta_{i}}+\sqrt{1+q^{2{\Delta_{i}}}-2q^{\Delta_{i}} \cos2\theta})~.
\end{equation}
By comparing this solution with~\eqref{eq:initial length exp val}, we can extract the effective inverse temperature \( \beta(\theta) \) for the microcanonical ensemble from the saddle point configuration, as defined in~\eqref{eq:micro inv temp}. This observation will be useful in the subsequent discussion of chaos bounds.

\subsection{Shockwave solutions from the DSSYK model}\label{ssec:sshockwaves 1p}
We define the double-scaled precursor operators in the same way as $\hat{\mathcal{O}}^{L}_{\Delta_{i}}(-t_i)$ and $\hat{\mathcal{O}}^{R}_{\Delta_{i}}(-t_i)$ in (\ref{eq:double scaled precursor shockwave}), however, \emph{in the Schrödinger picture} instead of Heisenberg; where, $t_{i}\in\mathbb{R}$ is a fixed insertion time of the precursor operators. Physically, the precursors evolve states, like $\ket{\Psi^{L/R}_{\Delta_{i}}}$ in (\ref{eq:rho LR})), backwards or forwards in time (depending on $t_i$), inserting a matter chord operator and evolving in the opposite time direction afterwards. This definition is motivated by the notion of timefolds in quantum circuits \cite{Stanford:2014jda}, which generates a partial cancellation in the value that circuit complexity can reach. From the bulk perspective, this corresponds to inserting a defect in the future and in the past of a two-sided AdS background \cite{Stanford:2014jda,Berkooz:2022fso}.

Note performing the above timefold in (\ref{eq:double scaled precursor shockwave}) shifts the overall time dependence of the wormhole density matrix (\ref{eq:rho LR}) in operator basis \cite{Berkooz:2022fso} as
\begin{equation}\label{eq:rho LR precursor}
    \hat{\rho}_{\Delta_{i}}=\rme^{-\frac{\beta}{4}\hH}\hO^R_{\Delta_i}(t_R-t_i)\rme^{-\frac{\beta}{2}\hH_L}\hO^L_{\Delta_i}(t_L-t_i)\rme^{-\frac{\beta}{4}\hH}~.
\end{equation}
So, the Heisenberg picture operators (\ref{eq:new operators Heisenberg wormhole}) get shifted by the backwards/forwards insertion time $t_i$. The initial conditions are then
\begin{equation}\label{eq:in cond single precursor}
\eval{\ell}_{t_L=t_R=t_i}=\ell_*(\theta)~,\qquad \eval{\dv{\ell}{t_{L/R}}}_{t_L=t_R=t_i}=0~,
\end{equation}
where, for concreteness, we will be interested in the case $t_i>0$ (the opposite one is similar).

\paragraph{Saddle point solutions} Note that the solutions of (\ref{eq:shockwave EOM}) take the same form as (\ref{eq:womrhole semiclassical answer many}) with a time shift $t_{L/R}\rightarrow t_{L/R}-t_i$, i.e.
\begin{equation}\label{eq:almost shockwave}
\begin{aligned}
\ell(t_L,t_R)=2\log \Biggl(&\qty(1-\frac{q^{\Delta_{i}}\rme^{-\ell_*}}{\sin^2\theta})\sinh(J\sin\theta (t_L-t_i))\sinh(J\sin\theta (t_R-t_i))\\
&+\cosh(J\sin\theta(t_L-t_R))\Biggr)+\ell_*~.
\end{aligned}
\end{equation}
In particular, we take the insertion time $t_i\gg \abs{t_{L/R}}$, then (\ref{eq:almost shockwave}) becomes
\begin{equation}\label{eq:shockwave approx}
\begin{aligned}
\ell(t_L,t_R)\simeq
\begin{cases}
2\log \biggl(\cosh((t_L-t_R)J\sin\theta)+\frac{\alpha_i}{2}\rme^{-\qty(t_L+t_R)J\sin\theta }\biggr)+\ell_*(\theta)&t_i\gg \abs{t_{L/R}}~,\\
    2\log \biggl(\cosh((t_L-t_R)J\sin\theta)+\frac{\alpha_i}{2}{\rme^{(t_L+t_R)J\sin\theta }}\biggr)+\ell_*(\theta)&-t_i\gg \abs{t_{L/R}}~,
\end{cases}
\end{aligned}
\end{equation}
where we denote
\begin{equation}\label{eq:alpha sw}
\alpha_{i}\equiv\qty(1-\frac{q^{\Delta_{i}}\rme^{-\ell_*(\theta)}}{\sin^2\theta})\frac{\rme^{2\abs{t_i}J\sin\theta}}{2}~.
\end{equation}

\paragraph{Shockwave limit} We can now make connection with the literature on shockwaves \cite{Shenker:2013pqa,Shenker:2013yza,Berkooz:2022fso}. The solutions of the form (\ref{eq:shockwave approx}) describe geodesic lengths in JT gravity seen as a dimensional reduction of the BTZ black hole in the s-wave sector \cite{Shenker:2013yza}. The strict shockwave limit requires that $\alpha_{i}$ is fixed. This can be achieved with (\ref{eq:alpha sw}) by considering the semiclassical limit (i.e. 
$\lambda\rightarrow0$) for light operators, and taking the insertion time $\abs{t_i}\rightarrow\infty$, while keeping
\begin{equation}\label{eq:alpha sw2}
\alpha_{i}\eqlambda\frac{{\Delta_{i}}\lambda}{2\sin\theta}\rme^{2\abs{t_i}J\sin\theta}\quad:\text{ fixed}~.
\end{equation}
The result is also in agreement with a chord diagram calculation (albeit in the triple-scaling limit) in \cite{Berkooz:2022fso}.\footnote{There are a few factors that differ from \cite{Berkooz:2022fso}. While (5.48) in \cite{Berkooz:2022fso} is computed in the canonical ensemble, it agrees with our (\ref{eq:alpha sw2}) when their parameter $T$ takes the role of half the dimensionless fake temperature $2\pi/\sin\theta$ and after rescaling $\lambda\rightarrow\lambda^{3/2}$. The latter one is attributed to the different scaling of the energy spectrum, $E=\frac{2}{\sqrt{1-q}}\cos\theta$, compared to our (\ref{eq:conserved energies}).} 

\paragraph{Scrambling properties}
We will consider two limiting cases for the dual geodesic length (\ref{eq:shockwave approx}).
\begin{itemize}
    \item \textbf{Exponential growth}: For concreteness, consider $t_i<0$. In the early time regime $\alpha_i\rme^{J\sin\theta(t_L+t_R)}\ll1$, we can approximate (\ref{eq:shockwave approx}) by
    \begin{equation}\label{eq:length like parameter}
\begin{aligned}
\ell\simeq&2\log\cosh(J\sin\theta(t_L-t_R))+\ell_*(\theta)\\
&+\alpha_i\sech(J\sin\theta(t_L-t_R))\rme^{J\sin\theta(t_L+t_R)}+\mathcal{O}(\alpha_i^2)~.    
\end{aligned}
\end{equation}
    While the exponential term above is suppressed at late enough times when $t_L=-t_R$; in the special case that $t_L=t_R:=t$ (\ref{eq:length like parameter}) can grow exponentially, i.e.
    \begin{equation}\label{eq:exp growth Krylov}
        \ell(t_L=t_R=t)=\ell_*(\theta)+\alpha_i\rme^{2J\sin\theta~t}+\mathcal{O}(\alpha_i^2)~.
    \end{equation}
    \item \textbf{Linear growth}: On the other hand, when $\alpha_i\rme^{J\sin\theta(t_L+t_R)}\gg1$,
    \begin{equation}\label{eq:sb 1}
    \ell(t_L,t_R)\simeq 2J\sin\theta\qty(t_L-2t_i+t_R-2t_{\rm sc})~.
\end{equation}
Here $t_{\rm sc}$ is the time scale when $\ell(t_L,t_R)$ (\ref{eq:almost shockwave}) changes from exponential to linear growth, i.e.
\begin{equation}
    \frac{\sin^2\theta-\rme^{-\ell_*}q^{\Delta_{i}}}{\sin^2\theta+\rme^{-\ell_*}q^{\Delta_{i}}}\rme^{2J\sin\theta~t_{\rm sc}^{(i)}}\simeq1~.
\end{equation}
This is simply solved by
\begin{equation}\label{eq:scrambling time}
    t_{\rm sc}^{(i)}\simeq-\frac{1}{2J\sin\theta}\log\qty(\frac{\sin^2\theta-q^{\Delta_{i}}\rme^{-\ell_*(\theta)}}{\sin^2\theta+q^{\Delta_{i}}\rme^{-\ell_*(\theta)}})\eqlambda-\frac{1}{2J\sin\theta}\log(\frac{\Delta_{i}\lambda}{2\sin\theta})~,
\end{equation}
where in the last step we are considering the leading order term in the semiclassical expansion with $\Delta_i\sim\mathcal{O}(1)$; however, we do not need to assume that $\hat{\mathcal{O}}_{\Delta_i}$ is necessarily light in the remainder of the text. Note that in this limiting case, we can express the fixed term (\ref{eq:alpha sw2}) as
\begin{equation}
    \alpha_i=\rme^{-2J\sin\theta\qty(t_i+t_{\rm sc}^{(i)})}~,
\end{equation}
which agrees with (4.3) of \cite{Stanford:2014jda}. Furthermore, one finds the same result with $t_i\rightarrow-t_i$ in the case \eqref{eq:shockwave approx}.
\item In contrast, when $t_R\neq t_L$ in the regime of {small backreaction} is substantially different. The term $\log(\cosh(J\sin\theta(t_L-t_R)))$ dominates (\ref{eq:shockwave approx}).
This indicates that general $t_L\neq t_R$ case (\ref{eq:KRylov complexity shockwave}) does not exhibit scrambling. In particular, when $t_L=-t_R\equiv t$, (\ref{eq:shockwave approx}) is just
    \begin{equation}\label{eq:ell 4 crossed function}
        \ell(t)\simeq2\log(\cosh(2J\sin\theta t)+\frac{\alpha_{i}}{2})+\ell_*(\theta)~,
    \end{equation}
    which indicates the function only grows parabolic at early times ($t\ll (J\sin\theta)^{-1}$), and linearly at late times (i.e. $t\gg(J\sin\theta)^{-1}$). Thus, although all the bulk geodesic distances (i.e. when $t_{L/R}$ are arbitrary) in the semiclassical limit can be interpreted as Krylov operator complexity; not all display the above scrambling transition, as seen by contrasting the $t_L=t_R$ case with $t_L=-t_R$.
\end{itemize}
Therefore, we can express (\ref{eq:shockwave approx}) as
\begin{equation}\label{eq:unified shockwave length}
    \ell(t_L,t_R)\simeq2\log\qty(\cosh(J\sin\theta(t_L-t_R))+\frac{1}{2}\rme^{J\sin\theta(\abs{t_L-t_i}+\abs{t_R-t_i}-2t_{\rm sc}^{(i)})})~.
\end{equation}

\subsection{Krylov operator complexity}\label{ssec:Krylov operator complexity}
In this section we explore the relation between the length operator and the Krylov operator complexity of the wormhole density matrix. To begin with, we note that each of the states decomposed in terms of the same time-independent basis, as
\begin{align}\label{eq:final states Krylov}
&\ket{\Psi^{L}_{{\Delta_{i}}}(\tau_{L,i},\tau_{i,L})}=\sum_n\Psi^{\Delta_i}_n(\tau_{L,i})\ket{K^{\Delta_{i}}_n}~,\\
&\ket{\Psi^{R}_{{\Delta_{i}}}(\tau_{i,R},\tau_{R,i})}=\sum_n\Psi^{{\Delta_i}}_n(\tau_{i,R})\ket{K^{\Delta_{i}}_n}~,
\end{align}
where $\tau_{i,j}=\frac{\beta}{4}+\rmi(t_i-t_j)$. This follows from the fact that the only difference between  $\tau_{L/R}$ in  $\ket{\Psi^{L}_{{\Delta_{i}}}(\tau_{L,i},\tau_{i,L})}$ and $\ket{\Psi^{R}_{{\Delta_{i}}}(\tau_{i,R},\tau_{R,i})}$ in (\ref{eq:final states Krylov}).

In particular, we consider the Choi–Jamiołkowski isomorphism \cite{jamiolkowski1972linear,choi1975completely} between 
\begin{equation}
\rme^{-\beta\hH_{L/R}}\hO^{L/R}_{\Delta_i}(t_{L/R}+t_i)\rme^{-\beta\hH_{L/R}}~,\quad\text{and}\quad\ket{\Psi^{L/R}_{\Delta_i}(\beta+\rmi (t_{L/R}+t_i),\beta-\rmi (t_{L/R}+t_i))}~,    
\end{equation}
to construct $\ket{K^{\Delta_{i}}_n}$ as the Krylov basis for the operator $\hO^{L/R}_{\Delta_i}(t_{L/R}+t_i)$, which is recovered through the Lanczos algorithm
    \begin{align}    b_{n+1}&\ket{K^{\Delta_{i}}_{n+1}}=(\hH_R-\hH_L)\ket{K_{n}^{\Delta_{i}}}-a_n\ket{K^{\Delta_{i}}_{n}}-b_n\ket{K^{\Delta_{i}}_{n-1}}~,\label{eq:Lanczos precursor}
\end{align}
where $a_n$ and $b_n$ are the Lanczos coefficients; and we initialize the algorithm with the state 
\begin{equation}\label{eq:initial Krylov state}
\ket{K^{\Delta_i}_0}=\ket{\Delta_{i};0,0}~.    
\end{equation}
Note that (\ref{eq:Lanczos precursor}) follows from the Liouvillian operator for a precursor operator $\hO^L_{\Delta_i}(t_i)$ being $\mathcal{L}=\hH_L-\hH_R$ (while $\mathcal{L}=\hH_R-\hH_L$ for $\hO^R_{\Delta_i}$), given that $\rme^{-\rmi \hat{\mathcal{L}}t}\hO_{\Delta_i}(t_i)=\hO_{\Delta_i}(t+t_i)$.
We also denote the observables in (\ref{eq:final states Krylov}) as
\begin{equation}
    \begin{aligned}\label{eq:amplitudes single shock}
&\Psi^{\Delta_i}_n(\tau_{L,i})=\bra{K_n^{\Delta_i}}\ket{\Psi^{L}_{{\Delta_{i}}}(\tau_{L,i},\tau_{i,L})}~,\\
&\Psi^{{\Delta_i}}_n(\tau_{i,R})=\bra{K^{\Delta_{i}}_n}\ket{\Psi^{R}_{{\Delta_{i}}}(\tau_{i,R},\tau_{R,i})}~.
\end{aligned}
\end{equation}
We then define the Krylov complexity operator and Krylov operator complexity as
\begin{equation}\label{eq:Krylov operator complexity definition}
    \hat{\mathcal{C}}_{\Delta_i}=\sum_nn\ket{K^{\Delta_i}_{n}}\bra{K^{\Delta_i}_{n}}~,\quad \mathcal{C}_{\Delta_i}(t_L,t_R)\equiv\frac{\Tr(\hat{\rho}_{\Delta_{i}}~\hat{\mathcal{C}})}{Z_{\Delta_i}(\tfrac{\beta}{2},\tfrac{\beta}{2})}~,
\end{equation}
where $Z_{\Delta_i}({\beta}/{2},{\beta}/{2})$ is in (\ref{eq:partition function 1p}).

The Krylov basis $\ket{K_n^{\Delta_i}}$ for $\hO_{\Delta_i}(t-t_i)$ follows in the same way as \cite{Aguilar-Gutierrez:2025pqp,Ambrosini:2024sre}, namely when $\beta=0$ (or when $t_L=t_R$ \cite{Aguilar-Gutierrez:2025pqp})
\begin{equation}\label{eq:explicit Krylov basis}
    \ket{K_n^{\Delta_i}}=c_n\sum_{k=0}^n(-1)^k\begin{pmatrix}
        n\\
        k
    \end{pmatrix}\ket{\Delta_i;k,n-k}~,\quad c_l\eqlambda\sqrt{\frac{2^{l+1}}{\lambda~l!(l+1)!}}~,
\end{equation}
where we used $\frac{\lambda^l(1-q)^l}{2^l(q^{1/2};q^{1/2})_l(q^{1/2};q^{1/2})_{l+1}}\eqlambda\frac{2^{l+1}}{\lambda l!(l+1)!}$ in the last equality; and
\begin{equation}\label{eq:Lanczos shocwaves}
    a_n=0~,\quad b_n\eqlambda\frac{2J}{\sqrt{\lambda}}\sqrt{[n/2]_q(1- q^{n/2+\Delta_{i}})}~.
\end{equation}
This follows from replacing (\ref{eq:a dagger a eta=-1}) in
\begin{subequations}\label{eq:a dagger a eta=-1}
    \begin{align}
    &(\hat{a}^\dagger_R-\hat{a}^\dagger_L)\sum_{k=0}^n(-1)^k\begin{pmatrix}
        n\\
        k
    \end{pmatrix}\ket{\Delta_{i};k,n-k}=\sum_{k=0}^{n+1}(-1)^k\begin{pmatrix}
        n+1\\
        k
    \end{pmatrix}\ket{\Delta_{i};k,n+1-k}~,\\
    &(\hat{a}_R-\hat{a}_L)\sum_{k=0}^n(-1)^k\begin{pmatrix}
        n\\
        k
    \end{pmatrix}\ket{\Delta_{i};k,n-k}=\sum_{k=0}^{n-1}A_k(n)(-1)^k\begin{pmatrix}
        n-1\\
        k
    \end{pmatrix}\ket{\Delta_{i};k,n-1-k}~,\\
    &\text{with}\quad A_k(n)=\frac{n[n-k]_q}{n-k}(1-q^{\Delta_{i}+k})+\frac{n[k+1]_q}{k+1}(1-q^{\Delta_{i}+n-1-k})~,\nonumber
\end{align}
\end{subequations}
and using the fact that the binomial factor $\begin{pmatrix}
        n\\
        k
    \end{pmatrix}$ localizes at $k=n/2$ for the type of ansatz in (\ref{eq:explicit Krylov basis}) \cite{Ambrosini:2024sre}. Note that, for general values of $\beta$, the Krylov basis and Lanczos coefficients depends explicitly on $\beta$.

We will now consider the semiclassical approximation (where $\lambda n_{L/R}$ is fixed as $\lambda\rightarrow0$) such that the total chord number is equal to the Krylov complexity operator \cite{Aguilar-Gutierrez:2025pqp} 
\begin{equation}\label{eq:total chord number}
    \hat{N}\eqlambda\sum_nn\ket{K^{\Delta_i}_n}\bra{K^{\Delta_i}_n}~.
\end{equation}
This indicates that the semiclassical Krylov operator complexity of the wormhole density matrix with a precursor (\ref{eq:rho LR precursor}) obtained as a saddle point of the path integral (\ref{eq:Hamilton PI}) when $-t_i\gg \abs{t_{L/R}}$ is the semiclassical expectation value of the total chord number in (\ref{eq:shockwave approx}):
\begin{equation}\label{eq:KRylov complexity shockwave}
    \mathcal{C}_{\Delta_i}(t_L,t_R)\eqlambda\frac{2}{\lambda} \log(\cosh(J(t_L-t_R)\sin\theta)+\frac{1}{4}\qty(1-\frac{q^{\Delta_{i}}\rme^{-\ell_*}}{\sin^2\theta})\rme^{J\sin\theta(t_L+t_R-2t_i)})+\ell_*(\theta)~.
\end{equation}
Thus, the effect of the precursor is to create a time shift in the evolution of Krylov operator complexity, or equivalently in the amplitudes $\Psi_n^{\Delta_i}$ (\ref{eq:amplitudes single shock}) in comparison with inserting a matter chord operator $\hO_{\Delta_i}(t_i)$. Moreover, note that during the regime of exponential growth for Krylov operator complexity in (\ref{eq:exp growth Krylov}), we observe that the Lyapunov growth for the total time variable $t_L+t_R=2t$ is $J\sin\theta$. This is indeed the same Lyapunov growth as an OTOC in the ``fake'' disk of the bulk dual of the DSSYK model (as we explain in Sec. \ref{ssec:probing chaos}). This saturates the bound on the Krylov operator complexity exponent in \cite{Parker:2018yvk} (see also \cite{Aguilar-Gutierrez:2025pqp}).

\paragraph{Alternative reference state}We stress that in deducing the above relation between the total chord number with a one-particle insertion and Krylov operator complexity of the Heisenberg picture operator $\hO_{\Delta_i}(t-t_i)$; we select as the initial state $\ket{\Delta_i;0,0}$ (\ref{eq:initial Krylov state}) in the Lanczos algorithm \eqref{eq:Lanczos precursor}. Thus, the result (\ref{eq:KRylov complexity shockwave}) is different from the Krylov complexity of a matter chord operator $\hO_{\Delta_i}(t)$, where the Lanczos algorithm is initialized by the precursor operator acting on the vacuum chord state
\begin{equation}
\label{eq:alt state 1}    \ket{K_0^{\Delta_1}}'=\frac{\rme^{-\frac{\beta}{4}\qty(\hH_{L}+\hH_{R})}\hO_{\Delta_1}(t_1)|\Omega)}{\sqrt{(\Omega|\hO_{\Delta_1}(t_1)\rme^{-\frac{\beta}{2}\qty(\hH_{L}+\hH_{R})}\hO_{\Delta_1}(t_1)|\Omega)}}~,
\end{equation}
where $t_1$ is a fixed parameter. However, given that the Liouvillian is the same as in (\ref{eq:Lanczos precursor}) but the reference state is a different one compared to (\ref{eq:initial Krylov state}). In principle, this could result in another valid observable that may be included in the holographic dictionary of the DSSYK model.
We discuss about this further in Sec. \ref{ssec:switchback Krylov}.

The analysis above suggests that the dual gravitational description can help identify the appropriate quantum information observable to include in the holographic dictionary. For example, one could have chosen an alternative definition of Krylov complexity based on~\eqref{eq:alt state 1}, which may or may not admit a smooth geometric interpretation. This guidance from the bulk is particularly valuable when the quantum complexity measure becomes more intricate, as in Sec.~\ref{ssec:switchback Krylov}, where the reference state involves multiple operator insertions. 

\subsection{Holographic dictionary}\label{ssec:holo dictionary}
We just found that the Krylov operator complexity for the density matrix (\ref{eq:rho LR precursor}) translates into the wormhole length of an AdS$_2$ black hole shockwave geometry \cite{Shenker:2013pqa}\footnote{While \cite{Shenker:2013pqa} considers a BTZ black hole instead of an AdS$_2$ black hole, their results on geodesic lengths in shockwave geometries are also applicable to JT gravity since there is no angular dependence, so that the s-wave reduction leads to the same answer.} dual to (\ref{eq:shockwave approx}):
\begin{equation}\label{eq:AdS shockwave length}
    L_{\rm AdS}(u_L,u_R)=2 \log(\cosh(\frac{\Phi_h}{2}(u_L-u_R))+\frac{\alpha_{\rm sw}}{2}\rme^{\pm\frac{\Phi_h}{2}(u_L+u_R)})+L_{\rm reg}~.
\end{equation}
Here $u_{L/R}$ denote the AdS$_2$ boundary times, where the sign in the exponent depends on the insertion time in the same way as in (\ref{eq:shockwave approx}); $\alpha_{\rm sw}$ is the shockwave shift parameter; $L_{\rm reg}$ is the regularized length; and $\Phi_h$ the value of the dilaton at the black hole horizon. Thus, from (\ref{eq:KRylov complexity shockwave}) and (\ref{eq:AdS shockwave length}) we can deduce the holographic dictionary of Krylov complexity $L_{\rm AdS}=\lambda~\mathcal{C}$: 
\begin{subequations}\label{eq:holo dictionary}
    \begin{align}
        &u_{L/R}=t_{L/R}~,\quad &&\Phi_h=2J\sin\theta~,\\
        &\alpha_{\rm sw}=\frac{1}{4}\qty(1-\frac{q^{\Delta_{i}}\rme^{-\ell_*}}{\sin^2\theta})\rme^{2\abs{t_i}J\sin\theta}~,\quad &&L_{\rm reg}=\ell_*(\theta)~.
    \end{align}
\end{subequations}
This gives a concrete realization of the bulk-to-boundary map of \cite{Lin:2022rbf}.

The shockwave geometries associated with the double-scaled precursor operators (\ref{eq:double scaled precursor shockwave}) in the DSSYK model are qualitatively very similar to those in AdS space. However, there are remnants of $q$-deformation and the microstructure that are encoded in how the length reacts to the matter insertion as well as its dependence on the energy parameters (\ref{eq:length0 Delta many}).In addition, by setting \( t_L = -t_R = t \), we can extract from the solution~\eqref{eq:AdS shockwave length} the scrambling time \( t_* = 2\pi / \beta_{\text{fake}} \), where we define the effective inverse temperature for the microcanonical ensemble of the dual AdS geometry as
\begin{equation}\label{eq:fake temperature}
\beta_{\text{fake}} = \frac{\pi}{J \sin\theta}~.
\end{equation}
This effective temperature, referred to as the ``fake'' temperature in~\cite{Lin:2023trc} and also identified in~\cite{Blommaert:2024ydx}, is closely tied to the Lyapunov exponent governing the out-of-time-order correlator (OTOC). In the next section, we will show how this temperature naturally emerges from a saddle point evaluation of the OTOC, and how it is connected to the ``fake'' disk geometry relevant for the scramblon dynamics described in~\cite{Lin:2023trc}.

\subsection{Probing chaos in the fake disk}\label{ssec:probing chaos}
We would like to interpret the results of effective temperature $\beta_{\text{fake}}$ in terms of chaos as probed by out-of-time-order correlators (OTOCs), and along the lines of fake disk introduced in \cite{Lin:2023trc}. For concreteness, we study the case where $-t_1\gg\abs{t_{L/R}}$ in this subsection, although the same analysis carries for $t_1\gg\abs{t_{L/R}}$.\footnote{We remind the reader that $-t_1$ is the time where we insert the matter operator $\mo_{\Delta_1}$, and evolve the state by the two Hamiltonian $H_{L/R}$ with time $t_{L/R}$ respectively.  }

\paragraph{Two-sided two-point (crossed four-point) correlation functions}
Let us first consider the two-sided two-point function in the one-particle chord space (i.e. a crossed four-point function) in Fig. \ref{fig:otocprecursor},
\begin{equation}
\begin{aligned}\label{eq:6j OTOC}
        {\Tr(\hat{\rho}_{\Delta_{1}}~q^{\Delta\hat{N}})}=&\int\qty(\prod_{j=1}^4\rmd\mu(\theta_j)\rme^{-\tau_jE_j})\left\{ \begin{array}{ccc} \Delta & \theta_1 & \theta_2 \\ \Delta_1 & \theta_4 & \theta_3 \end{array} \right\}_q\cdot\\
        &\qquad\qquad\cdot\gamma_{\Delta}(\te_1,\te_2)\gamma_{\Delta}(\te_3,\te_4) \gamma_{\Delta_1}(\te_1,\te_4)\gamma_{\Delta_1}(\te_2,\te_3)~.
\end{aligned}
\end{equation}
The one-shot wormhole density matrix (\ref{eq:rho LR precursor}) allows to identify the following time configuration:
\begin{equation}
\begin{aligned}
    &\tau_1=\frac{\beta}{4}-\rmi (t_L-t_1)~,\quad\tau_2=\frac{\beta}{4}+\rmi (t_L-t_1)~,\\
    &\tau_3=\frac{\beta}{4}+\rmi (t_R-t_1)~,\quad\tau_4=\frac{\beta}{4}-\rmi (t_R-t_1)~.
\end{aligned}
\end{equation}
We can apply our results to deduce the expectation value of total chord number in the semiclassical limit (\ref{eq:shockwave approx}) for the one-shot wormhole density matrix (\ref{eq:rho LR precursor}). Performing the explicit evaluation of (\ref{eq:6j OTOC}), with the saddle-point solution of length (\ref{eq:unified shockwave length}) leads to\footnote{\label{fnt:saddle points}Alternatively, we refer the reader for an explicit chord diagram evaluation of the crossed four-point function to \cite{Aguilar-Gutierrez:2025pqp} App E. We emphasize that the evaluation of $\Tr(\rho_{\Delta_i}\hat{N})$ is simpler to carry out through the path integral (\ref{eq:Hamilton PI}) rather than the chord amplitude \eqref{eq:6j OTOC} (and taking derivative with respect to $\Delta$). However, both approaches evaluate the same quantity, as we have shown in Secs. \ref{ssec:double scaled wormhole} and \ref{ssec:PI}.}
\begin{equation}
\frac{\Tr(\hat{\rho}_{\Delta_1}~q^{\Delta\hat{N}})}{Z_{\Delta_1}\qty(\tfrac{\beta}{2},\tfrac{\beta}{2})}\eqlambda\rme^{-\Delta~\ell(t_L,t_R)} =\frac{\rme^{-\Delta\ell_*(\theta)}}{\biggl(\cosh((t_L-t_R)J\sin\theta)+\frac{\alpha_1}{2}\rme^{(t_L+t_R)J\sin\theta }\biggr)^{2\Delta}}~,\label{eq:twosidedtwopoint}
\end{equation}
where $\ell_*$ is in \eqref{eq:initial length exp val}, while $\alpha_1$ appears in (\ref{eq:alpha sw}), and the saddle point value of energy $\theta$ is related to $\beta$ by the following relation: \footnote{There are other saddle point solution for the inverse temperature, but they are thermodynamically unstable; see App E in \cite{Aguilar-Gutierrez:2025pqp}.}
\begin{equation}\label{eq:micro inv temp}
\beta(\theta)=\frac{2\theta-\pi}{J\sin\theta} ~,   
\end{equation}
where the saddle point values are given by $\theta_{1\leq i\leq 4}=\theta+\mathcal{O}(\lambda)$ for light fields (i.e. $\Delta_i\sim\mathcal{O}(1)$ as $\lambda\rightarrow0$) \cite{Aguilar-Gutierrez:2025pqp}.

We stress that the reason that the saddle point solutions of the path integral (\ref{eq:Hamilton PI}) leads to crossed four-point functions is that we are evaluating expectation values with respect to states in the one-particle chord Hilbert space, which means 
\begin{equation}
\ell(t_L,t_R)\eqlambda\lambda\Tr[\rho_{\Delta_1}\hat{N}]/Z_\Delta(\beta/2,\beta/2)~.    
\end{equation}
This can then be exponentiated to recover the crossed four-point function \eqref{eq:twosidedtwopoint} (discussed in Sec. \ref{ssec:fixed energy wavefunctions}).

Focusing on the case $t_L=t_R$, the combination (\ref{eq:twosidedtwopoint}) can be interpreted as an OTOC of the form \cite{Berkooz:2022fso},
\begin{equation}\label{eq:explicit OTOC}
    \Tr(\rme^{-\frac{\beta}{4}\hH}\hO_{\Delta}(0)\rme^{-\frac{\beta}{4}\hH}\hO_{\Delta_1}(t-t_1)\rme^{-\frac{\beta}{4}\hH}\hO_{\Delta}(0)\rme^{-\frac{\beta}{4}\hH}\hO_{\Delta_1}(t-t_1))~,
\end{equation}
where the left side corresponds operators acting on the physical Hilbert space of the SYK model in the double-scaling limit, and the right side to double-scaled operators (which instead act on the auxiliary Hilbert space $\mH_m$).

Since we know how to evaluate the right-hand side of (\ref{eq:explicit OTOC}) in the double-scaling limit through (\ref{eq:twosidedtwopoint}), we note that the OTOC experiences a transition when $\alpha_1$ is fixed and $t$ increases, i.e.
\begin{equation}\label{eq:4pnt function}
\begin{aligned}
    \frac{\Tr(\hat{\rho}_{\Delta_1}~q^{\Delta\hat{N}})}{Z_{\Delta_1}\qty(\tfrac{\beta(\theta)}{2},\tfrac{\beta(\theta)}{2})}\simeq
    \begin{cases}
        
    \rme^{-\Delta\ell_*(\theta)}-\alpha_1\Delta\rme^{2J\sin\theta~t}&\text{for}~~(J\sin\theta)^{-1}\ll t\ll t_{\rm sc},\\
    \qty(\frac{2}{\alpha_1})^{2\Delta}\rme^{-4J\Delta\sin\theta(t-t_1-t^{(1)}_{\rm sc})-\Delta\ell_*(\theta)}&\text{for}~~t\gg t_{\rm sc}~,
    \end{cases}
\end{aligned}
\end{equation}
which indicates that for fixed $\alpha$ there is a \emph{scrambling} transition as $t$ decreases. In contrast, it can be seen from (\ref{eq:twosidedtwopoint}) that when $t_L=- t_R$, there occurs no transition of the type in (\ref{eq:4pnt function}). 
The Lyapunov exponent of the OTOC evaluated in the wormhole density matrix with the pair of precursor operator insertions is then
\begin{equation}\label{eq:Lyapunov exp}
    \lambda_{\rm L}=2J\sin\theta=\frac{2(2\theta-\pi)}{\beta(\theta)}=\frac{2\pi}{\beta_{\rm fake}}~,
\end{equation}
where to make direct contact with \cite{Lin:2023trc}, we have expressed the result in terms of the fake inverse temperature $\beta_{\text{fake}}= \beta/v$ (\ref{eq:fake temperature}) which is slightly larger than the physical length of the Euclidean circle $\beta$, with $v\in(0,1)$ related to $\theta$ as:
\be
\theta=\frac{\pi v}{2}+\frac{\pi}{2},\quad  \frac{\pi v}{J\beta}= \cos\frac{\pi v}{2}, \quad \lambda_{L}= \frac{2\pi v}{\beta} = \frac{2\pi}{\beta_{\text{fake}}}.  
\ee

\paragraph{Geometry of Sub-maximal Chaos} 
The expression~\eqref{eq:Lyapunov exp} shows that the theory, in its semi-classical regime, generally does not saturate the maximal Lyapunov exponent~\cite{Maldacena:2015waa,Xu:2019lhc} at the physical temperature \( \beta \). A geometric interpretation of the sub-maximal chaos exponent in~\eqref{eq:Lyapunov exp} is provided in~\cite{Lin:2023trc}, through a careful analysis of the semi-classical limit of the symmetry algebra of chord operators. In this limit, the algebra acts naturally as the near-horizon \( \mathfrak{sl}_2 \) algebra on the so-called fake disk, which contains the physical Euclidean circle as a proper subset, along with an extended “fake” region whose size depends on the temperature. The resulting scrambling behavior is governed by this algebra and therefore saturates the chaos bound with respect to the fake inverse temperature, which corresponds to the boundary length of the fake circle.

Our analysis of \eqref{eq:6j OTOC} and \eqref{eq:explicit OTOC} reveals that the quantum \( 6j \)-symbol encodes the same sub-maximal Lyapunov exponent. This observation is consistent with the original insight of~\cite{Berkooz:2018jqr}, which established a connection between OTOCs in DSSYK and quantum \( 6j \)-symbols across a broad range of parameters (see also~\cite{Berkooz:2022fso,Aguilar-Gutierrez:2025pqp}).  Moreover, the appearance of the \( 6j \)-symbol in our analysis suggests a possible interpretation of scrambling dynamics as a shockwave scattering process\footnote{We refer interested readers to~\cite{Lam_2018} for discussion of this perspective in Schwarzian quantum mechanics.} on a 'fake' spacetime, whose Euclidean counterpart is the fake disk. A notable feature of this description is that it regulates UV behavior in a manner compatible with the underlying symmetry structure. Specifically, correlators defined on the fake disk are naturally UV finite—even at infinite physical temperature.

We leave a more precise formulation of this idea, as well as an exploration of the corresponding real-time geometry, to future work.

\section{Multiple shockwaves and the switchback effect}\label{sec:mutiple shocks}
We will now turn to the states with multiple double-scaled precursor operators (\ref{eq:double scaled precursor shockwave}) insertions.

\paragraph{Outline}In Sec. \ref{ssec:wormholes many} we generalize the wormhole density matrices with many particle chords and its corresponding Heisenberg equations. In Sec. \ref{ssec:switchback many} we derive the switchback effect with two and greater number of particle states using our results on chord diagrams with one-particle insertions (Sec. \ref{sec:chord blocks}). We show that in the semiclassical limit and by considering the timefold conditions for shockwave solutions (initially proposed by \cite{Stanford:2014jda} in quantum circuits and a holographic complexity conjecture) we recover the switchback effect in the total chord number. At last, in Sec. \ref{ssec:switchback Krylov} we show that the result on the switchback effect can be interpreted in terms of Krylov complexity for the precursor operators, and spread complexity of the HH state between the timefold insertions under specific conditions.

\subsection{Multishot wormhole density matrices}\label{ssec:wormholes many}
 Now, we allow for $m$-particle chords in the analysis (i.e. $2m$ operators). We use the most general two-sided Hamiltonian in (\ref{eq:H many particles}) with $\ell_{0\leq i\leq m}$ and $P_{0\leq i\leq m}$ as canonical variables. The generalization of the wormhole density matrix with multiple precursors in operator (\ref{eq:rho LR}) or state basis (\ref{eq:rho LR2}) is respectively
\begin{subequations}
\begin{align}\label{eq:multiple precursor state}
    \hat{\rho}^{(\rm op.\,basis)}_{\tilde{\Delta}}\equiv&\qty(\rme^{-\tau_L^*\hH}\hO_{\Delta_1}(-t_1)\rme^{-\tau_L^*\hH}\hO_{\Delta_2}(-t_2)\cdots\hO_{\Delta_m}(-t_m))^\dagger\cdot\\
    &\cdot\rme^{-\tau_R^*\hH}\hO_{\Delta_1}(-t_1)\rme^{-\tau_R^*\hH}\hO_{\Delta_2}(-t_2)\cdots\hO_{\Delta_m}(-t_m)\nonumber\\
    \hat{\rho}^{(\rm st.\,basis)}_{\tilde{\Delta}}=&\ket{\Psi^{R}_{\tilde{\Delta}}\qty(\tau_{1,R},\rmi t_{2,1},\dots,\tau_{R,m})}\bra{\Psi^{L}_{\tilde{\Delta}}(\tau_{L,1},\rmi t_{1,2},\dots,\tau_{m,L})}
    ~,
\label{eq:womrhole density m particles}\end{align}
\end{subequations}
where we again denote $\tilde{\Delta}=\qty{\Delta_1,\dots,\Delta_m}$, $\tau_{i,j}\equiv\frac{\beta}{4}+\rmi t_{i,j}$ and $t_{i,j}=t_i-t_j$. Given that \eqref{eq:multiple precursor state} is prepared with multiple operator insertions, we refer to it as a multishoot density matrix, based on previous literature \cite{Berkooz:2022fso}.

We define the operators $\hat{\ell}(\tau_L,\tau_R)\equiv\sum_{i=0}^m\hat{\ell}_i(\tau_L,\tau_R)$, and $\hat{P}_{0/m}(\tau_L,\tau_R)$ in a similar way as in (\ref{eq:multiple precursor state}), with $0=L$, $m=R$. The overall $\rme^{-\tau_{L/R}\hH_{L/R}}$ factors remain in the same way as in the one-particle wormhole density matrix (\ref{eq:rho LR}). To be more explicit, the expectation values of the Heisenberg picture operators is given by
\begin{subequations}
\begin{align}\label{eq:H vs S}
    &\Tr(\hat{\rho}_{\tilde{\Delta}}\hat{\ell})=(\Omega|\hat{\mathcal{O}}^{L}_{\Delta_m}(-t_m)\cdots\hat{\mathcal{O}}^{L}_{\Delta_1}(-t_1)\hat{\ell}(\tau_L,\tau_R)\hat{\mathcal{O}}^{R}_{\Delta_1}(-t_1)\cdots\hat{\mathcal{O}}^{R}_{\Delta_m}(-t_m)|\Omega)~,\\
   &\Tr(\hat{\rho}_{\tilde{\Delta}}\rme^{-\rmi\hat{P}_{0/m}})=(\Omega|\hat{\mathcal{O}}^{L}_{\Delta_m}(-t_m)\cdots\hat{\mathcal{O}}^{L}_{\Delta_1}(t_1)\rme^{-\rmi\hat{P}_{0/m}(\tau_L,\tau_R)}\hat{\mathcal{O}}^{R}_{\Delta_1}(-t_1)\cdots\hat{\mathcal{O}}^{R}_{\Delta_m}(-t_m)|\Omega)~,
\end{align}
\end{subequations}
where we used the notation in (\ref{eq:double scaled precursor shockwave}), and the expectation value is taken with respect to the empty state $|\Omega)$ \cite{Xu:2024hoc}. In the following section, we compare it to multiple shockwave geometries studied in \cite{Shenker:2013yza}.

\subsection{Multiple-shocks from correlation functions}\label{ssec:switchback many}
\begin{figure}
    \centering
    \includegraphics[width=0.7\textwidth]{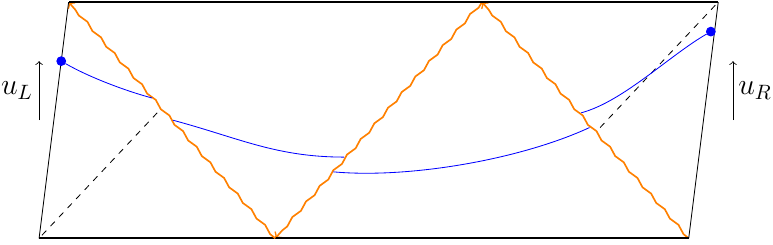}
    \caption{Example of a multiple shockwave geometry in conformal coordinates \cite{Shenker:2013yza} with $m=4$ shocks. Each pair of precursor operator insertions in the left and right chord sectors $\hO^{L/R}_{\Delta_i}(t_i)$ generates the individual shocks in the bulk (orange). The shocks are emitted in the future/past ``singularities'' by preparing wormhole density matrices (\ref{eq:multiple precursor state}) in the boundary theory \cite{Berkooz:2022fso}. The geodesic curved anchored at asymptotic AdS$_2$ boundaries (blue) at the times $u_{L/R}$ is generated by the operator $q^{\Delta\hat{N}}$ in the $(2m+2)$-correlation functions (\ref{eq:2m2 point function}).}
    \label{fig:multi_shock}
\end{figure}

We have found that the wormhole density matrix with a single precursor operator (\ref{eq:rho LR precursor}) generates a total chord number dual to a shockwave geodesic length (\ref{eq:AdS shockwave length}) that can be interpreted as Krylov operator complexity (\ref{eq:KRylov complexity shockwave}). We now seek to generate a multiple shockwave geometry in AdS$_2$ with total geodesic length \cite{Stanford:2014jda}. In boundary terms, the switchback effect is defined by the relation\footnote{Note that we are using $t_R\rightarrow-t_R$ in comparison with \cite{Stanford:2014jda}. We argued that the difference in sign occurs due to the construction of wormhole density matrices instead of a HH state with operator insertions in Sec. \ref{ssec:double scaled wormhole}.}
\begin{equation}\label{eq:switchback property0}
\centering\mathcal{C}_V\propto\abs{t_L-t_1}+\abs{t_1-t_2}+\dots\abs{t_m-t_R}-2m t_{\rm sc}+\mathcal{O}(1)~,\quad \abs{t_{i+1}-t_i}\gg t_{\rm sc}~,
\end{equation}
where it is assumed that all the precursor operators have the same weight, i.e. $\Delta_{1}=\dots=\Delta_m$ so that the scrambling time generated by each one is the same, meaning that: $t_{\rm sc}^{(1)}=\dots=t_{\rm sc}^{(m)}\equiv t_{\rm sc}$ \cite{Stanford:2014jda}, and $\mathcal{O}(1)$ are subleading terms in the approximation. $\mathcal{C}_V$ in \eqref{eq:switchback property0} is also used as a measure of quantum complexity dual to a maximal volume slice in an AdS black hole background with shockwave geometries \cite{Stanford:2014jda}. We stress that we work with wormhole density matrices in chord space (\ref{eq:womrhole density m particles}), based on the original work by \cite{Berkooz:2022fso}. The insertion of the precursor operators in the DSSYK results in bulk matter fields emitted in the past or future singularities of the dual black hole geometry, similar to the construction of partially thermal entangled states in \cite{Goel:2018ubv}. We illustrate the bulk configuration with the multiple shockwaves in Fig. \ref{fig:multi_shock}.

\paragraph{Six-point function}
To illustrate the general mechanism behind the switchback effect in the DSSYK model, we first consider the six-point function (Fig. \ref{fig:6pointchord}) previously derived in Sec. \ref{sec:chord blocks}. We assume the timefold condition $\abs{t_{2}-t_1}\gg t_{\rm sc}^{(1)}$, $t_{\rm sc}^{(2)}$ for the two precursor operators $\hat{\mathcal{O}}_{\Delta_1}(-t_1)$, $\hat{\mathcal{O}}_{\Delta_2}(-t_2)$; and we use the chord diagram derivation of the six-point function in energy basis $I_6$ (\ref{eq:I6 energy basis}) to write
\begin{equation}\label{eq: 6pnt function}
\begin{aligned}
    {\rm Tr}(q^{\Delta\hat{N}}\hat{\rho}_{\Delta_1,\Delta_2})=\int&\qty(\prod_{i=1}^6\rmd\mu(\theta_i)\rme^{-\tau_iE(\theta_i)})\frac{\langle \Delta; \theta_5, \theta_6 | q^{\Delta_2 \hat{N}} | \Delta; \theta_3, \theta_4 \rangle}{(\theta_3 | q^{\Delta \hat{n}} | \theta_4)}\cdot\\
    &\cdot \langle \Delta; \theta_3, \theta_4 | q^{\Delta_1 \hat{N}} | \Delta; \theta_1, \theta_2 \rangle~,
\end{aligned}
\end{equation}
where we use the same notation as (\ref{eq:multiple particles}) for the case $m=2$.

Using the explicit wormhole density matrix with two particle insertions ($m=2$)(\ref{eq:womrhole density m particles}, \ref{eq:multiple particles}), the evaluation of the crossed correlation function (\ref{eq: 6pnt function}) in the $m=2$ case follows from the analytic continuation:
\begin{equation}\label{eq:6pnt analytic cont}
\begin{aligned}
        &\tau_1=\frac{\beta}{4} +\rmi(t_1-t_L)~,&& \tau_2=\frac{\beta}{4}-\rmi(t_1-t_L)~,&& \tau_3=\rmi(t_1-t_2)~,\\ 
        &\tau_4=\rmi(t_2-t_1)~,&&\tau_5=\frac{\beta}{4} +\rmi(t_R-t_2)~,&&\tau_6=\frac{\beta}{4}-\rmi(t_R-t_2)~.
\end{aligned}
\end{equation}
When we evaluate the saddle point solutions to (\ref{eq: 6pnt function}), which can be put in the form
\begin{equation}\label{eq:oscillatory int}
    \int\rmd\mu(\theta_{3})\rmd\mu(\theta_{4})\rme^{-\tau_{3}E(\theta_{3})-\tau_{4}E(\theta_{4})}f(\theta_{3},\theta_{4},\dots)~,
\end{equation}
for $\abs{t_2-t_1}\gg t_{\rm sc}^{(1)}$, $t_{\rm sc}^{(2)}$; $f(\theta_{3},\theta_{4},\dots)$ represents the integrand that involves quantum $6j$ symbols in (\ref{eq: 6pnt function}). Note that this term will be highly oscillatory for general $\theta_{3,4}\in[0,\pi]$ in the $\lambda\rightarrow0$ limit. This means that the integral (\ref{eq:oscillatory int}) picks a saddle point when the Boltzmann factor is a constant, {{which in this case is}}
\begin{equation}\label{eq:saddle points 3,4}
    \theta_{3s,4s}=\pi/2~,\quad (\lambda\rightarrow0)~.
\end{equation}
{{This can be shown by performing the asymptotic analysis in the $\lambda\rightarrow0$ limit of \eqref{eq: 6pnt function} (see e.g. App. F in \cite{Aguilar-Gutierrez:2025pqp}) we find that the saddle point equation of motion at leading order (i.e. $\mathcal{O}(1/\lambda)$ in this case) is\footnote{{{Another $\mathcal{O}(1/\lambda)$ in \eqref{eq: 6pnt function} corresponds to the terms $\tau_{3,4}E(\theta_{3,4})$ in the exponential; however, given that it is an overall phase (i.e. an infinite temperature in \eqref{eq:6pnt analytic cont}), it does not modify the saddle point equations.}}}
$$\dv{S(\theta_{3,4})}{\theta_{3,4}}+\mathcal{O}(1)=0~,$$
where $S(\theta)=S_0-\frac{2}{\lambda}\qty(\theta_{L/R}-\frac{\pi}{2})^2$ (with $S_0$ being a constant) which can be recovered from the $\lambda$ expansion of the $\mu(\theta)\rightarrow\rme^{S(\theta)}$ factors in \eqref{eq: 6pnt function} \cite{Goel:2023svz}; which leads to \eqref{eq:saddle points 3,4}.}} 
This means that in the saddle point approximation (\ref{eq: 6pnt function}) becomes:
\begin{equation}\label{eq:saddle point integral}
\begin{aligned}
    {\rm Tr}(q^{\Delta\hat{N}}\hat{\rho}_{\Delta_1,\Delta_2})=&\frac{\mathcal{N}}{(\theta_{3s} | q^{\Delta \hat{n}} | \theta_{4s})}\int\prod_{i=5}^6\rmd\mu(\theta_i)\rme^{-\tau_iE(\theta_i)}\langle \Delta; \theta_5, \theta_6 | q^{\Delta_2 \hat{N}} | \Delta; \theta_{3s}, \theta_{4s} \rangle\cdot\\
&\cdot\int\prod_{i=1}^2\rmd\mu(\theta_i)\rme^{-\tau_iE(\theta_i)}\langle \Delta; \theta_{3s}, \theta_{4s} | q^{\Delta_1 \hat{N}} | \Delta; \theta_1, \theta_2 \rangle~,
\end{aligned}
\end{equation}
where $\mathcal{N}$ is an overall normalization after the saddle point approximation, namely 
\begin{equation}
\mathcal{N}=\qty(\eval{\int\rmd\theta\rme^{S(\theta)}}_{\theta=\te_s})^2~.    
\end{equation}
Then, (\ref{eq:saddle point integral}) is manifestly just the product of two-sided two-point correlation functions in the saddle point approximation for $\theta_{3,4}$. 

We can further simplify (\ref{eq:saddle point integral}) as
\begin{equation}\label{eq:Tr rho D1D2}
    \begin{aligned}
    {\rm Tr}(q^{\Delta\hat{N}}\hat{\rho}_{\Delta_1,\Delta_2})=&\mathcal{N}'\frac{\int\prod_{i=3}^6\rmd\mu(\theta_i)\rme^{-\tau_iE(\theta_i)}\langle \Delta_2; \theta_5, \theta_{3} | q^{\Delta \hat{N}} | \Delta_2; \theta_6, \theta_{4} \rangle}{\int\prod_{i=3}^4\rmd\mu(\theta_i)\rme^{-\tau_i E(\theta_i)}(\theta_{3} | q^{\Delta \hat{n}} | \theta_{4})}\cdot\\
&\cdot\int\prod_{i=1}^4\rmd\mu(\theta_i)\rme^{-\tau_iE(\theta_i)}\langle \Delta_1; \theta_{2}, \theta_{4} | q^{\Delta \hat{N}} | \Delta_1; \theta_1,\theta_{3}\rangle~,
\end{aligned}
\end{equation}
where we have applied (\ref{eq:swtich property}) in both integrands in (\ref{eq:saddle point integral}) and we have restored the Boltzmann factors $\rme^{-\tau_3 E(\theta_{3})}$, $\rme^{-\tau_4 E(\theta_{4})}$ that lead to the saddle point values for $\theta_{3s,4s}$ (\ref{eq:saddle points 3,4}) in (\ref{eq:saddle point integral}), and $\mathcal{N}'$ is a $\Delta$ independent constant. {{The explicit evaluation of the two and four-point functions in \eqref{eq:Tr rho D1D2} appear in \cite{Aguilar-Gutierrez:2025pqp} App. F and \cite{Goel:2023svz} Sec 2 respectively; where one finds $\theta_{3,4}=\pi/2$ for the saddle point at infinite temperature with the corresponding Boltzmann factors $\rme^{-\tau_{3,4}E(\theta_{3,4})}$ in the integrands. Furthermore, the analysis in the previous works indicates that this holds for arbitrary $\Delta$. }} 

Next, to evaluate (\ref{eq:Tr rho D1D2}) we consider the semiclassical four-point function (\ref{eq:two-sided correlator}), as well as the semiclassical one-sided two-point function (see e.g. \cite{Goel:2023svz,Blommaert:2024ydx})
\begin{equation}\label{eq:2pnt semiclassical}
    \frac{(\tfrac{\beta}{2}-\rmi t|q^{\Delta\hat{n}}|\tfrac{\beta}{2}-\rmi t)}{Z_{\rm DSSYK}(\beta)}\eqlambda\qty(\frac{\sin\theta}{\cosh(J\sin\theta t)})^{2\Delta}~,
\end{equation}
where $|\tfrac{\beta}{2}-\rmi t)$ is an analytic continuation of the HH state (\ref{eq:HH state beta}) with real time evolution. In the late time limit (\ref{eq:2pnt semiclassical}) becomes:
\begin{equation}\label{eq:late time 1 particle}
    \lim_{t\rightarrow\infty}(\tfrac{\beta}{2}-\rmi t|q^{\Delta\hat{n}}|\tfrac{\beta}{2}-\rmi t)\propto \exp(-2J\Delta\sin\theta~\abs{t})~.
\end{equation}
Moreover, given that $\hat{n}$ is the Krylov complexity operator for an initial state $|\Omega)$ with complex time evolution $\tau=\frac{\beta}{2}+\rmi t$, then \eqref{eq:late time 1 particle} is the generating function of the spread complexity of the HH state, i.e.
\begin{equation}\label{eq:HH spread complexity}
    \mathcal{C}(t)=-\eval{\frac{1}{\lambda}\dv{\Delta}~\frac{(\tfrac{\beta}{2}-\rmi t|q^{\Delta\hat{n}}|\tfrac{\beta}{2}-\rmi t)}{Z_{\rm DSSYK}(\beta)}}_{\Delta=0}\eqlambda\frac{2}{\lambda}\log\qty(\frac{\cosh(J\sin\theta t)}{\sin\theta})~.
\end{equation}
We can now carry out the evaluation of the six-point correlation function (\ref{eq:6pnt analytic cont}) in the large backreaction and late time limit with (\ref{eq:unified shockwave length}) and (\ref{eq:2pnt semiclassical}), which results in:
\begin{subequations}\label{eq:correlators approx}
    \begin{align}
    &\int\prod_{i=3}^6\rmd\mu(\theta_i)\rme^{-\tau_iE(\theta_i)}\langle \Delta_2; \theta_5, \theta_{3} | q^{\Delta \hat{N}} | \Delta_2; \theta_6, \theta_{4} \rangle\appropto \rme^{-2J\Delta\sin\theta~(\abs{t_R-t_2}+\abs{t_1-t_2}-2t_{\rm sc}^{(2)})}~,\\
&\int\prod_{i=1}^4\rmd\mu(\theta_i)\rme^{-\tau_iE(\theta_i)}\langle \Delta_1; \theta_{2}, \theta_{4} | q^{\Delta \hat{N}} | \Delta_1; \theta_1,\theta_{3}\rangle\appropto \rme^{-2J\Delta\sin\theta~(\abs{t_L-t_1}+\abs{t_2-t_1}-2t_{\rm sc}^{(1)})}~,\\
&\int\prod_{i=3}^4\rmd\mu(\theta_i)\rme^{-\tau_iE(\theta_i)}\langle\theta_{3} | q^{\Delta \hat{n}} | \theta_{4})\appropto \exp[-2J\Delta\sin\theta~\abs{t_1-t_2}] ~,
\end{align}
\end{subequations}
where $t_{\rm sc}^{(i)}$ appears in (\ref{eq:scrambling time}). This means that
\begin{equation}
    \eval{-\frac{1}{\lambda}\dv{\Delta}~{\rm Tr}(q^{\Delta\hat{N}}\hat{\rho}_{\Delta_1,\Delta_2})}_{\Delta=0}\eqlambda \mathcal{N}'J\sin\theta\bigg(\abs{t_L-t_1}+\abs{t_1-t_2}+\abs{t_R-t_2}-2\qty(t^{(1)}_{\rm sc}+t^{(2)}_{\rm sc})\bigg)~.
\end{equation}
Thus, the result agrees with the switchback effect (\ref{eq:switchback property0}) when $m=2$ and $\Delta_1=\Delta_2$.

\paragraph{General $2m+2$-point function with crossed four-point sub-diagrams}Consider the correlation function with $m$ particle insertions in energy basis from (\ref{eq:2m2 point function}),
\begin{equation}\label{eq:general m}
    \begin{aligned}
        \Tr(\hat{\rho}_{\tilde{\Delta}}q^{\Delta\hat{N}})=\int&\prod_{i=1}^{2m+2}\qty(\rmd\mu(\theta_i)\rme^{-\tau_i E(\theta_i)})\langle \Delta; \theta_1, \theta_2 | q^{\Delta_1 \hat{N}}\cdot \\
        \cdot&\left[ \prod_{j=1}^{m-1} \frac{ | \Delta; \theta_{2j+1}, \theta_{2j+2} \rangle \langle \Delta; \theta_{2j+1}, \theta_{2j+2} | }{(\theta_{2j+1} | q^{\Delta \hat{n}} | \theta_{2j+2})} q^{\Delta_{j+1} \hat{N}} \right] | \Delta; \theta_{2m+1}, \theta_{2m+2} \rangle~,
    \end{aligned}
\end{equation}
where
\begin{equation}
\begin{aligned}
    &\tau_1=\frac{\beta}{4}-\rmi(t_L-t_1)~,\quad \tau_2=\tau_1^*,\quad \tau_{2i-1}=\rmi(t_{i-1}-t_{i})~,\quad (2\leq i\leq m)~,\\
    &\tau_{2i}=\tau_{2i-1}^*~,\quad \tau_{2m+1}=\frac{\beta}{4}+\rmi(t_R-t_{m})~,\quad \tau_{2m+1}=\tau_{2m+1}^*~.
    \end{aligned}
\end{equation}
The diagram is shown in Fig. \ref{fig:2n2_amplitude}. We can make further simplifications, for instance, (\ref{eq:general m}) with (\ref{eq:swtich property}) can be rewritten as
\begin{equation}\label{eq:2n2 integral}
    \begin{aligned}
        &\Tr(\hat{\rho}_{\tilde{\Delta}}q^{\Delta\hat{N}})=\int\prod_{i=1}^{2m+2}\qty(\rmd\mu(\theta_i)\rme^{-\tau_i E(\theta_i)})\langle \Delta_m; \theta_{2m+1}, \theta_{2m-1} | q^{\Delta \hat{N}}\ket{\Delta_m; \theta_{2m+2}, \theta_{2m}}\cdot \\
        &\qquad\cdot\left[ \prod_{i=1}^{m-1} \frac{ \langle \Delta_{i+1}; \theta_{2i+1}, \theta_{2i+3} | q^{\Delta\hat{N}}| \Delta_{i+1}; \theta_{2i+2}, \theta_{2i+4} \rangle}{(\theta_{2i+1} | q^{\Delta \hat{n}} | \theta_{2i+2})} \right] \langle \Delta_{1}; \theta_{1}, \theta_{3} |q^{\Delta \hat{N}}| \Delta_{1}; \theta_{2}, \theta_{4} \rangle~.
\end{aligned}
\end{equation}
Following the same argument as in the six-point function case, the $\lambda\rightarrow0$ limit and the timefold condition 
\begin{equation}\label{eq:timefold cond}
\abs{t_{i+1}-t_{i}}\gg t_{\rm sc}^{(i)}~,    
\end{equation}
imply that the Boltzmann factors $\rme^{-\tau_iE(\theta_i)}$ generate rapidly oscillating integrands of the $\theta_i$ variables when $i=3,4,\dots,2m-1,2m$. This means that the integrals in this limit can be evaluated with saddle point values $\theta_i\simeq\theta_{is}$($=\pi/2$) for $i=3,4,\dots,2m-1,2m$. Then, (\ref{eq:2n2 integral}) factorizes into a product of two and crossed four-point correlation functions, i.e.
\begin{equation}\label{eq:manifest many particle correlation function}
    \begin{aligned}
        &\Tr(\hat{\rho}_{\tilde{\Delta}}q^{\Delta\hat{N}})=\mathcal{N}\int\prod_{j=1,2m+1}^{2m+2}\qty(\rmd\mu(\theta_j)\rme^{-\tau_j E(\theta_j)})\langle \Delta_m; \theta_{(2m+1)s}, \theta_{(2m-1)s} | q^{\Delta \hat{N}}\ket{\Delta_m; \theta_{(2m+2)s}, \theta_{(2m)s}}\cdot \\
        &\qquad\cdot\left[ \prod_{i=1}^{m-1} \frac{ \langle \Delta_{i+1}; \theta_{2i+1}, \theta_{2i+3} | q^{\Delta\hat{N}}| \Delta_{i+1}; \theta_{2i+2}, \theta_{2i+4} \rangle}{(\theta_{2i+1} | q^{\Delta \hat{n}} | \theta_{2i+2})} \right]_s \langle \Delta_{1}; \theta_{1}, \theta_{3s} |q^{\Delta \hat{N}}| \Delta_{1}; \theta_{2}, \theta_{4s} \rangle~,
\end{aligned}
\end{equation}
where, again, $\mathcal{N}$ is a constant, and the subindex $s$ denotes saddle point value. By analogy to (\ref{eq:Tr rho D1D2}), the factors in (\ref{eq:manifest many particle correlation function}) for $\abs{t_{i+1}-t_{i}}\gg t_{\rm sc}^{(i)}$ and $\lambda\rightarrow0$ can be evaluated as
\begin{subequations}\allowdisplaybreaks
\begin{align}
\int\Bigg(\prod_{j=2m}^{2m+2}&\rmd\mu(\theta_j)\rme^{-\tau_jE(\theta_j)}\Bigg)\langle \Delta_m; \theta_{2m+1}, \theta_{2m-1} | q^{\Delta_{w} \hat{N}}\ket{\Delta_m; \theta_{2m+2}, \theta_{2m}}\\
&\appropto \rme^{-2J\Delta\sin\theta~(\abs{t_R-t_{m}}+\abs{t_{m}-t_{m-1}}-2t_{\rm sc}^{(m)})}~,\nonumber\\
\int\Bigg(\prod_{j=2i+1}^{2i+4 }&\rmd\mu(\theta_j)\rme^{-\tau_jE(\theta_j)}\Bigg)\langle \Delta_i; \theta_{2i+1}, \theta_{2i+3} | q^{\Delta\hat{N}}| \Delta_i; \theta_{2i+2}, \theta_{2i+4} \rangle\\
&\appropto \rme^{-2J\Delta\sin\theta~(\abs{t_{i+1}-t_{i}}+\abs{t_{i}-t_{i-1}}-2t_{\rm sc}^{(i)})}~,\nonumber\\
\int\Bigg(\prod_{j=1}^{4}&\rmd\mu(\theta_j)\rme^{-\tau_jE(\theta_j)}\Bigg)\langle \Delta_{1}; \theta_{3}, \theta_1 |q^{\Delta \hat{N}}| \Delta_{1}; \theta_{4}, \theta_2 \rangle\\
&\appropto \rme^{-2J\Delta\sin\theta~(\abs{t_{L}-t_{1}}+\abs{t_{1}-t_{2}}-2t_{\rm sc}^{(1)})}~,\nonumber\\
\int\Bigg(\prod_{j=i}^{i+1}\rmd\mu(\theta_j)&\rme^{-\tau_jE(\theta_j)}\Bigg)(\theta_{i} | q^{\Delta \hat{n}} | \theta_{i+1})\appropto \exp[-2J\Delta\sin\theta~\abs{t_{i+1}-t_i}] ~.
\end{align}\label{eq:contributions many particle saddle}
\end{subequations}
Thus, collecting all the contributions (\ref{eq:contributions many particle saddle}) for (\ref{eq:manifest many particle correlation function}) we recover
\begin{equation}
\begin{aligned}
    -\frac{1}{\lambda}\dv{\Delta}~{\rm Tr}&\eval{\qty(q^{\Delta\hat{N}}\hat{\rho}_{\tilde{\Delta}})}_{\Delta=0}\simeq\frac{(0|\rme^{-\frac{\beta}{4}\hH}\hat{\mathcal{O}}_{\Delta_1}(t_L-t_1)\rme^{-\frac{\beta}{4}\hH}q^{\Delta\hat{N}}\rme^{-\frac{\beta}{4}\hH}\hat{\mathcal{O}}_{\Delta_1}(t_1-t_2)\rme^{-\frac{\beta}{4}\hH}|0)\dots}{(0|\rme^{-\qty(\frac{\beta}{2}+\rmi(t_2-t_1))}q^{\Delta\hat{N}}\rme^{-\qty(\frac{\beta}{2}+\rmi(t_1-t_2))}|0)\dots }\cdot\\
&\cdot\frac{\dots(0|\rme^{-\frac{\beta}{4}\hH}\hat{\mathcal{O}}_{\Delta_m}(t_{m-1}-t_m)\rme^{-\frac{\beta}{4}\hH}q^{\Delta\hat{N}}\rme^{-\frac{\beta}{4}\hH}\hat{\mathcal{O}}_{\Delta_m}(t_{m}-t_R)\rme^{-\frac{\beta}{4}\hH}|0)}{\dots(0|\rme^{-\qty(\frac{\beta}{2}+\rmi(t_{m}-t_{m-1}))}q^{\Delta\hat{N}}\rme^{-\qty(\frac{\beta}{2}+\rmi(t_{m-1}-t_{m}))}|0)}~,
\end{aligned}
\end{equation}
which in the late time limit (i.e. $(\abs{t_L-t_1}\gg t_{\rm sc}^{(1)}$, $(\abs{t_R-t_m}\gg t_{\rm sc}^{(m)}$ and (\ref{eq:timefold cond})) reproduces the switchback effect (\ref{eq:switchback property0}) for the total chord number:
\begin{equation}\label{eq:total chord number shocks}
\begin{aligned}
\Tr\qty(\hat{N}\hat{\rho}_{\tilde{\Delta}})\eqlambda J\mathcal{N}\sin\theta\bigg(\abs{t_L-t_1}+\abs{t_1-t_2}+\dots+\abs{t_R-t_m}-2\sum_{i=1}^m t^{(i)}_{\rm sc}\bigg)~.
\end{aligned}
\end{equation}

\subsection{The switchback effect and Krylov complexity}\label{ssec:switchback Krylov}
\begin{figure}
    \centering
    \includegraphics[width=0.4\textwidth]{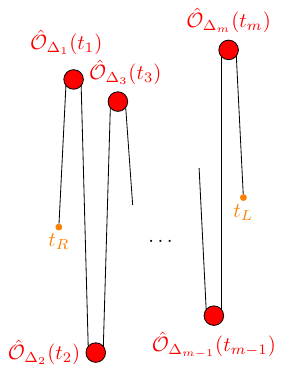}
    \caption{Timefold evaluation of Krylov operator complexity with several precursor operators. We consider the time segment between the boundary times $t_{L/R}$ (orange) and the timefold times $t_i$, with $i\in{1,\dots m}$, i.e. where the precursors $\hO^{L/R}_{\Delta_i}(t_i)$ (red dots) are inserted. Each segment is associated with the Krylov operator complexity of $\hO_{\Delta_i}(t_{L/R}-t_i)$ (\ref{eq:KRylov complexity shockwave}). The sum in the semiclassical limit generates the total chord number (\ref{eq:timefold Krylov}). The resulting measure thus has a bulk interpretation as the total geodesic length with multiple shockwave insertions (\ref{eq:switchback property0}). Figure based on \cite{Stanford:2014jda}.}
    \label{fig:timefold complexity}
\end{figure}

As noted above, the expectation value of the total chord number with $2m$ precursor operator insertions (\ref{eq:general m}) under the timefold condition \eqref{eq:timefold cond} results and in the semiclassical limit results in a sum of the expectation value of the chord number with one (or zero) precursor particle insertion each. Moreover, the terms with a one-particle insertion can be interpreted as Krylov complexity for a precursor operator $\hO_{\Delta_i}$ with a timeshift $t\rightarrow t-t_i$ at late times, as we saw in Sec. \ref{ssec:Krylov operator complexity}. Meanwhile, the terms without particle insertion are related to the spread complexity of the HH state (\ref{eq:HH spread complexity}). This means that at late times and under the timefold condition (\ref{eq:timefold cond}) in the semiclassical limit, the expectation value of the total chord number operator can be expressed as a linear combination of the Krylov complexity operators for each precursor $\hO^{L/R}_{\Delta_i}(t_{L/R}+t_i)$ in the Heisenberg picture and the spread complexity of the HH state \eqref{eq:HH spread complexity} between the insertion times, i.e.
\begin{equation}\label{eq:timefold Krylov}
    \begin{aligned}
        \eval{\Tr\qty(\hat{N}\rho_{\tilde{\Delta}})}_{\abs{t_{i+1}-t_{i}}\gg t_{\rm sc}^{(i)}}\eqlambda\mathcal{N}\bigg(&\mathcal{C}_{\Delta_1}(t_L-t_1,t_2-t_1)+\sum_{i=2}^{m-1}\mathcal{C}_{\Delta_{i}}(t_{i-1}-t_i,t_{i+1}-t_i)\\
        &+\mathcal{C}_{\Delta_m}(t_m-t_{m-1},t_R-t_m)-\sum_{i=1}^{m-1}\mathcal{C}(t_{i+1},t_i)\bigg)~.
    \end{aligned}
\end{equation}
%\jx{Should there be a sum over $i$ in the last $\mathcal{C}(t_{i+1},t_i)$ term?}
This indicates that there is a natural generalization of Krylov complexity for multiple $2m$ precursor operators $\hO_{\Delta_1}(t_1)$, \dots, $\hO_{\Delta_2}(t_m)$ given by the sum over the Krylov complexity for each contributing operator and subtracting the spread complexity of the HH state evolving between the timefold insertions. As seen in (\ref{eq:timefold Krylov}) one should use the time interval between the insertion of one precursor operator with respect to the left boundary time $t_L$; and adding up the contributions between the timefold times $t_1$ to $t_m$ until reaching the time interval t$_R$. This is illustrated in Fig. \ref{fig:timefold complexity}. 
Following this procedure, we perform the time shift $t\rightarrow t-t_i$ for each contribution in the Krylov operator complexity corresponding to each time interval. In order for this evaluation to match the bulk geodesic length \eqref{eq:swtich property}, we also have to subtract the spread complexity associated to evolution between the timefold insertion times $t_{1\leq i\leq m}$. Then, the total chord number in the semiclassical limit can then be identified with sum of the Krylov complexity for each $\hO^{L/R}_{\Delta_i}(t_{L/R}-t_i)$ in the time interval between their insertion and the two-sided Hamiltonian times $t_{L/R}$, as defined above.

\paragraph{Alternative reference state}Note that the measure above is in principle different from the Krylov operator complexity of a product of matter chord $\hO_{\Delta_1}(t)\hO_{\Delta_2}(t)$, where we select\footnote{We omit the $L/R$ sub-indices due to the lack of ambiguity.}
\begin{equation}\label{eq:initial Krylov 2 particles}
    \ket{K^{\Delta_1,\Delta_2}_0}'=\frac{\rme^{-\frac{\beta}{4}\hH}\hO_{\Delta_1}(t_1)\hO_{\Delta_2}(t_2)\rme^{-\frac{\beta}{4}\hH}|\Omega)}{\sqrt{(\Omega|\rme^{-\frac{\beta}{4}\hH}\hO_{\Delta_2}(t_2)\hO_{\Delta_1}(t_1)\rme^{-\frac{\beta}{2}\hH}\hO_{\Delta_1}(t_1)\hO_{\Delta_2}(t_2)\rme^{-\frac{\beta}{4}\hH}|\Omega)}}~,
\end{equation}
as initial state in the Lanczos algorithm, given that this results in a different Krylov basis, similar to the $m=1$ case in Sec. \ref{ssec:Krylov operator complexity}.

As noted in Sec.~\ref{ssec:Krylov operator complexity}, identifying a natural quantum information observable corresponding to the total chord number becomes nontrivial when multiple matter chord insertions are present. However, by leveraging its connection to shockwave geometries and the structure of DSSYK correlation functions, we were able to propose a quantum complexity measure for states with multiple matter operator insertions. In the semiclassical limit, the sum of individual contributions in \eqref{eq:timefold Krylov} reproduces the dynamics of total chord number expectation value.

\section{Conclusion and Some Future Prospectives}\label{sec:discussion}
To summarize, this work provides three new series of results relating chord diagrams with particle insertions and on the holography of the DSSYK model.
\begin{itemize}
    \item We introduced the notion of a chord intertwiner and interpreted it as a general mechanism by which bulk states emerge from correlations among boundary states. This perspective enables us to reformulate various analytic results—previously derived through $q$-combinatorial methods—in terms of an isometric factorization map. This map, in turn, provides a straightforward framework for evaluating chord diagrams with multiple matter chord crossings.

    \item We established an effective theory of DSSYK in the semi-classical regime at finite energy and temperature, extending earlier analysis in the triple-scaling limit by allowing heavy chord operator insertions. In particular, we provided a bulk interpretation of multiple matter chord operator insertions in terms of a multi-shockwave geometry. Our path integral techniques enabled us to compute the semiclassical expectation value of the total chord number in four-point and higher-point functions evaluated within wormhole density matrices.
    
    \item We found that this total chord number reproduces the geodesic lengths previously identified by Shenker and Stanford~\cite{Shenker:2013pqa,Shenker:2013yza}, albeit with notable modifications. For instance, the state prepared by the path integral appears to naturally live on the ``fake'' disk geometry introduced in~\cite{Lin:2023trc}, with an effective boundary length \( \beta_{\text{fake}} \) slightly larger than the physical inverse temperature \( \beta \).  Following the ideas of~\cite{Lam_2018}, the appearance of the quantum \( 6j \)-symbol in the evaluation of the OTOC suggests that the scrambling dynamics should be interpreted as a shockwave scattering process in a ``fake'' spacetime geometry, whose Euclidean counterpart is the fake disk. Our semiclassical analysis demonstrates how submaximal chaotic behavior—with Lyapunov exponent \( \lambda_L \) set by the fake inverse temperature—can be derived from the structure of the quantum \( 6j \)-symbol.
    
    \item We derived a factorization property for a family of higher-point correlation functions relevant to the study of the switchback effect. By applying the previously established isometric factorization map, we showed that these correlators decompose into a product of crossed four-point functions, each accompanied by appropriate normalization factors. In the semiclassical limit and within the timefold regime, this decomposition reproduces the switchback effect. Furthermore, in the triple-scaling limit, our results reduce to those obtained in~\cite{Lam_2018}.
    \item We also proved that the switchback effect appears in the expectation value of the total chord number of the DSSYK in semi-classical limit and with the timefold condition (\ref{eq:timefold cond}). This expectation value, under specific restrictions, can be expressed as a sum of Krylov operator complexity for the precursor operator insertions, and including factors of spread complexity for the HH state between the timefold insertions. We proposed a natural measure of Krylov complexity for multiple precursor operators under these conditions, namely (\ref{eq:timefold Krylov}), which reproduces the semiclassical wormhole length in JT or sine-dilaton gravity. Moreover, this result provides a relation between the total chord number and a specific quantum complexity measure, which is a priori not clear in the configurations that we study. The correspondence to a bulk geodesic effectively promotes the Krylov complexity of multiple precursor operators to a notion of holographic complexity in the DSSYK model and its bulk dual.
\end{itemize}
{{One should also note that the relation between lengths and chords has been pointed out since the early literature on DSSYK \cite{Berkooz:2018jqr}, albeit in the triple-scaling limit. The contribution in this work is to make a more precise connection between them in the semiclassical doubled-scaled limit and with arbitrary number of operator insertions. Furthermore, while the connection between the fake disk and its connection with OTOCs appeared since \cite{Lin:2023trc} (Sec 4.6), our results  provide a new perspective that supports their analysis with arbitrary number of operator insertions. In contrast to \cite{Lin:2023trc} this work does not rely on a perturbative expansion in the length and Hamiltonians after taking the semiclassical limit. Furthermore, the derivation in this manuscript clarifies the role of the quantum 6j symbols in modifying the period the thermal circle that gives rise to the fake disk.}}

We conclude with a list of related future directions
\subsection{Outlook}\label{ssec:outlook}

    \paragraph{Probing Inelastic scattering with Negativity in Length} It was observed in~\cite{Gu_2019} that the commutator \( [V, W] \) captures the inelastic component of the final state produced by two-body scattering of the particles \( V \) and \( W \). This idea was further developed in~\cite{Lin:2023trc}, where it was shown that in the semiclassical regime of DSSYK, analogous inelastic effects arise and can be understood as the creation of folded wormholes at late times due to the time evolution of the two particles. In the chord Hilbert space, these folded wormhole configurations correspond to states in which the expectation value of the chord number becomes negative. Thus, the emergence of negativity under dynamical time evolution serves as a probe of inelastic scattering effects. In the context of string theory, such commutator states are associated with configurations involving long strings~\cite{Stanford:2014jda}.

Given that the family of correlation functions involving multiple matter chord insertions (see Fig.~\ref{fig:2n2_amplitude}) can be computed using chord inner products within the one-particle Hilbert space \( \mathcal{H}_1 \)—where both analytic control and an effective semiclassical description are available—it would be interesting to study the dynamical evolution of intermediate chord numbers (or lengths) and investigate the origin of negativity and its precise relation to inelastic scattering processes.
       
\paragraph{Characterizing Sub-maximal Chaos with Operator Algebras} 
Scramblon dynamics (also referred to as modular perturbations) has been linked to the structure of operator algebras through the concept of half-sided modular inclusion. This setup involves a one-parameter family of unitaries with a positive generator and is known to lead to exponential growth under modular time evolution~\cite{Borchers:2000pv}. As argued in~\cite{Maldacena:2015waa}, this growth is bounded, and the bound is saturated in the case of maximal chaos. The algebraic structure thereby encodes a form of maximal modular chaos.

In DSSYK, the appearance of sub-maximal chaos suggests the presence of a similar but modified operator algebra, now associated with the ``fake'' disk geometry. In this context, the usual \( \mathfrak{sl}_2 \) symmetry is mildly deformed by the physical temperature. We expect that this submaximal behavior is governed by an underlying operator-algebraic structure on the fake disk, potentially realizing a generalized version of modular inclusion. It would be interesting to formulate this structure explicitly and explore whether the associated modular flow captures the transition from maximal to submaximal chaos in an algebraic way.

    \paragraph{Subregion-subalgebra duality and Entanglement Wedge Reconstruction} Our findings regarding the chord intertwiner suggest the existence of a notion of bulk subregions whose information is encoded in corresponding boundary subregions. Specifically, bulk wavefunctions defined on larger spatial patches can be constructed from boundary states defined on correspondingly larger boundary subregion ~\eqref{eq:geodesics intertwining}. This observation points toward a possible realization of subregion duality and entanglement wedge reconstruction in the DSSYK model, which aligns with the framework developed in~\cite{Dong:2016eik}. We do not demonstrate this in the current paper explicitly because a full treatment would require local bulk excitations  to be reconstructible via HKLL-like methods on the associated boundary subregion. 

To solidate this idea, one needs to identify a suitable notion of boundary locality, so that the concept of a boundary subregion becomes physically meaningful. Based on such a structure, one can then investigate whether operator algebras\footnote{The operator algebra might be change before and after scrambling time in shockwave geometries, similar to AdS black hole spacetimes \cite{Chandrasekaran:2022eqq}.} associated with boundary subregions encode information about corresponding bulk regions—potentially realizing an entanglement wedge in the geometry of the DSSYK holographic dual. Currently, a complete bulk dual to DSSYK valid across its full parameter regime is not known. In this context, the existence of a subregion–subalgebra correspondence could serve as a consistency criterion for candidate dual descriptions. Moreover, establishing entanglement wedge reconstruction in DSSYK (with matter) would offer a pathway toward defining a sharp notion of generalized entropy and understanding how the horizon value of the dilaton emerges from the underlying chord structure.

   \paragraph{Multiparticle factorization} Many of our results rely on the factorization of the one-particle sector \( \mathcal{H}_1 \) into a tensor product of two zero-particle sectors. It would be interesting to extend this framework to the multi-particle setting using the coproduct structure of the chord algebra~\cite{Lin:2023trc}, which governs how multi-particle states can be constructed from one-particle building blocks. Moreover, since the correlation function associated with a given chord diagram is independent of the choice of slicing, it would be insightful to reinterpret our \( 2m+2 \)-point functions as expectation values of the operator \( q^{\hat{N}_{\text{tot}} \Delta} \) between \( m \)-particle states. The fact that multi-particle states decompose into one-particle states, together with the isometric factorization map, suggests the existence of an underlying boundary construction for bulk multi-particle states. We expect that this perspective could lead to new techniques for evaluating higher-point correlation functions involving multiple matter crossings.

\paragraph{Uncovering complexity in dS space}{Our findings are based on the Schrödinger equation for the chord states with discrete chord number sectors. The bulk interpretation of our results under these conditions are consistent with sine-dilaton gravity in the sense that one may interpret that the shockwaves propagate in an effective AdS$_2$ black hole geometry \cite{Blommaert:2024ydx}. However, if we use a different set of constraints, such as those in \cite{Narovlansky:2023lfz,Narovlansky:2025tpb}, where instead, on imposes energy constraints for a pair of decoupled DSSYK models; then we expect to be able to probe the dS$_3$ space hologram of the pair of DSSYK models. It would be very interesting to study how our results can be used to study the cosmological horizon of dS space with shockwave geometries. For instance, there has been a lot of attention to holographic complexity in dS spacetimes, including the CV conjecture and several others \cite{Baiguera:2023tpt,Aguilar-Gutierrez:2024rka,Baiguera:2024xju,Aguilar-Gutierrez:2023zqm,Aguilar-Gutierrez:2023tic,Jorstad:2022mls,Auzzi:2023qbm,Chapman:2021eyy,Faruk:2025bed}. It would be very useful to develop our methods to describe shockwaves in Schwarzschild-dS$_3$ space considering constraints in the one-particle chord space. One can then study whether Krylov operator complexity is still related to the CV conjecture in a different holographic description of the DSSYK model. In order to accomplish this, it would be worth revisiting the DSSYK results on operator growth in \cite{xu2024chord} and provide them with bulk interpretation that elucidates the properties of dS$_3$ space.} This might allow a first principles derivation of holographic complexity for dS space that one might compare to results based on the CV conjecture in the literature \cite{Aguilar-Gutierrez:2024rka,Baiguera:2024xju,Aguilar-Gutierrez:2023zqm,Jorstad:2022mls,Chapman:2021eyy}. While this has been previously attempted in \cite{Aguilar-Gutierrez:2024nau}, it would be worth considering how this treatment changes once double-scaled precursor operators are inserted, in a similar way to our analysis.

\paragraph{Exploring Krylov complexity for a product of precursor operators}
It would be interested to further study the Krylov complexity of several precursor operators, which was a crucial point to argue that the switchback effect is also present in Krylov operator complexity (Sec. \ref{ssec:switchback Krylov}). One could select as the initial state in the Lanczos algorithm (\ref{eq:initial Krylov 2 particles}) (or its generalization with more precursor operators \eqref{eq:initial Krylov 2 particles}),
in contrast to the one we presented in (\ref{eq:timefold Krylov}) as a sum over the precursor operator contributions. While the latter one has a geometric interpretation as the minimal geodesic length, the former one is in principle another natural choice to generate the Krylov basis for a product of precursor operators, which results in an apparently different observable. Given that the Krylov complexity for the product of operator insertions can be expressed in a gauge invariant matter, it should correspond to another valid element in the holographic dictionary of the DSSYK model, which may then have a geometric description in the bulk. We hope to explore this further in future work.

\paragraph{Switchback effect in higher dimensions}In order to associate the switchback effect for Krylov operator complexity in the DSSYK one has to properly define the Lanczos algorithm when there are multiple operator insertions. We expect this is a general feature, independent of the theory under consideration. This means that if one attempts to generalize our results in higher dimensional holographic conformal field theories (CFTs), one will find the same type of indeterminacy for the initial state to feed in the Lanczos algorithm. There are known expressions for OTOCs in higher dimensional CFTs; and the analogous six-point function to the one we considered is also known \cite{Haehl:2021tft}. Therefore, it is a very good future prospect to compare our results to the ones obtained in higher dimensional holographic CFTs.\footnote{SEAG acknowledges past discussions in this direction with H. Camargo, V. Jahnke, KY. Kim and M. Nishida.} This should allow us to show there is a range of applicability of the techniques developed in this work for more intricate models. In additions, there has been a recent proposal that the rate of change of Krylov state complexity is equal to the proper radial momentum of a point particle \cite{Caputa:2024sux}. So far, backreaction in the original \cite{Caputa:2024sux} proposal by has been treated only with ETW branes \cite{Caputa:2025dep}. It would be fascinating to find out how this relation is modified in the presence of shockwave backreaction, and explore how the switchback effect, corresponding to alternating insertions of shockwaves, may modify the proposed correspondence pointed out in \cite{Caputa:2024sux}. We hope more can be done using the techniques in this work.

\section*{Acknowledgments}
 We thank Andreas Blommaert, Xuchen Cao, Xi Dong, Elliott Gesteau, Jeremy v.d Heijden,  Henry W. Lin, Thomas G. Mertens, Alexey Milekhin, Onkar Parrikar, Adrián Sánchez-Garrido, Tim Schuhmann, Erik P. Verlinde, Masataka Watanabe and Mengyang Zhang for helpful discussions; especially Hugo Camargo for comments on a previous draft. SEAG is also grateful for collaboration on related topics with Hugo Camargo, Viktor Janke, Keun-Young Kim, and Mitsuhiro Nishida; to the High Energy Theory group at UC Santa Barbara for hospitality, which allowed this collaboration to start, and the QISS consortium for travel support. 
It is a pleasure to acknowledge the Yukawa Institute for Theoretical Physics at Kyoto University, and KU Leuven, where this work was presented and completed during the YITP-I-25-01 on ``Black Hole, Quantum Chaos and Quantum Information'', and ``The Holographic Universe'' workshops respectively.
SEAG is supported by the Okinawa Institute of Science and Technology Graduate University. The work of J.X. is supported by the U.S. Department of Energy, Office
of Science, Office of High Energy Physics, under Award Number DE-SC0011702. J.X. acknowledges the support by the Graduate Division Dissertation Fellowship and the Physics Department Graduate Fellowship at UCSB.  This project/publication was also made possible through the support of the ID\#62312 grant from the John Templeton Foundation, as part of the ‘The Quantum Information Structure of Spacetime’ Project (QISS), as well as Grant ID\# 62423 from the John Templeton Foundation. The opinions expressed in this project/publication are those of the author(s) and do not necessarily reflect the views of the John Templeton Foundation.

\appendix
\section{Notation}\label{app:notation}
\begin{itemize}
\item $q=\rme^{-\lambda}\in[0,1]$: Intersection weight of $H$-chords.
\item $\Delta$: Weight of the matter chord operator $\hat{\mathcal{O}}_{\Delta}$.
    \item $\mathcal{H}_m$ (\ref{eq:Fock space with matter}): Chord Hilbert space with $m$ particles.
    \item In general, hatted quantities represent operators; unhatted their classical value.
    \item $\hH_{L/R}$ (\ref{eq:H many particles}): Left/right chord Hamiltonian.
    \item $\ket{\tilde{\Delta};n_0,n_1,\cdots,n_m}$: States in chord number basis, where $n_0$ is the number of open ($H$-)chords to the left of all matter chords, $n_1$ the number of open chords between the first two particles; and continuing up to the $n_m$-particle sector.
\item $|\Omega)$: Empty chord state.
\item $|n)$ zero-particle state in chord number basis. We denote tensor products as $|m,n)\equiv|m)\otimes|n)$.
\item $|\theta)$: zero-particle state in energy basis.
\item $|s)$ (\ref{eq:DS coherent state}): zero-particle state in coherent basis.
\item $\ket{\Delta;n_{L},n_{R}}$: One-particle state in chord number basis.
\item $\ket{\Delta;\theta_{L},\theta_{R}}$: One-particle state in energy basis.
\item $\ket{\Delta;s_1,s_2}$: One-particle state in coherent basis.
\item $\phi_{E(\te)}\left(n\right)$ (\ref{eq:wavefunctions E basis}):  Wavefunctions in the energy basis without matter insertion.
\item $\phi^{\Delta}_{E_L E_R}(m_L, m_R)$ (\ref{eq:wavef Energy 1 part}): Bulk wavefunction with one-particle insertion.
\item $d\mu(\theta)$ (\ref{eq:measure theta}): Integration measure in energy basis.
\item $E(\theta)$ (\ref{eq:conserved energies}): DSSYK energy eigenvalues in $\mathcal{H}_0$, parameterized by $\theta\in[0,\pi]$.
\item $|\beta)$ (\ref{eq:HH state beta}): HH state without particle insertions.
\item $\phi_\beta(n) := (n|\beta)$ (\ref{eq:thermal wave}): Thermal wavefunction.
\item $Z_{\mathrm{DSSYK}}(\beta)$ (\ref{eq:DSSYK Z}): Thermal partition function without matter insertion.
\item $H_{n }\left(\cos\theta;q\right)$ (\ref{eq:qHermite rec}): q-Hermite polynomials.
\item $I_{n_{1},n_{2},n_{3}}$ (\ref{eq:I coefficients}): Components of chord intertwiner in number basis.
\item $ I^{\Delta}_{n_L, n_R; m_L, m_R}$ (\ref{eq:fusion-3}): Components of the one-particle chord intertwiner in number basis.
\item $\Gamma^{\Delta_1} =\gamma_{\Delta_1}(\te_1,\te_2)\gamma_{\Delta_1}(\te_3,\te_4)$: Normalization of the matter chord operator in the energy basis: the configuration of arguments \( \theta_i \) depend on the energy parameters adjacent to the points where the matter chord intersects the boundary.
\item $\me_0$ (\ref{eq:fusion-2}): Entangler map without particles.
\item $|\mf_{mn})$ (\ref{eq:Fmn state}): intertwiner state.
\item $|\mf_{st})$ (\ref{eq:def-F-no-matter}) Coherent intertwiner state; while $|\mf_{st}^{\Del})=q^{\Delta\hat{N}}|\mf_{st})$ (\ref{eq:qDelta F coherent})
\item $\me_{\Delta}$ (\ref{eq:one particle entangler}): Entangler map with one-particle insertion.
\item $\hh_{L/R}$ (\ref{eq:action-H}): Modified two-sided Hamiltonian with one-particle insertion.
\item $\hfa_{L/R}^\dagger$, $\hfa_{L/R}$ (\ref{eq:mathfrak a dagger}): Creation/annihilation operators for $\hH_{L/R}$.
\item $\ha_{L/R}^\dagger$, $\ha_{L/R}$ (\ref{eq:a adagger hLR}): Creation/annihilation operators for $h_{L/R}$.
\item $ \zeta_{n_{L},n_{R}}$ (\ref{eq:recursion-Q}): coefficients relating one-particle chord states in energy vs chord number basis.
\item $a_{n_{L},n_{R}}$ (\ref{eq:anLnR relation}): Rescaled coefficients $ \zeta_{n_{L},n_{R}}$.
\item $ Q_{n_{L},n_{R}}\left(\cos\te_{L},\cos\te_{R};q,q^{\Del}\right)$ (\ref{App:Q-def}): Bi-variate generating function.
    \item $\tilde{\Delta}=\qty{\Delta_1,\dots,\Delta_m}$: Shorthand for denoting multiple operators $\hO_{\Delta_i}$.
    \item $\ket{\Psi^{L/R}_{\tilde{\Delta}}(\tau_0,\dots,\tau_m)}$ (\ref{eq:multiple particles}): State generated by the double-scaled algebra operators.
\item $Z_{\Delta_i}(\beta_L,\beta_R)$ (\ref{eq:partition function 1p}): Partition function with one-particle insertion.
    \item $\tau_{L/R}=\frac{\beta_{L/R}}{2}+\rmi t_{L/R}$; and $\tau_{i,j}\equiv\frac{\beta}{2}+\rmi(t_i-t_j)$.
\item $\hat{\rho}_{\Delta_i}(\tau_L,\tau_R)$ (\ref{eq:rho LR}): Single-shot wormhole density matrix.
\item $\hat{\rho}_{\tilde{\Delta}}(\tau_L,\tau_R)\equiv\ket{\Psi^{R}_{\tilde{\Delta}}\qty(\tau_{1,R},\tau_{2,1},\dots,\tau_{R,m})}\bra{\Psi^{L}_{\tilde{\Delta}}(\tau_{L,1},\tau_{1,2},\dots,\tau_{m,L})}$ (\ref{eq:multiple precursor state}): Multi-shot wormhole density matrix.
\item $I_{2m+2}$ (\ref{eq:2m2 point function}): $(2m+2)$-point function in energy basis.
\item $
    {\Tr[\hat{\rho}_{\tilde{\Delta}}q^{\Delta\hat{N}}]_q}$ with $\tilde{\Delta}=\qty{\Delta_1,\dots,\Delta_m}$ (\ref{eq:general m}): Unnormalized $(2m+2)$-point correlation function (\ref{eq:general m}).
\item $\left\{ \begin{array}{ccc}
\Delta_1 & \theta_1 & \theta_2 \\
\Delta_2 & \theta_4 & \theta_3
\end{array} \right\}_q$ (\ref{eq:6j energy basis}): Quantum \( 6j \)-symbol in energy basis.
\item $\left\{ \begin{array}{ccc}
\Delta_1 & n_1 & n_2 \\
\Delta_2 & n_4 & n_3
\end{array} \right\}_q$ (\ref{eq:6j chord number basis}): \( 6j \)-symbol in chord number basis.
\item $ \left\{ \begin{array}{ccc}
\Delta_1 & s_1 & s_2 \\
\Delta_2 & s_4 & s_3
\end{array} \right\}_q$ (\ref{eq:6j symbol coherent state basis}): \( 6j \)-symbol in coherent state basis.
\item $\hat{b}_{L/R}^\dagger$, $\hat{b}_{L/R}$ (\ref{eq:DS creation annih}): Creation and annihilation operators of the matter operators $\hat{\mathcal{O}}^{L/R}_{\Delta}$.
\item $\ket{K^{\Delta_i}_{n}}$: Krylov basis for the operator $\hO_{\Delta_i}(t-t_i)$ \eqref{eq:explicit Krylov basis}.
\item $\Psi^{\Delta_i}_n(\tau_{L,i})=\bra{K_n^{\Delta_i}}\ket{\Psi^{L}_{{\Delta_{i}}}(\tau_{L,i},\tau_{i,L})}$ (\ref{eq:amplitudes single shock}): Amplitudes.
\item $\hF$ (\ref{eq:isometric linear map}): Isometric linear map $\mH_1\rightarrow \mH_0\otimes \mH_0$.
\item $\hat{\mathcal{C}}=\sum_nn\ket{K^{\Delta_i}_n}\bra{K^{\Delta_i}_n}$ (\ref{eq:Krylov operator complexity definition}): Krylov complexity operator.
\item $\mathcal{C}\equiv\frac{\Tr(\hat{\rho}_{\Delta_{i}}~\hat{\mathcal{C}})}{Z_{\Delta_i}(\beta/2,\beta/2)}$ (\ref{eq:Krylov operator complexity definition}): Krylov operator complexity.
\item $\hat{N}$ (\ref{eq:total chord number}): Total chord number with $m$-particle insertions.
\item Lanczos algorithm (\ref{eq:Lanczos precursor}):$$b_{n+1}^{L/R}\ket{K^{\Delta_i}_{n+1}}= (\hH_L-\hH_R)\ket{K_{n}}-a^{L/R}_n\ket{K^{\Delta_i}_{n}}-b_n\ket{K^{\Delta_i}_{n-1}}$$
\item $a_n$ and $b_n$ (\ref{eq:Lanczos shocwaves}): Lanczos coefficients.
\item $\ket{K_n^{\Delta_i}}$ (\ref{eq:explicit Krylov basis}): Krylov basis for the left and right chord sectors.
\item $t_{\rm sc}^{(i)}\eqlambda-\frac{1}{2J\sin\theta}\log\frac{\Delta_i\lambda}{2\sin\theta}$ (\ref{eq:scrambling time}): Scrambling time.
\item $\alpha_{i}=\qty(1-\frac{q^{\Delta_{i}}\rme^{-\ell_*(\theta)}}{\sin^2\theta})\frac{\rme^{2t_iJ\sin\theta}}{2}$ (\ref{eq:alpha sw}): Parameter in the boundary theory.
\item $\ell_*(\theta)$ (\ref{eq:length0 Delta many}): Initial length.
\item $\hat{\ell}(\tau_L,\tau_R)$, $\hat{P}_{0/m}(\tau_L,\tau_R)$
(\ref{eq:new operators Heisenberg wormhole}): Length and conjugate momentum operators in the Heisenberg picture, with complex time where real part indicates temperature, and the imaginary part corresponds to real time evolution.
\item $\hat{\ell}$, $\hat{P}_{0/m}$ (\ref{eq:equiv to H and S pictures}): operators in the Schrödinger picture.
\item $\ell(t_L,t_R)$, $P_{L/R}(t_L,t_R)$ (\ref{eq:solutions shockwave}): Saddle point solutions of the path integral (\ref{eq:Hamilton PI}).
\item $\hat{\mathcal{O}}^{L}_{\Delta_{i}}(t_i)=\rme^{\rmi t_i \hH_{L}}\hat{\mathcal{O}}^{L}_{\Delta_{i}}\rme^{-\rmi t_i\hH_{L}}$, $\hat{\mathcal{O}}^{R}_{\Delta_{i}}(t_i)=\rme^{-\rmi t_i \hH_{R}}\hat{\mathcal{O}}^{R}_{\Delta_{i}}\rme^{\rmi t_i\hH_{R}}$ (\ref{eq:double scaled precursor shockwave}): Double-scaled precursor operators.
\item $\tilde{t}_{L/R}\equiv J\sqrt{\frac{1-q}{\lambda}}t_{L/R}$ (\ref{eq:shockwave EOM}): Rescaled time in the equation of motion.
\item $\beta(\theta)=(2\theta-\pi)/(J\sin\theta)$ (\ref{eq:micro inv temp}): Inverse temperature for the thermodynamically stable saddle point solutions at order $O(\lambda^0)$ in the semi-classical limit.
\item $L_{\rm AdS}(u_L,u_R)$ (\ref{eq:AdS shockwave length}): AdS$_2$ black hole minimal geodesic length.
\item $u_{L/R}$ Asymptotic boundary times.
\item $L_{\rm reg}$ (\ref{eq:AdS shockwave length}): regularized asymptotic boundary length.
\item $\alpha_{\rm sw}$ (\ref{eq:AdS shockwave length}): Shockwave shift parameter.
\item $\beta_{\rm fake}=2\pi/(J\sin\theta)$ (\ref{eq:fake temperature}): Fake inverse temperature.
\item $\lambda_{\rm L}$ (\ref{eq:Lyapunov exp}): OTOC Lyapunov exponent.
\item $\mathcal{C}_V$ (\ref{eq:switchback property0}): CV conjecture.
\end{itemize}
\paragraph{Acronyms}
\begin{itemize}[noitemsep]
\item (A)dS: (Anti-)de Sitter
\item CAny: Complexity=Anything
\item CFT: Conformal field theory
\item CV: Complexity=Volume
\item (DS)SYK: double-scaled Sachdev–Ye–Kitaev model
\item HH: Hartle-Hawking
\item JT: Jackiw-Teitelboim
\item OTOC: Out-of-time-order correlator
\item SC: Scrambling
\item TFD: Thermofield double
\item UV: Ultraviolet
\end{itemize}

\global\long\def\bra{\langle}
\global\long\def\ket{\rangle}
\section{Details on the Coefficients in (\ref{eq:overlap})} \label{app:recursion}
In this section, we present a detailed derivation of \eqref{eq:overlap}. We first review the definition of the bi-variate generating function \( Q_{m,n} \) and express it in terms of expectation values of the entangler map \( \me_\Delta \), as explored in \cite{Okuyama:2024yya,xu2024chord}. We then evaluate \eqref{eq:overlap} by converting it into an operator expression where most of the orthogonal relations become manifest, demonstrating how the one-particle energy density naturally arises.

\subsection{The bi-variate generating function and its recursion relation}
The bi-variate generating function introduced in the main text is defined in terms of the \( q \)-Hermite polynomial as:
\be\label{App:Q-def}
Q_{n_{L},n_{R}}\left(\cos\te_{L},\cos\te_{R};q,q^{\Del}\right)=\sum_{p=0}^{\infty}\frac{q^{\Del p}}{\left(q;q\right)_{p}}\frac{H_{n_{L}+p}\left(\cos\te_{L};q\right)H_{n_{R}+p}\left(\cos\te_{R};q\right)}{(\te_L|q^{\Del\hat{n}}|\te_{R})}.
\ee 
The term $(\te_L|q^{\Del\hat{n}}|\te_{R})$ in the denominator is a normalization factor which is not important for the following discussion of recursion relation.  The \( q \)-Hermite polynomials satisfy the recursion relation:
\be\label{eq:qHermite rec}
2xH_{n}\left(x;q\right)=H_{n+1}\left(x;q\right)+(1-q^{n})H_{n-1}(x;q),\quad H_0(x;q)=1,\quad H_{1}(x;q)=2x.
\ee
From this definition, we verify that the \( Q \)-polynomials satisfy the recursion:
\be
2\cos\te_{L}Q_{n_{L},n_{R}}=Q_{n_{L}+1,n_{R}}+q^{\Del+n_{L}}Q_{n_{L},n_{R}+1}+\left(1-q^{n_{L}}\right)Q_{n_{L}-1,n_{R}}.
\ee 
For simplicity, we omit the explicit arguments of the polynomials in what follows, keeping only the subscript indices. Readers should keep in mind that \( H_{n_{L/R}} \) depends on \( \cos\te_{L/R} \), respectively. Multiplying \( 2\cos\te_L \) to both sides of \eqref{App:Q-def}, we obtain:
\be
2\cos\te_{L}\sum_{p=0}^{\infty}\frac{q^{\Del p}H_{n_{L}+p}H_{n_{R}+p}}{\left(q;q\right)_{p}(\te_L|q^{\Del\hat{n}}|\te_{R})}=\sum_{p=0}^{\infty}\frac{q^{\Del p}(H_{n_{L}+p+1}+(1-q^{p+n_{L}})H_{n_{L}+p-1})}{\left(q;q\right)_{p}(\te_L|q^{\Del\hat{n}}|\te_{R})}H_{n_{R}+p},
\ee
where we have applied the recursion relation for \( H_{n_L} \) (\ref{eq:qHermite rec}) on the right-hand side. The first term directly gives \( Q_{n_L+1,n_R} \). The coefficient in the second term inside the parentheses can be rewritten as:
\be
1-q^{n_L+p}= (1-q^{n_L}) + q^{n_L}(1-q^p).
\ee
The first term contributes:
\be
(1-q^{n_{L}})\sum_{p=0}^{\infty}\frac{q^{\Del p}}{\left(q;q\right)_{p}}\frac{H_{n_{L}+p-1}H_{n_{R}+p}}{(\te_{L}| q^{\Del\hat{n}}|\te_{R})}=(1-q^{n_{L}})Q_{n_{L}-1,n_{R}}~.
\ee
The second term contributes:
\be
\sum_{p=0}^{\infty}\frac{q^{\Del p}}{\left(q;q\right)_{p}}(q^{n_{L}}-q^{p+n_{L}})\frac{H_{n_{L}+p-1}H_{n_{R}+p}}{(\te_{L}| q^{\Del\hat{n}}|\te_{R})}	=q^{\Del+n_{L}}\sum_{p=1}^{\infty}\frac{q^{\Del(p-1)}}{\left(q;q\right)_{p-1}}\frac{H_{n_{L}+p-1}H_{n_{R}+p}}{(\te_{L}| q^{\Del\hat{n}}|\te_{R})}~.
\ee
Shifting \( p \to p+1 \) yields:
\be
\sum_{p=0}^{\infty}\frac{q^{\Del p}}{\left(q;q\right)_{p}}(q^{n_{L}}-q^{p+n_{L}})\frac{H_{n_{L}+p-1}H_{n_{R}+p}}{(\te_{L}| q^{\Del\hat{n}}|\te_{R})}= q^{\Delta+ n_L} Q_{n_L,n_{R}+1}.
\ee
Thus, we conclude:
\begin{align}
2\cos\te_{L}Q_{n_{L},n_{R}}=Q_{n_{L}+1,n_{R}}+q^{\Del+n_{L}}Q_{n_{L},n_{R}+1}+\left(1-q^{n_{L}}\right)Q_{n_{L}-1,n_{R}}~,\\
2\cos\te_{R}Q_{n_{L},n_{R}}=Q_{n_{L},n_{R}+1}+q^{\Del+n_{R}}Q_{n_{L}+1,n_{R}}+\left(1-q^{n_{R}}\right)Q_{n_{L},n_{R}-1}~.
\end{align}
\subsection{Calculating the Overlap}
The bi-variate generating function $\{Q_{m,n}\}$ and its complex conjugate can be represented compactly in terms of the intertwiner operator $\me$ operators as follows: 
\be\label{app:Q-def}
\begin{split}
&\frac{Q_{m_{L},m_{R}}\left(\cos\te_{L},\cos\te_{R}|q,q^{\Delta}\right)}{\sqrt{\left(q;q\right)_{m_{L}}\left(q;q\right)_{m_{R}}}}	=\frac{(m_{L},m_{R}|\me_{\Delta}|\theta_{L},\theta_{R})}{(\vp_L|q^{\Delta \hat{n}}|\vp_{R})}, \\ 
&\frac{Q_{n_{L},n_{R}}^{*}\left(\cos\vp_{L},\cos\vp_{R}|q,q^{\Delta}\right)}{\sqrt{\left(q;q\right)_{n_{L}}\left(q;q\right)_{n_{R}}}}	=\frac{(\te_{L},\te_{R}|\me_{\Delta}^{\da}|n_{L},n_{R})}{(\vp_L|q^{\Delta \hat{n}}|\vp_{R})},
\end{split}
\ee
where $*$ in the second equation means complex conjugation.  The sum over chord numbers can be implemented using the resolution of identity:
\be
\begin{split}
& \sum_{m_{L},m_{R}=0}^{\infty}(\te_{L},\te_{R}|\me_{\Del}^{-1}|m_{L},m_{R})(m_{L},m_{R}|\me_{\Del}|\vp_{L},\vp_{R})	=\frac{\del\left(\te_{L}-\vp_{L}\right)\del\left(\te_{R}-\vp_{R}\right)}{\mu\left(\te_{1}\right)\mu\left(\te_{2}\right)},\\
&\sum_{n_{L},n_{R}=0}^{\infty}(\te_{L},\te_{R}|\me_{\Del}^{\da}|n_{L},n_{R})(n_{L},n_{R}|\me_{\Del}^{\da-1}|\vp_{L},\vp_{R})	=\frac{\del\left(\vp_{L}-\te_{L}\right)\del\left(\vp_R-\te_{R}\right)}{\mu\left(\te_{1}\right)\mu\left(\te_{2}\right)}.
\end{split}
\ee
Thus, plugging into the definition of $|\Delta;\te_L,\te_R\rangle$, we conclude with \eqref{eq:inner-product-2}.

\section{Details on \texorpdfstring{$q$}{}-Calculus}\label{app:q-calculus}
We present some details in the derivation of results in the main text. Most of these derivations rely on the properties of $q$-coherent states and the
Askey-Wilson integral \cite{Askey1985SomeBH}. 
\paragraph{One-particle coherent states}
We review the derivation of the one-particle chord inner product. According to the isometric factorization map, the inner product between chord number eigenstates is given by
\begin{equation}
\langle \Delta; m_L, m_R | \Delta; n_L, n_R \rangle = \int_0^{\pi} d\mu(\theta_L) d\mu(\theta_R) (\theta_L | q^{\Delta \hat{n}} | \theta_R) \cdot I_{m_L, m_R; n_L, n_R}^{\Delta}(\theta_L, \theta_R)
\end{equation}
where
\begin{equation}
I_{m_L, m_R; n_L, n_R}^{\Delta}(\theta_L, \theta_R) = (m_L, m_R | \me_\Delta^{-\dagger} | \theta_L, \theta_R)(\theta_L, \theta_R | \me_\Delta^{-1} | n_L, n_R)~.
\end{equation}
is the overlap of states obtained via the factorization map with fixed boundary energy. The matter density is
\begin{equation}
(\theta_L | q^{\Delta \hat{n}} | \theta_R) = \sum_{k=0}^\infty \frac{q^{\Delta k}}{(q;q)_k} H_k(\cos \theta_L | q) H_k(\cos \theta_R | q) = \frac{(q^{2\Delta}; q)_\infty}{(q^{\Delta} \rme^{\pm i \theta_L \pm i \theta_R}; q)_\infty}.
\end{equation}
To diagonalize \( \me_\Delta^{-1} \), we introduce coherent states:
\begin{equation}
|s_L, s_R) := \sum_{n_L, n_R = 0}^\infty \frac{s_L^{n_L} s_R^{n_R}}{\sqrt{(q;q)_{n_L} (q;q)_{n_R}}} |n_L, n_R),
\end{equation}
which satisfy \( \ha_{L/R} |s_L, s_R) = s_{L/R} |s_L, s_R) \), and hence
\begin{equation}
\me_\Delta^{-1} |s_L, s_R) = (q^{\Delta} s_L s_R; q)_\infty |s_L, s_R).
\end{equation}
The overlap with energy eigenstates follows from the generating function of \( q \)-Hermite polynomials:
\begin{equation}
(\theta | s) = \sum_{n=0}^\infty \frac{s^n H_n(\cos \theta | q)}{(q;q)_n} = \frac{1}{(s e^{\pm i \theta}; q)_\infty}.
\end{equation}
We define
\begin{align}
I_{s_L, s_R; t_L, t_R}^{\Delta}(\theta_L, \theta_R) &= \sum_{m_L, m_R, n_L, n_R = 0}^\infty \frac{s_L^{m_L} s_R^{m_R} t_L^{n_L} t_R^{n_R}}{\sqrt{(q;q)_{m_L}(q;q)_{m_R}(q;q)_{n_L}(q;q)_{n_R}}} I_{m_L, m_R; n_L, n_R}^{\Delta} \\
&= (s_L, s_R | \me_\Delta^{-\dagger} | \theta_L, \theta_R)(\theta_L, \theta_R | \me_\Delta^{-1} | t_L, t_R),
\end{align}
which evaluates to
\begin{equation}
I_{s_L, s_R; t_L, t_R}^{\Delta}(\theta_L, \theta_R) = \frac{(q^{\Delta} s_L s_R, q^{\Delta} t_L t_R; q)_\infty}{(s_L e^{\pm i \theta_L}, t_L e^{\pm i \theta_L}, s_R e^{\pm i \theta_R}, t_R e^{\pm i \theta_R}; q)_\infty}.
\end{equation}
Using the Askey-Wilson integral \cite{Ismail:1987}, we compute:
\begin{align}\label{eq:AW integral}
&\int d\mu(\theta_R) \frac{(q^{2\Delta};q)_\infty}{(q^{\Delta} e^{\pm i \theta_L \pm i \theta_R}, s_R e^{\pm i \theta_R}, t_R e^{\pm i \theta_R}; q)_\infty} = \frac{(q^{2\Delta} s_R t_R; q)_\infty}{(q^{\Delta} s_R e^{\pm i \theta_L}, q^{\Delta} t_R e^{\pm i \theta_L}, s_R t_R; q)_\infty}~,\\
&\int d\mu(\theta_L) \frac{(q^{2\Delta} s_R t_R, q^{\Delta} s_L s_R, q^{\Delta} t_L t_R; q)_\infty}{(q^{\Delta} s_R e^{\pm i \theta_L}, q^{\Delta} t_R e^{\pm i \theta_L}, s_L e^{\pm i \theta_L}, t_L e^{\pm i \theta_L}; q)_\infty} = \frac{(q^{2\Delta} s_L s_R t_L t_R; q)_\infty}{(s_L t_L, q^{\Delta} s_R t_L, q^{\Delta} s_L t_R; q)_\infty}~.
\end{align}
Thus,
\begin{equation}
\langle \Delta; s_L, s_R | \Delta; t_L, t_R \rangle = \frac{(q^{2\Delta} s_L s_R t_L t_R; q)_\infty}{(q^{\Delta} s_R t_L, q^{\Delta} s_L t_R, s_L t_L, s_R t_R; q)_\infty},
\end{equation}
where the coherent one-particle states are defined as
\begin{equation}
|\Delta; t_L, t_R \rangle = \sum_{n_L, n_R = 0}^\infty \frac{t_L^{n_L} t_R^{n_R}}{\sqrt{(q;q)_{n_L} (q;q)_{n_R}}} |\Delta; n_L, n_R \rangle.
\end{equation}
Expanding the coherent inner product in powers of \( s_{L/R}, t_{L/R} \), the coefficients yield the inner products between chord number eigenstates.

\paragraph{The \( 6j \) Symbol in Coherent Basis} \label{app:6j-derivation}
We compute the \( 6j \)-symbol in the coherent basis here. Let
\begin{equation}
|s) = \sum_{n=0}^{\infty} \frac{s^n}{\sqrt{(q;q)_n}} |n),
\end{equation}
be the coherent state in \( \mathcal{H}_0 \). One can show its overlap with the energy eigenstate is given by
\begin{equation}
(\theta|s) = \sum_{n=0}^\infty \frac{s^n H_n(\cos\theta; q)}{(q;q)_n} = \frac{1}{(s e^{\pm i \theta}; q)} = (s|\theta),
\end{equation}
where in the last equality we have assumed \( s \) to be real. Combined with \eqref{eq:6j chord number basis} and \eqref{eq:6j symbol coherent state basis}, we find
\begin{equation}
\left\{ \begin{array}{ccc}
\Delta_{1} & s_1 & s_2 \\
\Delta_{2} & s_4 & s_3
\end{array} \right\}_q
= \int \prod_{i=1}^{4} d\mu(\theta_i)\, (s_1, s_2 | \theta_1, \theta_2)\, \Gamma^{\Delta_1} \Gamma^{\Delta_2}
\left\{ \begin{array}{ccc}
\Delta_{1} & \theta_1 & \theta_2 \\
\Delta_{2} & \theta_3 & \theta_4
\end{array} \right\}_q (\theta_3, \theta_4 | s_3, s_4).
\end{equation}
We evaluate the integration over \( \theta_3, \theta_4 \) first, where
\begin{equation}
\begin{aligned}
I_1 &= \int d\mu(\theta_3)\, d\mu(\theta_4)\,
\left\{ \begin{array}{ccc}
\Delta_{1} & \theta_1 & \theta_2 \\
\Delta_{2} & \theta_3 & \theta_4
\end{array} \right\}_q (\theta_3, \theta_4 | s_3, s_4) \\
&= (\theta_1 | q^{\Delta_1 \hat{n}} | \theta_2)
(\theta_1, \theta_2 | \me_{\Delta_1}^{-1} q^{\Delta_2(\hat{n}_1 + \hat{n}_2)} \me_{\Delta_1} | s_3, s_4)~.
\end{aligned}
\end{equation}
In the second line, we have used the operator expression \eqref{eq:6j energy basis} and integrated out \( \theta_3, \theta_4 \), which is just a resolution of identity. Now, using the commutation relation, we find
\begin{equation}
\me_{\Delta_1}^{-1} q^{\Delta_2(\hat{n}_1 + \hat{n}_2)} \me_{\Delta_1}
= q^{\Delta_2(\hat{n}_1 + \hat{n}_2)} \frac{(q^{2\Delta_2 + \Delta_1} a_L a_R; q)_\infty}{(q^{\Delta_1} a_L a_R; q)_\infty}~.
\end{equation}
The second factor is diagonal in the coherent basis, namely $\ha_{L/R}|s_L,s_R)=s_{L/R} |s_L,s_R)$; therefore, we find
\begin{equation}
I_1 = (\theta_1 | q^{\Delta_1 \hat{n}} | \theta_2)
\frac{(q^{2\Delta_2 + \Delta_1} s_3 s_4; q)_\infty}{(q^{\Delta_2} s_3 e^{\pm i \theta_1}, q^{\Delta_2} s_4 e^{\pm i \theta_2}, q^{\Delta_1} s_3 s_4; q)_\infty}.
\end{equation}
Now, we proceed to evaluate the remaining integral
\begin{equation}
\left\{ \begin{array}{ccc}
\Delta_{1} & s_1 & s_2 \\
\Delta_{2} & s_4 & s_3
\end{array} \right\}_q = \int_{0}^{\pi} d\mu(\theta_1)\, d\mu(\theta_2)\, I_1.
\end{equation}
We find that this amounts to
\begin{equation}
\begin{aligned}
\left\{ \begin{array}{ccc}
\Delta_{1} & s_1 & s_2 \\
\Delta_{2} & s_4 & s_3
\end{array} \right\}_q
= \int&\frac{d\mu(\theta_1)\, d\mu(\theta_2)(q^{2\Delta_2 + \Delta_1} s_3 s_4; q)_\infty}{(q^{\Delta_2} s_3 e^{\pm i \theta_1}, q^{\Delta_2} s_4 e^{\pm i \theta_2}, q^{\Delta_1} s_3 s_4; q)_\infty} \frac{(q^{2\Delta_1}; q)_\infty}{(q^{\Delta_1} e^{\pm i \theta_1 \pm i \theta_2}; q)_\infty}\cdot\\
& \quad \cdot \frac{1}{(s_1 e^{\pm i \theta_1}, s_2 e^{\pm i \theta_2}; q)_\infty},
\end{aligned}
\end{equation}
where the integral can be performed using the Askey–Wilson integral. Integrating over \( \theta_1 \) leads to
\begin{equation}
\begin{aligned}
\left\{ \begin{array}{ccc}
\Delta_{1} & s_1 & s_2 \\
\Delta_{2} & s_4 & s_3
\end{array} \right\}_q
= \int_{0}^{\pi} &d\mu(\theta_2)\,
\frac{1}{(q^{\Delta_1 + \Delta_2} s_3 e^{\pm i \theta_2},
q^{\Delta_1} s_1 e^{\pm i \theta_2},
q^{\Delta_2} s_4 e^{\pm i \theta_2},
s_2 e^{\pm i \theta_2}; q)_\infty}, \\
& \quad \cdot
\frac{(q^{\Delta_2 + 2\Delta_1} s_1 s_3, q^{2\Delta_2 + \Delta_1} s_3 s_4; q)_\infty}{(q^{\Delta_2} s_1 s_3, q^{\Delta_1} s_3 s_4; q)_\infty}~.
\end{aligned}
\end{equation}
Further integrating over \( \theta_2 \), we conclude that
\begin{equation}
\left\{ \begin{array}{ccc}
\Delta_{1} & s_1 & s_2 \\
\Delta_{2} & s_4 & s_3
\end{array} \right\}_q
= \frac{(q^{2\Delta_1 + 2\Delta_2} s_1 s_2 s_3 s_4; q)_\infty}
{(q^{2\Delta_1 + \Delta_2} s_1 s_3,
q^{\Delta_1 + 2\Delta_2} s_3 s_4,
q^{\Delta_1 + \Delta_2} s_3 s_2,
q^{\Delta_1 + \Delta_2} s_1 s_4,
q^{\Delta_1} s_1 s_2,
q^{\Delta_2} s_2 s_4; q)_\infty}.
\end{equation}

\bibliographystyle{JHEP}
\bibliography{references.bib}

@article{Belin:2021bga,
    author = "Belin, Alexandre and Myers, Robert C. and Ruan, Shan-Ming and S\'arosi, G\'abor and Speranza, Antony J.",
    title = "{Does Complexity Equal Anything?}",
    eprint = "2111.02429",
    archivePrefix = "arXiv",
    primaryClass = "hep-th",
    reportNumber = "CERN-TH-2021-181, YITP-22-02",
    doi = "10.1103/PhysRevLett.128.081602",
    journal = "Phys. Rev. Lett.",
    volume = "128",
    number = "8",
    pages = "081602",
    year = "2022"
}

@article{Penington:2023dql,
    author = "Penington, Geoff and Witten, Edward",
    title = "{Algebras and States in JT Gravity}",
    eprint = "2301.07257",
    archivePrefix = "arXiv",
    primaryClass = "hep-th",
    month = "1",
    year = "2023"
}

@article{Ambrosini:2024sre,
    author = "Ambrosini, Marco and Rabinovici, Eliezer and S\'anchez-Garrido, Adri\'an and Shir, Ruth and Sonner, Julian",
    title = "{Operator K-complexity in DSSYK: Krylov complexity equals bulk length}",
    eprint = "2412.15318",
    archivePrefix = "arXiv",
    primaryClass = "hep-th",
    month = "12",
    year = "2024"
}

@article{Heller:2024ldz,
    author = "Heller, Michal P. and Papalini, Jacopo and Schuhmann, Tim",
    title = "{Krylov spread complexity as holographic complexity beyond JT gravity}",
    eprint = "2412.17785",
    archivePrefix = "arXiv",
    primaryClass = "hep-th",
    month = "12",
    year = "2024"
}

@article{Blommaert:2025avl,
    author = "Blommaert, Andreas and Levine, Adam and Mertens, Thomas G. and Papalini, Jacopo and Parmentier, Klaas",
    title = "{Wormholes, branes and finite matrices in sine dilaton gravity}",
    eprint = "2501.17091",
    archivePrefix = "arXiv",
    primaryClass = "hep-th",
    month = "1",
    year = "2025"
}

@article{Parker:2018yvk,
    author = "Parker, Daniel E. and Cao, Xiangyu and Avdoshkin, Alexander and Scaffidi, Thomas and Altman, Ehud",
    title = "{A Universal Operator Growth Hypothesis}",
    eprint = "1812.08657",
    archivePrefix = "arXiv",
    primaryClass = "cond-mat.stat-mech",
    doi = "10.1103/PhysRevX.9.041017",
    journal = "Phys. Rev. X",
    volume = "9",
    number = "4",
    pages = "041017",
    year = "2019"
}

@article{Balasubramanian:2022tpr,
    author = "Balasubramanian, Vijay and Caputa, Pawel and Magan, Javier M. and Wu, Qingyue",
    title = "{Quantum chaos and the complexity of spread of states}",
    eprint = "2202.06957",
    archivePrefix = "arXiv",
    primaryClass = "hep-th",
    doi = "10.1103/PhysRevD.106.046007",
    journal = "Phys. Rev. D",
    volume = "106",
    number = "4",
    pages = "046007",
    year = "2022"
}

@article{Nandy:2024htc,
    author = "Nandy, Pratik and Matsoukas-Roubeas, Apollonas S. and Mart\'\i{}nez-Azcona, Pablo and Dymarsky, Anatoly and del Campo, Adolfo",
    title = "{Quantum Dynamics in Krylov Space: Methods and Applications}",
    eprint = "2405.09628",
    archivePrefix = "arXiv",
    primaryClass = "quant-ph",
    reportNumber = "RIKEN-iTHEMS-Report-24",
    month = "5",
    year = "2024"
}

@article{Blommaert:2024whf,
    author = "Blommaert, Andreas and Levine, Adam and Mertens, Thomas G. and Papalini, Jacopo and Parmentier, Klaas",
    title = "{An entropic puzzle in periodic dilaton gravity and DSSYK}",
    eprint = "2411.16922",
    archivePrefix = "arXiv",
    primaryClass = "hep-th",
    month = "11",
    year = "2024"
}

@article{Lin:2023trc,
    author = "Lin, Henry W. and Stanford, Douglas",
    title = "{A symmetry algebra in double-scaled SYK}",
    eprint = "2307.15725",
    archivePrefix = "arXiv",
    primaryClass = "hep-th",
    doi = "10.21468/SciPostPhys.15.6.234",
    journal = "SciPost Phys.",
    volume = "15",
    number = "6",
    pages = "234",
    year = "2023"
}

@article{Verlinde:2024znh,
    author = "Verlinde, Herman",
    title = "{Double-scaled SYK, chords and de Sitter gravity}",
    eprint = "2402.00635",
    archivePrefix = "arXiv",
    primaryClass = "hep-th",
    doi = "10.1007/JHEP03(2025)076",
    journal = "JHEP",
    volume = "03",
    pages = "076",
    year = "2025"
}

@article{Narovlansky:2025tpb,
    author = "Narovlansky, Vladimir",
    title = "{Towards a microscopic description of de Sitter dynamics}",
    eprint = "2506.02109",
    archivePrefix = "arXiv",
    primaryClass = "hep-th",
    month = "6",
    year = "2025"
}

@article{Maldacena:2015waa,
    author = "Maldacena, Juan and Shenker, Stephen H. and Stanford, Douglas",
    title = "{A bound on chaos}",
    eprint = "1503.01409",
    archivePrefix = "arXiv",
    primaryClass = "hep-th",
    doi = "10.1007/JHEP08(2016)106",
    journal = "JHEP",
    volume = "08",
    pages = "106",
    year = "2016"
}

@article{Blommaert:2025rgw,
    author = "Blommaert, Andreas and Levine, Adam",
    title = "{Sphere amplitudes and observing the universe's size}",
    eprint = "2505.24633",
    archivePrefix = "arXiv",
    primaryClass = "hep-th",
    month = "5",
    year = "2025"
}

@article{Yuan:2024utc,
    author = "Yuan, Haiming and Ge, Xian-Hui and Kim, Keun-Young",
    title = "{Pole-skipping in two-dimensional de Sitter spacetime and double-scaled SYK model in the classical limit}",
    eprint = "2408.12330",
    archivePrefix = "arXiv",
    primaryClass = "hep-th",
    month = "8",
    year = "2024"
}

@article{Mahapatra:2025fpx,
    author = "Mahapatra, Paramesh and Rathi, Hemant and Roychowdhury, Dibakar",
    title = "{Holographic central charge for double scaled SYK}",
    eprint = "2502.13884",
    archivePrefix = "arXiv",
    primaryClass = "hep-th",
    month = "2",
    year = "2025"
}

@article{Brown:2015bva,
    author = "Brown, Adam R. and Roberts, Daniel A. and Susskind, Leonard and Swingle, Brian and Zhao, Ying",
    title = "{Holographic Complexity Equals Bulk Action?}",
    eprint = "1509.07876",
    archivePrefix = "arXiv",
    primaryClass = "hep-th",
    doi = "10.1103/PhysRevLett.116.191301",
    journal = "Phys. Rev. Lett.",
    volume = "116",
    number = "19",
    pages = "191301",
    year = "2016"
}

@article{Brown:2015lvg,
    author = "Brown, Adam R. and Roberts, Daniel A. and Susskind, Leonard and Swingle, Brian and Zhao, Ying",
    title = "{Complexity, action, and black holes}",
    eprint = "1512.04993",
    archivePrefix = "arXiv",
    primaryClass = "hep-th",
    doi = "10.1103/PhysRevD.93.086006",
    journal = "Phys. Rev. D",
    volume = "93",
    number = "8",
    pages = "086006",
    year = "2016"
}

@article{Almheiri:2024xtw,
    author = "Almheiri, Ahmed and Goel, Akash and Hu, Xu-Yao",
    title = "{Quantum gravity of the Heisenberg algebra}",
    eprint = "2403.18333",
    archivePrefix = "arXiv",
    primaryClass = "hep-th",
    doi = "10.1007/JHEP08(2024)098",
    journal = "JHEP",
    volume = "08",
    pages = "098",
    year = "2024"
}

@article{Xu:2019lhc,
    author = "Xu, Tianrui and Scaffidi, Thomas and Cao, Xiangyu",
    title = "{Does scrambling equal chaos?}",
    eprint = "1912.11063",
    archivePrefix = "arXiv",
    primaryClass = "cond-mat.stat-mech",
    doi = "10.1103/PhysRevLett.124.140602",
    journal = "Phys. Rev. Lett.",
    volume = "124",
    number = "14",
    pages = "140602",
    year = "2020"
}

@article{Couch:2016exn,
    author = "Couch, Josiah and Fischler, Willy and Nguyen, Phuc H.",
    title = "{Noether charge, black hole volume, and complexity}",
    eprint = "1610.02038",
    archivePrefix = "arXiv",
    primaryClass = "hep-th",
    reportNumber = "UTTG-16-16",
    doi = "10.1007/JHEP03(2017)119",
    journal = "JHEP",
    volume = "03",
    pages = "119",
    year = "2017"
}

@article{Chapman:2021jbh,
    author = "Chapman, Shira and Policastro, Giuseppe",
    title = "{Quantum computational complexity from quantum information to black holes and back}",
    eprint = "2110.14672",
    archivePrefix = "arXiv",
    primaryClass = "hep-th",
    doi = "10.1140/epjc/s10052-022-10037-1",
    journal = "Eur. Phys. J. C",
    volume = "82",
    number = "2",
    pages = "128",
    year = "2022"
}

@article{Myers:2024vve,
    author = "Myers, Robert C. and Ruan, Shan-Ming",
    title = "{Complexity Equals (Almost) Anything}",
    eprint = "2403.17475",
    archivePrefix = "arXiv",
    primaryClass = "hep-th",
    reportNumber = "YITP-24-34",
    month = "3",
    year = "2024"
}

@article{Sekino:2025bsc,
    author = "Sekino, Yasuhiro and Susskind, Leonard",
    title = "{Double-Scaled SYK, QCD, and the Flat Space Limit of de Sitter Space}",
    eprint = "2501.09423",
    archivePrefix = "arXiv",
    primaryClass = "hep-th",
    month = "1",
    year = "2025"
}

@article{Aguilar-Gutierrez:2024oea,
    author = "Aguilar-Gutierrez, Sergio E.",
    title = "{$T^2$ deformations in the double-scaled SYK model: Stretched horizon thermodynamics}",
    eprint = "2410.18303",
    archivePrefix = "arXiv",
    primaryClass = "hep-th",
    month = "10",
    year = "2024"
}

@article{Faruk:2025bed,
    author = "Faruk, Mir Mehedi and Rost, Facundo and van der Schaar, Jan Pieter",
    title = "{Quasinormal Modes and the Switchback Effect in Schwarzschild-de Sitter}",
    eprint = "2501.01388",
    archivePrefix = "arXiv",
    primaryClass = "hep-th",
    month = "1",
    year = "2025"
}

@article{Narovlansky:2023lfz,
    author = "Narovlansky, Vladimir and Verlinde, Herman",
    title = "{Double-scaled SYK and de Sitter holography}",
    eprint = "2310.16994",
    archivePrefix = "arXiv",
    primaryClass = "hep-th",
    doi = "10.1007/JHEP05(2025)032",
    journal = "JHEP",
    volume = "05",
    pages = "032",
    year = "2025"
}

@article{JACKIW1985343,
title = {Lower dimensional gravity},
journal = {Nuclear Physics B},
volume = {252},
pages = {343-356},
year = {1985},
issn = {0550-3213},
doi = {https://doi.org/10.1016/0550-3213(85)90448-1},
url = {https://www.sciencedirect.com/science/article/pii/0550321385904481},
author = {Roman Jackiw}
}

@article{Rahman:2024iiu,
    author = "Rahman, Adel A. and Susskind, Leonard",
    title = "{$p$-Chords, Wee-Chords, and de Sitter Space}",
    eprint = "2407.12988",
    archivePrefix = "arXiv",
    primaryClass = "hep-th",
    month = "7",
    year = "2024"
}

@article{Rahman:2024vyg,
    author = "Rahman, Adel A. and Susskind, Leonard",
    title = "{Infinite Temperature is Not So Infinite: The Many Temperatures of de Sitter Space}",
    eprint = "2401.08555",
    archivePrefix = "arXiv",
    primaryClass = "hep-th",
    month = "1",
    year = "2024"
}

@article{Susskind:2022bia,
    author = "Susskind, Leonard",
    title = "{De Sitter Space, Double-Scaled SYK, and the Separation of Scales in the Semiclassical Limit}",
    eprint = "2209.09999",
    archivePrefix = "arXiv",
    primaryClass = "hep-th",
    month = "9",
    year = "2022"
}

@article{Milekhin:2023bjv,
    author = "Milekhin, Alexey and Xu, Jiuci",
    title = "{Revisiting Brownian SYK and its possible relations to de Sitter}",
    eprint = "2312.03623",
    archivePrefix = "arXiv",
    primaryClass = "hep-th",
    month = "12",
    year = "2023"
}

@article{Okuyama:2025hsd,
    author = "Okuyama, Kazumi",
    title = "{De Sitter JT gravity from double-scaled SYK}",
    eprint = "2505.08116",
    archivePrefix = "arXiv",
    primaryClass = "hep-th",
    month = "5",
    year = "2025"
}

@article{Verlinde:2024zrh,
    author = "Verlinde, Herman and Zhang, Mengyang",
    title = "{SYK Correlators from 2D Liouville-de Sitter Gravity}",
    eprint = "2402.02584",
    archivePrefix = "arXiv",
    primaryClass = "hep-th",
    month = "2",
    year = "2024"
}

@article{Aguilar-Gutierrez:2025pqp,
    author = "Aguilar-Gutierrez, Sergio E.",
    title = "{Building the Holographic Dictionary of the DSSYK from Chords, Complexity {\&} Wormholes with Matter}",
    eprint = "2505.22716",
    archivePrefix = "arXiv",
    primaryClass = "hep-th",
    month = "5",
    year = "2025"
}

@article{Chandrasekaran:2022eqq,
    author = "Chandrasekaran, Venkatesa and Penington, Geoff and Witten, Edward",
    title = "{Large N algebras and generalized entropy}",
    eprint = "2209.10454",
    archivePrefix = "arXiv",
    primaryClass = "hep-th",
    doi = "10.1007/JHEP04(2023)009",
    journal = "JHEP",
    volume = "04",
    pages = "009",
    year = "2023"
}

@article{Goel:2018ubv,
    author = "Goel, Akash and Lam, Ho Tat and Turiaci, Gustavo J. and Verlinde, Herman",
    title = "{Expanding the Black Hole Interior: Partially Entangled Thermal States in SYK}",
    eprint = "1807.03916",
    archivePrefix = "arXiv",
    primaryClass = "hep-th",
    doi = "10.1007/JHEP02(2019)156",
    journal = "JHEP",
    volume = "02",
    pages = "156",
    year = "2019"
}

@article{Tietto:2025oxn,
    author = "Tietto, Damiano and Verlinde, Herman",
    title = "{A microscopic model of de Sitter spacetime with an observer}",
    eprint = "2502.03869",
    archivePrefix = "arXiv",
    primaryClass = "hep-th",
    month = "2",
    year = "2025"
}

@article{Dong:2016eik,
    author = "Dong, Xi and Harlow, Daniel and Wall, Aron C.",
    title = "{Reconstruction of Bulk Operators within the Entanglement Wedge in Gauge-Gravity Duality}",
    eprint = "1601.05416",
    archivePrefix = "arXiv",
    primaryClass = "hep-th",
    reportNumber = "NSF-KITP-16-005",
    doi = "10.1103/PhysRevLett.117.021601",
    journal = "Phys. Rev. Lett.",
    volume = "117",
    number = "2",
    pages = "021601",
    year = "2016"
}

@article{Chapman:2021eyy,
    author = "Chapman, Shira and Galante, Dami\'an A. and Kramer, Eric David",
    title = "{Holographic complexity and de Sitter space}",
    eprint = "2110.05522",
    archivePrefix = "arXiv",
    primaryClass = "hep-th",
    doi = "10.1007/JHEP02(2022)198",
    journal = "JHEP",
    volume = "02",
    pages = "198",
    year = "2022"
}

@article{Auzzi:2023qbm,
    author = "Auzzi, Roberto and Nardelli, Giuseppe and Ungureanu, Gabriel Pedde and Zenoni, Nicolo",
    title = "{Volume complexity of dS bubbles}",
    eprint = "2302.03584",
    archivePrefix = "arXiv",
    primaryClass = "hep-th",
    reportNumber = "OU-HET-1171",
    doi = "10.1103/PhysRevD.108.026006",
    journal = "Phys. Rev. D",
    volume = "108",
    number = "2",
    pages = "026006",
    year = "2023"
}

@article{Jorstad:2022mls,
    author = "J\o{}rstad, Eivind and Myers, Robert C. and Ruan, Shan-Ming",
    title = "{Holographic complexity in dS$_{d+1}$}",
    eprint = "2202.10684",
    archivePrefix = "arXiv",
    primaryClass = "hep-th",
    reportNumber = "YITP-22-15",
    doi = "10.1007/JHEP05(2022)119",
    journal = "JHEP",
    volume = "05",
    pages = "119",
    year = "2022"
}

@article{Aguilar-Gutierrez:2023tic,
    author = "Aguilar-Gutierrez, Sergio E. and Patra, Ayan K. and Pedraza, Juan F.",
    title = "{Entangled universes in dS wedge holography}",
    eprint = "2308.05666",
    archivePrefix = "arXiv",
    primaryClass = "hep-th",
    reportNumber = "IFT-UAM/CSIC-23-95",
    doi = "10.1007/JHEP10(2023)156",
    journal = "JHEP",
    volume = "10",
    pages = "156",
    year = "2023"
}

@article{Aguilar-Gutierrez:2023zqm,
    author = "Aguilar-Gutierrez, Sergio E. and Heller, Michal P. and Van der Schueren, Silke",
    title = "{Complexity equals anything can grow forever in de Sitter space}",
    eprint = "2305.11280",
    archivePrefix = "arXiv",
    primaryClass = "hep-th",
    doi = "10.1103/PhysRevD.110.066009",
    journal = "Phys. Rev. D",
    volume = "110",
    number = "6",
    pages = "066009",
    year = "2024"
}

@article{Baiguera:2024xju,
    author = "Baiguera, Stefano and Berman, Rotem",
    title = "{The cosmological switchback effect. Part II}",
    eprint = "2406.04397",
    archivePrefix = "arXiv",
    primaryClass = "hep-th",
    doi = "10.1007/JHEP08(2024)086",
    journal = "JHEP",
    volume = "08",
    pages = "086",
    year = "2024"
}

@article{Aguilar-Gutierrez:2024rka,
    author = "Aguilar-Gutierrez, Sergio E. and Baiguera, Stefano and Zenoni, Nicol\'o",
    title = "{Holographic complexity of the extended Schwarzschild-de Sitter space}",
    eprint = "2402.01357",
    archivePrefix = "arXiv",
    primaryClass = "hep-th",
    reportNumber = "OU-HET-1220",
    doi = "10.1007/JHEP05(2024)201",
    journal = "JHEP",
    volume = "05",
    pages = "201",
    year = "2024"
}

@article{Baiguera:2023tpt,
    author = "Baiguera, Stefano and Berman, Rotem and Chapman, Shira and Myers, Robert C.",
    title = "{The cosmological switchback effect}",
    eprint = "2304.15008",
    archivePrefix = "arXiv",
    primaryClass = "hep-th",
    doi = "10.1007/JHEP07(2023)162",
    journal = "JHEP",
    volume = "07",
    pages = "162",
    year = "2023"
}

@article{Caputa:2024sux,
    author = "Caputa, Pawel and Chen, Bowen and McDonald, Ross W. and Sim\'on, Joan and Strittmatter, Benjamin",
    title = "{Spread Complexity Rate as Proper Momentum}",
    eprint = "2410.23334",
    archivePrefix = "arXiv",
    primaryClass = "hep-th",
    reportNumber = "YITP-24-137",
    month = "10",
    year = "2024"
}

@article{Caputa:2025dep,
    author = "Caputa, Pawel and Di Giulio, Giuseppe",
    title = "{Local Quenches from a Krylov Perspective}",
    eprint = "2502.19485",
    archivePrefix = "arXiv",
    primaryClass = "hep-th",
    reportNumber = "YITP-25-26",
    month = "2",
    year = "2025"
}

@article{Aguilar-Gutierrez:2024nau,
    author = "Aguilar-Gutierrez, Sergio E.",
    title = "{Towards complexity in de Sitter space from the double-scaled Sachdev-Ye-Kitaev model}",
    eprint = "2403.13186",
    archivePrefix = "arXiv",
    primaryClass = "hep-th",
    doi = "10.1007/JHEP10(2024)107",
    journal = "JHEP",
    volume = "10",
    pages = "107",
    year = "2024"
}

@article{Gaiotto:2024kze,
    author = "Gaiotto, Davide and Verlinde, Herman",
    title = "{SYK-Schur duality: Double scaled SYK correlators from $N=2$ supersymmetric gauge theory}",
    eprint = "2409.11551",
    archivePrefix = "arXiv",
    primaryClass = "hep-th",
    month = "9",
    year = "2024"
}

@misc{HVTalks1,
author= "Verlinde, H.",
title= "{SYK gravity duality and 3D de Sitter holography}",
howpublished = {\url{ https://qmap.ucdavis.edu/qiqg5-scientific-program}},
note = {Talk given at the ‘QGQC5 conference’, UC Davis, U.S.A.} ,
month=" August",
year="2019"
  }

@misc{HVTalks2,
author= "Verlinde, H.",
title= "{ Comments on SYK and 3D de Sitter Holography }",
note = {Talk given at the ‘Franqui Symposium’, Brussels, Belgium} ,
month=" November",
year="2019"
  }

@misc{HVTalks3,
author= "Verlinde, H.",
title= "{SYK Gravity Duality and 3D de Sitter Holography }",
note = {Talk given at the ‘Quantum Gravity on Southern Cone’ conference, Argentina} ,
month=" December",
year="2019"
  }

@misc{HVTalks4,
author= "Verlinde, H.",
title= "{A Duality between SYK and 3D de Sitter}",
note = {Talk given at the ‘SYK models and Gauge Theory’ workshop, Weizmann Institute, Israel } ,
month="December",
year="2019"
  }

@article{Milekhin:2024vbb,
    author = "Milekhin, Alexey and Xu, Jiuci",
    title = "{On scrambling, tomperature and superdiffusion in de Sitter space}",
    eprint = "2403.13915",
    archivePrefix = "arXiv",
    primaryClass = "hep-th",
    month = "3",
    year = "2024"
}

@article{Rahman:2023pgt,
    author = "Rahman, Adel A. and Susskind, Leonard",
    title = "{Comments on a Paper by Narovlansky and Verlinde}",
    eprint = "2312.04097",
    archivePrefix = "arXiv",
    primaryClass = "hep-th",
    month = "12",
    year = "2023"
}

@article{Rahman:2022jsf,
    author = "Rahman, Adel A.",
    title = "{dS JT Gravity and Double-Scaled SYK}",
    eprint = "2209.09997",
    archivePrefix = "arXiv",
    primaryClass = "hep-th",
    month = "9",
    year = "2022"
}

@article{Lin:2022nss,
    author = "Lin, Henry and Susskind, Leonard",
    title = "{Infinite Temperature's Not So Hot}",
    eprint = "2206.01083",
    archivePrefix = "arXiv",
    primaryClass = "hep-th",
    month = "6",
    year = "2022"
}

@article{Berkooz:2018qkz,
    author = "Berkooz, Micha and Narayan, Prithvi and Simon, Joan",
    title = "{Chord diagrams, exact correlators in spin glasses and black hole bulk reconstruction}",
    eprint = "1806.04380",
    archivePrefix = "arXiv",
    primaryClass = "hep-th",
    doi = "10.1007/JHEP08(2018)192",
    journal = "JHEP",
    volume = "08",
    pages = "192",
    year = "2018"
}

@article{Rabinovici:2023yex,
    author = "Rabinovici, E. and S\'anchez-Garrido, A. and Shir, R. and Sonner, J.",
    title = "{A bulk manifestation of Krylov complexity}",
    eprint = "2305.04355",
    archivePrefix = "arXiv",
    primaryClass = "hep-th",
    doi = "10.1007/JHEP08(2023)213",
    journal = "JHEP",
    volume = "08",
    pages = "213",
    year = "2023"
}

@article{Baiguera:2025dkc,
    author = "Baiguera, Stefano and Balasubramanian, Vijay and Caputa, Pawel and Chapman, Shira and Haferkamp, Jonas and Heller, Michal P. and Halpern, Nicole Yunger",
    title = "{Quantum complexity in gravity, quantum field theory, and quantum information science}",
    eprint = "2503.10753",
    archivePrefix = "arXiv",
    primaryClass = "hep-th",
    reportNumber = "YITP-25-39",
    month = "3",
    year = "2025"
}

@article{Blommaert:2023opb,
    author = "Blommaert, Andreas and Mertens, Thomas G. and Yao, Shunyu",
    title = "{Dynamical actions and q-representation theory for double-scaled SYK}",
    eprint = "2306.00941",
    archivePrefix = "arXiv",
    primaryClass = "hep-th",
    doi = "10.1007/JHEP02(2024)067",
    journal = "JHEP",
    volume = "02",
    pages = "067",
    year = "2024"
}

@article{TEITELBOIM198341,
title = {Gravitation and hamiltonian structure in two spacetime dimensions},
journal = {Physics Letters B},
volume = {126},
number = {1},
pages = {41-45},
year = {1983},
issn = {0370-2693},
doi = {https://doi.org/10.1016/0370-2693(83)90012-6},
url = {https://www.sciencedirect.com/science/article/pii/0370269383900126},
author = {Claudio Teitelboim}
}

@article{Goel:2023svz,
    author = "Goel, Akash and Narovlansky, Vladimir and Verlinde, Herman",
    title = "{Semiclassical geometry in double-scaled SYK}",
    eprint = "2301.05732",
    archivePrefix = "arXiv",
    primaryClass = "hep-th",
    doi = "10.1007/JHEP11(2023)093",
    journal = "JHEP",
    volume = "11",
    pages = "093",
    year = "2023"
}

@article{Susskind:2021esx,
    author = "Susskind, Leonard",
    title = "{Entanglement and Chaos in De Sitter Space Holography: An SYK Example}",
    eprint = "2109.14104",
    archivePrefix = "arXiv",
    primaryClass = "hep-th",
    doi = "10.22128/jhap.2021.455.1005",
    journal = "JHAP",
    volume = "1",
    number = "1",
    pages = "1--22",
    year = "2021"
}

@article{Berkooz:2024lgq,
    author = "Berkooz, Micha and Mamroud, Ohad",
    title = "{A cordial introduction to double scaled SYK}",
    eprint = "2407.09396",
    archivePrefix = "arXiv",
    primaryClass = "hep-th",
    doi = "10.1088/1361-6633/ada889",
    journal = "Rept. Prog. Phys.",
    volume = "88",
    number = "3",
    pages = "036001",
    year = "2025"
}

@article{Greenberger:1989tfe,
    author = "Greenberger, Daniel M. and Horne, Michael A. and Zeilinger, Anton",
    title = "{Going Beyond Bell\textquoteright{}s Theorem}",
    eprint = "0712.0921",
    archivePrefix = "arXiv",
    primaryClass = "quant-ph",
    doi = "10.1007/978-94-017-0849-4_10",
    journal = "Fundam. Theor. Phys.",
    volume = "37",
    pages = "69--72",
    year = "1989"
}

@article{Gesteau:2024rpt,
    author = "Gesteau, Elliott and Liu, Hong",
    title = "{Toward stringy horizons}",
    eprint = "2408.12642",
    archivePrefix = "arXiv",
    primaryClass = "hep-th",
    reportNumber = "MIT-CTP/5751",
    month = "8",
    year = "2024"
}

@article{Borchers:2000pv,
    author = "Borchers, H. J.",
    title = "{On revolutionizing quantum field theory with Tomita's modular theory}",
    doi = "10.1063/1.533323",
    journal = "J. Math. Phys.",
    volume = "41",
    pages = "3604--3673",
    year = "2000"
}

@article{Kolchmeyer:2023gwa,
    author = "Kolchmeyer, David K.",
    title = "{von Neumann algebras in JT gravity}",
    eprint = "2303.04701",
    archivePrefix = "arXiv",
    primaryClass = "hep-th",
    doi = "10.1007/JHEP06(2023)067",
    journal = "JHEP",
    volume = "06",
    pages = "067",
    year = "2023"
}

@article{Gu_2019,
   title={On the relation between the magnitude and exponent of OTOCs},
   volume={2019},
   ISSN={1029-8479},
   url={http://dx.doi.org/10.1007/JHEP02(2019)075},
   DOI={10.1007/jhep02(2019)075},
   number={2},
   journal={Journal of High Energy Physics},
   publisher={Springer Science and Business Media LLC},
   author={Gu, Yingfei and Kitaev, Alexei},
   year={2019},
   month=feb }

@article{Berkooz:2024ifu,
    author = "Berkooz, Micha and Frumkin, Ronny and Mamroud, Ohad and Seitz, Josef",
    title = "{Twisted times, the Schwarzian and its deformations in DSSYK}",
    eprint = "2412.14238",
    archivePrefix = "arXiv",
    primaryClass = "hep-th",
    doi = "10.1007/JHEP05(2025)080",
    journal = "JHEP",
    volume = "05",
    pages = "080",
    year = "2025"
}

@article{Berkooz:2022mfk,
    author = "Berkooz, Micha and Isachenkov, Misha and Isachenkov, Mikhail and Narayan, Prithvi and Narovlansky, Vladimir",
    title = "{Quantum groups, non-commutative AdS$_{2}$, and chords in the double-scaled SYK model}",
    eprint = "2212.13668",
    archivePrefix = "arXiv",
    primaryClass = "hep-th",
    doi = "10.1007/JHEP08(2023)076",
    journal = "JHEP",
    volume = "08",
    pages = "076",
    year = "2023"
}

@article{Lam_2018,
   title={Shockwave S-matrix from Schwarzian quantum mechanics},
   volume={2018},
   ISSN={1029-8479},
   url={http://dx.doi.org/10.1007/JHEP11(2018)182},
   DOI={10.1007/jhep11(2018)182},
   number={11},
   journal={Journal of High Energy Physics},
   publisher={Springer Science and Business Media LLC},
   author={Lam, Ho Tat and Mertens, Thomas G. and Turiaci, Gustavo J. and Verlinde, Herman},
   year={2018},
   month=nov }

@article{Belin:2022xmt,
    author = "Belin, Alexandre and Myers, Robert C. and Ruan, Shan-Ming and S\'arosi, G\'abor and Speranza, Antony J.",
    title = "{Complexity equals anything II}",
    eprint = "2210.09647",
    archivePrefix = "arXiv",
    primaryClass = "hep-th",
    reportNumber = "CERN-TH-2022-159; YITP-22-101",
    doi = "10.1007/JHEP01(2023)154",
    journal = "JHEP",
    volume = "01",
    pages = "154",
    year = "2023"
}

@article{xu2024chord,
    author = "Xu, Jiuci",
    title = "{On Chord Dynamics and Complexity Growth in Double-Scaled SYK}",
    eprint = "2411.04251",
    archivePrefix = "arXiv",
    primaryClass = "hep-th",
    month = "11",
    year = "2024"
}

@article{Yang18,
    author = "Yang, Zhenbin",
    title = "{The Quantum Gravity Dynamics of Near Extremal Black Holes}",
    eprint = "1809.08647",
    archivePrefix = "arXiv",
    primaryClass = "hep-th",
    doi = "10.1007/JHEP05(2019)205",
    journal = "JHEP",
    volume = "05",
    pages = "205",
    year = "2019"
}

@misc{lin2022hot,
      title={Infinite Temperature's Not So Hot}, 
      author={Henry Lin and Leonard Susskind},
      year={2022},
      eprint={2206.01083},
      archivePrefix={arXiv},
      primaryClass={hep-th},
      url={https://arxiv.org/abs/2206.01083}, 
}

@article{Iliesiu:2024cnh,
    author = "Iliesiu, Luca V. and Levine, Adam and Lin, Henry W. and Maxfield, Henry and Mezei, M\'ark",
    title = "{On the non-perturbative bulk Hilbert space of JT gravity}",
    eprint = "2403.08696",
    archivePrefix = "arXiv",
    primaryClass = "hep-th",
    doi = "10.1007/JHEP10(2024)220",
    journal = "JHEP",
    volume = "10",
    pages = "220",
    year = "2024"
}

@article{Ismail:1987,
    author = "Mourad E.H. Ismail and Dennis Stanton and Gérard Viennot",
    title = "{The Combinatorics of q-Hermite polynomials and the Askey—Wilson Integral}",
    journal = "European Journal of Combinatorics",
    year = "1987",
    Number = "4",
    Pages = "379-392",
    Doi= "https://doi.org/10.1016/S0195-6698(87)80046-X",
    Issn = "0195-6698",
    Url = "https://www.sciencedirect.com/science/article/pii/S019566988780046X"
}

@article{Okuyama:2024yya,
    author = "Okuyama, Kazumi",
    title = "{Doubled Hilbert space in double-scaled SYK}",
    eprint = "2401.07403",
    archivePrefix = "arXiv",
    primaryClass = "hep-th",
    month = "1",
    year = "2024"
}

@article{Xu:2024gfm,
    author = "Xu, Jiuci",
    title = "{On Chord Dynamics and Complexity Growth in Double-Scaled SYK}",
    eprint = "2411.04251",
    archivePrefix = "arXiv",
    primaryClass = "hep-th",
    month = "11",
    year = "2024"
}

@article{Berkooz:2018jqr,
    author = "Berkooz, Micha and Isachenkov, Mikhail and Narovlansky, Vladimir and Torrents, Genis",
    title = "{Towards a full solution of the large N double-scaled SYK model}",
    eprint = "1811.02584",
    archivePrefix = "arXiv",
    primaryClass = "hep-th",
    doi = "10.1007/JHEP03(2019)079",
    journal = "JHEP",
    volume = "03",
    pages = "079",
    year = "2019"
}

@article{Stanford:2014jda,
    author = "Stanford, Douglas and Susskind, Leonard",
    title = "{Complexity and Shock Wave Geometries}",
    eprint = "1406.2678",
    archivePrefix = "arXiv",
    primaryClass = "hep-th",
    doi = "10.1103/PhysRevD.90.126007",
    journal = "Phys. Rev. D",
    volume = "90",
    number = "12",
    pages = "126007",
    year = "2014"
}

@article{Blommaert:2024ydx,
    author = "Blommaert, Andreas and Mertens, Thomas G. and Papalini, Jacopo",
    title = "{The dilaton gravity hologram of double-scaled SYK}",
    eprint = "2404.03535",
    archivePrefix = "arXiv",
    primaryClass = "hep-th",
    month = "4",
    year = "2024"
}

@article{Shenker:2013pqa,
    author = "Shenker, Stephen H. and Stanford, Douglas",
    title = "{Black holes and the butterfly effect}",
    eprint = "1306.0622",
    archivePrefix = "arXiv",
    primaryClass = "hep-th",
    reportNumber = "SU-ITP-13-08",
    doi = "10.1007/JHEP03(2014)067",
    journal = "JHEP",
    volume = "03",
    pages = "067",
    year = "2014"
}

@article{Haehl:2021tft,
    author = "Haehl, Felix M. and Streicher, Alexandre and Zhao, Ying",
    title = "{Six-point functions and collisions in the black hole interior}",
    eprint = "2105.12755",
    archivePrefix = "arXiv",
    primaryClass = "hep-th",
    doi = "10.1007/JHEP08(2021)134",
    journal = "JHEP",
    volume = "08",
    pages = "134",
    year = "2021"
}

@article{Askey1985SomeBH,
  title={Some basic hypergeometric orthogonal polynomials that generalize Jacobi polynomials},
  author={Richard A. Askey and James A. Wilson},
  journal={Memoirs of the American Mathematical Society},
  year={1985},
  volume={54},
  pages={0-0},
  url={https://api.semanticscholar.org/CorpusID:120050879}
}

@article{Lin:2022rbf,
    author = "Lin, Henry W.",
    title = "{The bulk Hilbert space of double scaled SYK}",
    eprint = "2208.07032",
    archivePrefix = "arXiv",
    primaryClass = "hep-th",
    doi = "10.1007/JHEP11(2022)060",
    journal = "JHEP",
    volume = "11",
    pages = "060",
    year = "2022"
}

@article{Berkooz20,
   title={The double scaled limit of super-symmetric SYK models},
   volume={2020},
   ISSN={1029-8479},
   url={http://dx.doi.org/10.1007/JHEP12(2020)110},
   DOI={10.1007/jhep12(2020)110},
   number={12},
   journal={Journal of High Energy Physics},
   publisher={Springer Science and Business Media LLC},
   author={Berkooz, Micha and Brukner, Nadav and Narovlansky, Vladimir and Raz, Amir},
   year={2020},
   month=dec }

@article{Okuyama:2024more,
    author = "Okuyama, Kazumi",
    title = "{More on doubled Hilbert space in double-scaled SYK}",
    eprint = "2404.02833",
    archivePrefix = "arXiv",
    primaryClass = "hep-th",
    doi = "10.1016/j.physletb.2024.138858",
    journal = "Phys. Lett. B",
    volume = "855",
    pages = "138858",
    year = "2024"
}

@article{Xu:2024hoc,
    author = "Xu, Jiuci",
    title = "{Von Neumann Algebras in Double-Scaled SYK}",
    eprint = "2403.09021",
    archivePrefix = "arXiv",
    primaryClass = "hep-th",
    month = "3",
    year = "2024"
}

@article{Susskind:2014rva,
    author = "Susskind, Leonard",
    title = "{Computational Complexity and Black Hole Horizons}",
    eprint = "1403.5695",
    archivePrefix = "arXiv",
    primaryClass = "hep-th",
    doi = "10.1002/prop.201500092",
    journal = "Fortsch. Phys.",
    volume = "64",
    pages = "24--43",
    year = "2016",
    note = "[Addendum: Fortsch.Phys. 64, 44--48 (2016)]"
}

@article{jamiolkowski1972linear,
  title={Linear transformations which preserve trace and positive semidefiniteness of operators},
  author={Jamio{\l}kowski, Andrzej},
  journal={Reports on mathematical physics},
  volume={3},
  number={4},
  pages={275--278},
  year={1972},
  publisher={Elsevier}
}

@article{choi1975completely,
  title={Completely positive linear maps on complex matrices},
  author={Choi, Man-Duen},
  journal={Linear algebra and its applications},
  volume={10},
  number={3},
  pages={285--290},
  year={1975},
  publisher={Elsevier}
}

@article{Berkooz:2022fso,
    author = "Berkooz, Micha and Brukner, Nadav and Ross, Simon F. and Watanabe, Masataka",
    title = "{Going beyond ER=EPR in the SYK model}",
    eprint = "2202.11381",
    archivePrefix = "arXiv",
    primaryClass = "hep-th",
    doi = "10.1007/JHEP08(2022)051",
    journal = "JHEP",
    volume = "08",
    pages = "051",
    year = "2022"
}

@article{Shenker:2013yza,
    author = "Shenker, Stephen H. and Stanford, Douglas",
    title = "{Multiple Shocks}",
    eprint = "1312.3296",
    archivePrefix = "arXiv",
    primaryClass = "hep-th",
    reportNumber = "SU-ITP-13-24",
    doi = "10.1007/JHEP12(2014)046",
    journal = "JHEP",
    volume = "12",
    pages = "046",
    year = "2014"
}
\end{document}